\documentclass[journal]{IEEEtran}
\ifCLASSINFOpdf
\else
\fi
\usepackage{graphicx}
\usepackage[english]{babel}
\usepackage{blindtext}
\usepackage{algorithm}
\usepackage{algpseudocode}
\usepackage{bm}
\usepackage{multirow}
\usepackage{amsmath}
\usepackage{mathtools}
\usepackage{mathrsfs}
\usepackage{theorem}
\usepackage{amsfonts,amssymb}
\usepackage{multicol}

\usepackage{extarrows}

\newcommand\lncs{\text{LNCS}}
\newcommand\crypto{\text{CRYPTO}~}
\newcommand\eurocrypt{\text{EUROCRYPT}~}
\newcommand\auscrypt{\text{AUSCRYPT}~}
\newcommand\asiacrypt{\text{ASIACRYPT}~}
\newcommand\tcc{\text{TCC}~}
\newcommand\fse{\text{FSE}~}
\newcommand\ches{\text{CHES}~}

\newcommand\sac{\text{SAC}~}

\newcommand\indocrypt{\text{INDOCRYPT}~}
\newcommand\arrangespace{\vspace{+1em}}
\newcommand\arrangelittlespace{\vspace{+0.5em}}
\newcommand\codeindent{\ \ \ }

\newcommand\e{\textsf{E}}
\newcommand\kac{\textsf{KAC}}

\newcommand\kaf{\textsf{KAF}}
\newcommand\kafsg{\ensuremath{\kaf^{f,\ga}}\xspace}
\newcommand\kafsp{\ensuremath{\kaf^{P,\ga}}\xspace}
\newcommand\kafsf{\ensuremath{\kaf^{F,\ga}}\xspace}
\newcommand\kafv{\textsf{KAFv}}
\newcommand\lucifer{\textsf{Luc}}
\newcommand\kafvsg{\ensuremath{\kafv^{f,\ga^*}}\xspace}
\newcommand\kafvsp{\ensuremath{\kafv^{P,\ga^*}}\xspace}
\newcommand\kafvsf{\ensuremath{\kafv^{F,\ga^*}}\xspace}
\newcommand\kafw{\textsf{KAFw}}
\newcommand\kafwsg{\ensuremath{\kafw^{f,(\wf,\ga)}}\xspace}
\newcommand\kafwsp{\ensuremath{\kafw^{P,(\wf,\ga)}}\xspace}
\newcommand\kafwsf{\ensuremath{\kafw^{F,(\wf,\ga)}}\xspace}
\newcommand\rk{\textsf{RK}}
\newcommand\D{\Delta}
\newcommand\kd{k_{\D}}
\newcommand\Zn{\{0,1\}^n}
\newcommand\xor{\oplus}

\newcommand\ga{\gamma}
\newcommand\wf{w}    % KDF for the whitening keys
\newcommand\sample{\xleftarrow{\$}}   % to simply the command.

\newcommand\functionset{\mathcal{F}(n)}
\newcommand\permutationset{\mathcal{P}(n)}
\newcommand\blockcipherset{\mathcal{BC}(n,2n)}

\newcommand\bft{\mathbf{t}}   % to represent a "T"uple

\newcommand\badf{\textsf{B}}
\newcommand\badfi{\textsf{B1}}
\newcommand\badfii{\textsf{B2}}
\newcommand\con{\ensuremath{CON}}

\newcommand\xornotation{\textsf{XOR}}

\newcommand\dom{\mathcal{X}(\tau)}
\newcommand\rng{\mathcal{Y}(\tau)}

\newtheorem{theorem}{Theorem}
\newtheorem{lemma}{Lemma}
\newtheorem{corollary}{Corollary}

\newtheorem{definition}{Definition}

\renewcommand{\IEEEQED}{\IEEEQEDopen}

\usepackage[pdfstartview=FitH,
bookmarksnumbered=true,bookmarksopen=true,
colorlinks=true,pdfborder=001,citecolor=blue,
linkcolor=blue,urlcolor=blue]{hyperref}

\renewcommand\theenumi{\roman{enumi}}
\renewcommand\labelenumi{(\theenumi)}

\begin{document}
\title{Understanding the Related-Key Security of\\Feistel Ciphers from a Provable Perspective}
%\thanks{In submission to IEEE. Copyright may be transferred without notice, after which this version may no longer be accessible.}


\author{Chun Guo % <-this % stops a space
	\thanks{Chun Guo is with the Department
		of ICTEAM/ELEN/Crypto Group, Universit\'{e} catholique de Louvain, Louvain-la-Neuve e-mail: (chun.guo.sc@gmail.com). Copyright (c) 2017 IEEE. Personal use of this material is permitted.  However, permission to use this material for any other purposes must be obtained from the IEEE by sending a request to pubs-permissions@ieee.org.}% <-this % stops a space
	% <-this % stops a space
	\thanks{Manuscript received May xx, 2018; revised xxxxxx.}}


\markboth{Journal of \LaTeX\ Class Files,~Vol.~14, No.~8, August~2018}%
{Shell \MakeLowercase{\textit{et al.}}: Bare Demo of IEEEtran.cls for IEEE Journals}



\maketitle

\begin{abstract}
	We initiate the provable related-key security treatment for	models of \emph{practical} Feistel ciphers. In detail, we consider
	Feistel networks with four whitening keys
	$\wf_i(k)$, $i=0,1,2,3$, and round-functions of the form
	$f(\ga_j(k)\oplus X)$, where $k$ is the master-key, $\wf_i$ and $\ga_j$ are efficient transformations, and $f$ is a \emph{public} ideal function or permutation accessible by the adversary.
	We investigate key-schedule conditions that are sufficient for security against XOR-induced related-key attacks up to
	$2^{n/2}$ adversarial queries. When the key-schedules are
	\emph{non-linear}, we prove security for 4 rounds. When only
	\emph{affine} key-schedules are used, we prove security for 6
	rounds. These also imply secure tweakable Feistel ciphers in
	the Random Oracle model.

	By shuffling the key-schedules, our model unifies both the
	DES-like structure (known as \emph{Feistel-2} scheme in the cryptanalytic community, a.k.a. \emph{key-alternating Feistel} due
	to Lampe and Seurin, FSE 2014) and the Lucifer-like model
	(previously analyzed by Guo and Lin, TCC 2015). This allows us
	to derive concrete implications on these two (more
	common) models, and helps understanding their related-key security difference.
\end{abstract}

\begin{IEEEkeywords}
blockcipher, provable security, indistinguishability, related-key, Feistel cipher, key-alternating paradigm.
\end{IEEEkeywords}

% \fi
\IEEEpeerreviewmaketitle

\section{Introduction}
\label{section:introduction}

%\IEEEPARstart{F}eistel-like
Feistel-like blockciphers consist of several iterative applications of a simple Feistel
permutation
\begin{align}
\Phi_{G_{k_i}}(W_L\|W_R)=W_R\|W_L\xor{G_{k_i}(W_R)}
\label{eq:defn-luby-rackoff-round}
\end{align}
for a keyed function $G:\{0,1\}^{\kappa}\times\Zn\rightarrow\Zn$ on
$n$-bit strings, yielding a $2n$-bit blockcipher~\cite{1451934}. Such ciphers
and their generalizations constitute a half proportion of
modern blockciphers, including some most popular designs such as
DES~\cite{DESDesign}, Lucifer~\cite{LuciferDesign}, GOST~\cite{GOSTDesign}, and NSA's SIMON
family~\cite{simonspeckDesign}. This has made it the object of a very large (and still increasing) amount of analyses.

In information-theoretic model, the round-function $G$ would
be assumed somewhat random. Without additional hardness
assumption, provable security is limited to {\it at most} $2^n$ queries~\cite{patarin04}, which is much smaller than $2^{2n}$, the domain-size of the Feistel ciphers. Despite this limitation as well as the gap between the strong assumption on $G$ and the weak round-functions
in practical ciphers, this approach excludes any possibility
of generic attacks and supplies insights into the cipher
structures. Therefore, it has found applications in both Feistel ciphers~\cite{LRFeistel88,patarin04,GFSCRYPTO10,KAFFSE2014,RKAFeistelFSE2014,roicequalJoC2014} and their counterpart Key-Alternating Ciphers (\kac
s)~\cite{KAprpEC12,KAtightboundEC2014,KAtightboundCrypto2016,IffSKA5CRYTO17}.\footnote{\kac s are blockciphers that alternatively apply key-additions and keyless permutations, i.e., $\kac_{k_0,k_1,\ldots,k_t}^{P_1,\ldots,P_t}(M)=k_t\xor P_t(\ldots(k_1\xor P_1(k_0\xor M)))$.}

%%\subsubsection{Related-Key Attacks.}
%%
%%Related-key attacks were introduced in early

Related-Key Attacks (RKAs) were independently introduced by Biham~\cite{Biham94RKAttack} and Knudsen~\cite{Knu92RKLOKI} in early 1990s, and was later formalized by Bellare and Kohno~\cite{rksecurity03}. In this setting, the adversary is allowed to query the blockcipher under multiple secret keys that satisfy adversary-chosen relations. The presence of such related-keys may be the consequence of a protocol-level
key update~\cite{IK043gpp}, or the user key being tampered  by
fault injections~\cite{DBLP:conf/spw/AndersonK97}. The adversarial goal is to either recover the secret key(s), or to distinguish the related-key oracles from independent random permutations~\cite{rksecurity03}.

%%%\subsubsection{Related-Key Attacks (RKAs)}
%%%
%%%were introduced in early
%%%1990s~\cite{Biham94RKAttack,Knu92RKLOKI}, and soon became one
%%%of the main topic of the community. In this setting, the
%%%adversary is allowed to query the blockcipher under keys that
%%%satisfy chosen relations to the targeted secret key, and the
%%%goal is to either recover the secret key, or to distinguish the
%%%cipher oracles from independent random
%%%permutations~\cite{rksecurity03}.

%%
%%In fact, in this setting, remarkable weakness has been found on
%%a lot of non-reduced blockciphers, with the past standard
%%DES~\cite{complementingDES1976} and 3-DES, the current standard
%%AES-192 and AES-256~\cite{rkAESAsiacrypt09}, the Russian
%%standard
%%GOST~\cite{complementingFSE2010,complementFeistelFSE2013}, and
%%the 3GPP encryption algorithm KASUMI~\cite{rkKASUMIJoc2014} as
%%examples. Sometimes RKA weakness results in marvelous attacks
%%on high-level symmetric primitives, e.g. forgery attack on
%%3-DES-based RMAC~\cite{KK03RMAC}.

RKAs can be classified according to the adversary-chosen relations between the keys.
Likely, the most important category is the so-called XOR-induced Related-Key Attack
($\xor$-RKA)~\cite{seqiffSKA4EUROCRYPT15}, i.e., RKA that
allows the adversary to XOR any constant of its choice to the
secret user key. Such RKAs are important for at least three
reasons. First, they arise naturally in a number of contexts,
such as the f8 and f9 protocols of the 3GPP
standard~\cite{IK043gpp}. Second, from a theoretical point of
view, they are the simplest kind of attacks to have the
completeness property~\cite{GL10}, namely, for any keys
$k,k'\in\Zn$, there exists $\D\in\Zn$ such that $k\xor\D=k'$.

Last---but most importantly,---$\xor$-RKAs are the most relevant to cryptanalytic practice. Most practical ciphers mix the keys into the state via the XOR operation. As commented in~\cite{KHP12IT}, for such targets $\xor$-RKAs are inherent to the majority of differential-based attacks, as XOR key-relations leave the chance of canceling the state difference with the (chosen) round-key difference (this phenomena was named {\it local collision}~\cite{rkAESAsiacrypt09}) and extending differentials without decreasing their probabilities. Due to this, $\xor$-RKAs have been the most widely used attack model in symmetric cryptanalysis (as another example, the powerful related-key boomerang and rectangle attacks were in the $\xor$-RKA form when firstly introduced~\cite{rkBoomerangEC2005}). And they have given rise to a plenty of prominent results, including very efficient (distinguishing) attacks on many Feistel ciphers that will be mentioned in the next subsection, a practical-time attack on the 3GPP encryption algorithm KASUMI~\cite{rkKASUMIJoc2014}, and a forgery attack on 3-DES-based RMAC~\cite{KK03RMAC}. And their variants break full AES-192 and AES-256~\cite{rkAESAsiacrypt09} and 10-round AES-256 in practical-time~\cite{rkAESEurocrypt10}.\footnote{These variants assumed XORing constants into the round-keys, and are thus called {\it related sub-key attacks}.} The mentioned attack on RMAC is also a notable example of RKA weakness resulting in more disastrous attacks on high-level primitives, showing that pursuing RKA security is not purely theoretical.

\arrangelittlespace

%%%---actually, it was commented in~\cite{KHP12IT}	that XOR relations are common and inherent to the majority of differential-based related-key attacks.

%%%The so-called XOR-induced Related-Key Attack ($\xor$-RKA)~\cite{seqiffSKA4EUROCRYPT15}, i.e., RKA that allows the adversary to XOR any constant of its choice to the secret user key,

%In this paper, we ask if Feistel ciphers could be provably
%secure against XOR-induced Related-Key Attacks ($\xor$-RKA)?
%This question is obviously fundamental due to the widely
%use of Feistel ciphers, and due to the deep influence of
%$\xor$-RKAs (as mentioned).
%Hopefully, the answers may help design RKA-secure Feistel
%ciphers, and further tweakable blockciphers based on Feistel
%networks (as there is a strong relation between RKA-secure
%ciphers and tweakable ciphers~\cite{seqiffSKA4EUROCRYPT15}).

\noindent\textbf{Our Question.}
With the above, $\xor$-RKAs deserve special attention on the theoretical side. Recall that such a provable security requires that with a secret key $k$, the $q$ blockcipher instances $\e_{k\xor\D_1},\ldots,\e_{k\xor\D_q}$ queried by the attacker with distinct chosen constants $\D_1,\ldots,\D_q$ are indistinguishable from $q$ independent random permutations. Such security has been established for \kac s~\cite{seqiffSKA4EUROCRYPT15,FP15RKEM} and their tweakable variants~\cite{XPXTweakableEMBartIC16}. It's then natural to ask: under which conditions could Feistel ciphers be provably secure against $\xor$-RKAs?

In fact, to a large extent, our motivation also stems from practice: certain structural features cause remarkable $\xor$-RKA
weakness in a lot of Feistel ciphers in reality. The most
well-known example must be the complementation property in
DES~\cite{complementingDES1976}, i.e. $\text{DES}_{\overline{k}}(\overline{M})=\overline{\text{DES}_k(M)}$， where $\overline{X}$ is the bit-by-bit complementation of $X$. This non-random behavior also exists in its variants 3-DES~\cite{barker2004recommendation} and DESL~\cite{DESLDesign}. This not only cinches efficient related-key distinguishers on
DES, but also reduces its effective key-length by 1 bit in the
traditional single-key attack setting. Although appearing
harmless, it has been long asked how to overcome~\cite{Davies1981DES}. Other marvelous examples include $\xor$-RKAs
on GOST with \emph{very low complexity} described in~\cite{Ko2004} and~\cite{complementFeistelFSE2013}, and \emph{very efficient} distinguisher on the SHA-3 candidate
based on {\it Lesamnta}~\cite{complementingFSE2010}. In all, it
appears that the components (e.g. key-schedules) of Feistel
ciphers have to be carefully designed in order to achieve RKA
security. This is sharply contrast to the \kac~model, for which even the simplest idea $k\xor P(k\xor P(k\xor P(k\xor M)))$ already buys some level of
security (see~\cite{FP15RKEM}). A better understanding of Feistel ciphers in the RKA setting is thus crucial.

%%Indeed, we have not noticed
%%such {\it severe} RKA weakness in any practical \kac s (attacks
%%in e.g.~\cite{rkAESAsiacrypt09} require much higher
%%complexities).

We have noticed two works that partially addressed our
question. The first work of Barbosa and Farshim proved that the famous Luby-Rackoff model with round-keys rightfully reused is RKA secure~\cite{RKAFeistelFSE2014}. Such models are Feistel
networks using a pseudorandom function (PRF) $G_{k_i}$ as the round-function~\cite{LRFeistel88}, and have been extensively studied, with~\cite{patarin04} and~\cite{GFSCRYPTO10} to name a few. Unfortunately, this model overlooks many structural properties, e.g. the complementation property, and this leaves a huge gap between model and reality. In addition, it's arguably too strong to model the round-function as a PRF secure against RKAs---while the practice-motivated model $G_{k_i}(W_R)=f(k_i\xor W_R)$ may be a PRF when $f$ is not too weak, it's {\it never} an RKA-secure PRF. A comprehensive discussion is given later in page \pageref{page:comparison-to-BF14}. In all, in the RKA setting, Luby-Rackoff results appear less convincing.   \label{page:insufficiency-for-luby-rackoff}

The second work of Guo and Lin
proved that a Lucifer-like Feistel structure (will be clarified
	later: see Eq. (\ref{eq:round-function-of-kafv}), or Eq. (\ref{eq:lucifer-entire}) in Appendix \ref{subsec:lucifer-to-kafv}) could be indifferentiable from ideal
ciphers~\cite{iffKAFTCC15}, which implies $\xor$-RKA security
by~\cite{FP15RKEM}. But their extremely weak bound $q^{30}/2^n$ appears meaningless.

With these considerations, we'd like to bridge theory and reality: we'd like to {\it find a model that could well capture the
	structural features}---including the known RKA weakness---of
practical Feistel ciphers, and {\it then study under which
	constraints the model could achieve $\xor$-RKA security}.
Hopefully, this will serve invaluable insights, and help
address the challenge of designing RKA secure Feistel
ciphers---and further tweakable Feistel ciphers, as RKA-secure
ciphers and tweakable blockciphers~\cite{LRWTBCJoC11} are strongly
related~\cite{rksecurity03}.

\arrangelittlespace

\noindent\textbf{A Unified Model for Feistel Ciphers in Reality.}
Practical Feistel ciphers usually employ keyless transformations
for round-functions, and mix the keys into the structure via
efficient group operations (usually xor). In addition,
whitening keys may be used. This naturally motivates modeling
the keyless round-functions as {\it public} (random) functions
or permutations $f_i$, {\it explicitly xoring the round-keys somewhere}, and eventually adding whitening keys.

In detail, we consider Feistel networks in which the state at
round $i$ is updated according to
\begin{align}
W_L\|W_R\mapsto W_R\|W_L\xor \underline{f_i(k_i\xor W_R)},
\label{eq:round-function-of-kafw}
\end{align}
and four $n$-bit whitening keys $(wk_0,wk_1,wk_2,wk_3)$ are used. Among
them, $wk_0\|wk_1$ is used as the pre-whitening key, while
$wk_2\|wk_3$ is the post-whitening key. Its special case \emph{without whitening keys} was named {\it
Key-Alternating Feistel (\kaf)} by Lampe and
Seurin~\cite{KAFFSE2014}. Thus we name our model {\it
Key-Alternating Feistel with Whitening keys (\kafw)}.

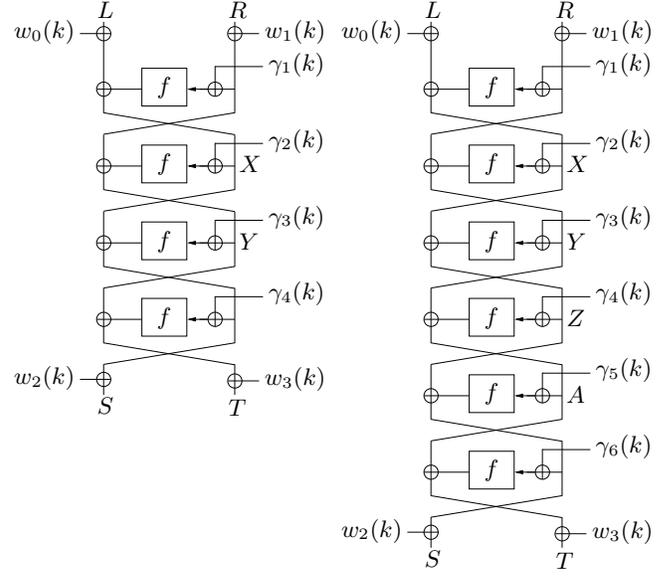
\begin{figure}
	\centering
	\setlength{\unitlength}{1bp}%
	\begin{picture}(252.39, 227.27)(0,0)
	\put(0,0){\includegraphics{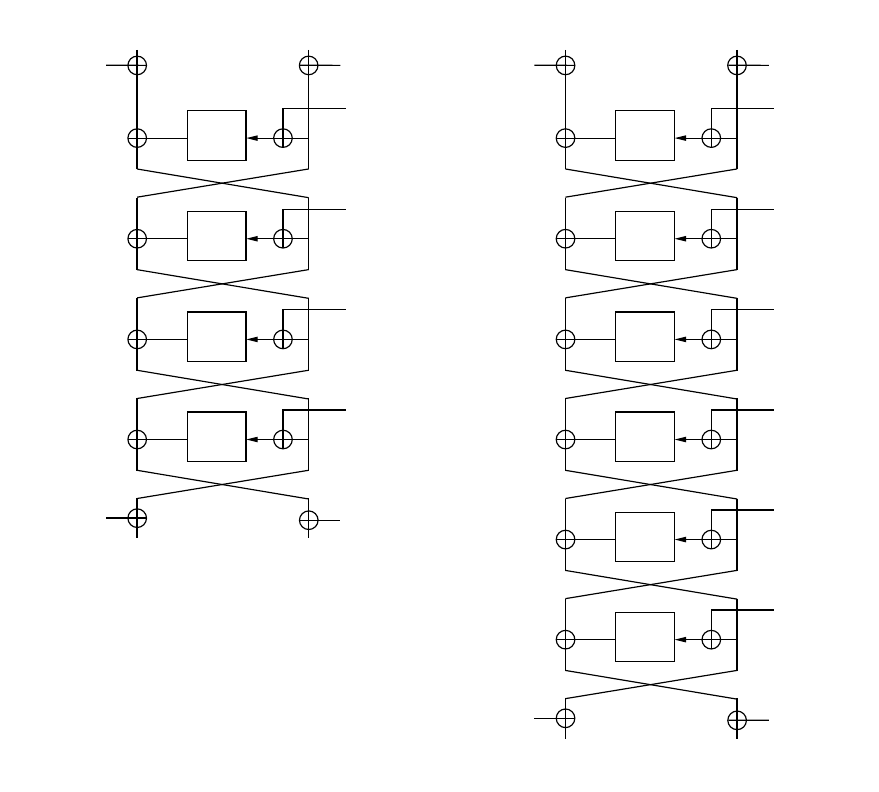}}
	\put(209.76,214.93){\fontsize{8.54}{10.24}\selectfont $R$}
	\put(182.95,186.56){\fontsize{8.54}{10.24}\selectfont $f$}
	\put(160.41,214.93){\fontsize{8.54}{10.24}\selectfont $L$}
	\put(213.92,156.79){\fontsize{8.54}{10.24}\selectfont $X$}
	\put(209.91,7.51){\fontsize{8.54}{10.24}\selectfont $T$}
	\put(160.41,8.10){\fontsize{8.54}{10.24}\selectfont $S$}
	\put(225.02,195.37){\fontsize{8.54}{10.24}\selectfont $\ga_1(k)$}
	\put(225.02,166.41){\fontsize{8.54}{10.24}\selectfont $\ga_2(k)$}
	\put(225.02,79.91){\fontsize{8.54}{10.24}\selectfont $\ga_5(k)$}
	\put(225.02,51.16){\fontsize{8.54}{10.24}\selectfont $\ga_6(k)$}
	\put(223.58,207.52){\fontsize{8.54}{10.24}\selectfont $\wf_1(k)$}
	\put(129.03,207.52){\fontsize{8.54}{10.24}\selectfont $\wf_0(k)$}
	\put(129.03,19.49){\fontsize{8.54}{10.24}\selectfont $\wf_2(k)$}
	\put(223.58,18.90){\fontsize{8.54}{10.24}\selectfont $\wf_3(k)$}
	\put(182.95,157.58){\fontsize{8.54}{10.24}\selectfont $f$}
	\put(213.92,70.15){\fontsize{8.54}{10.24}\selectfont $A$}
	\put(182.95,70.95){\fontsize{8.54}{10.24}\selectfont $f$}
	\put(182.95,42.12){\fontsize{8.54}{10.24}\selectfont $f$}
	\put(86.40,214.93){\fontsize{8.54}{10.24}\selectfont $R$}
	\put(59.59,186.56){\fontsize{8.54}{10.24}\selectfont $f$}
	\put(37.05,214.93){\fontsize{8.54}{10.24}\selectfont $L$}
	\put(90.56,156.79){\fontsize{8.54}{10.24}\selectfont $X$}
	\put(86.55,65.15){\fontsize{8.54}{10.24}\selectfont $T$}
	\put(37.05,65.74){\fontsize{8.54}{10.24}\selectfont $S$}
	\put(101.66,195.37){\fontsize{8.54}{10.24}\selectfont $\ga_1(k)$}
	\put(101.66,166.41){\fontsize{8.54}{10.24}\selectfont $\ga_2(k)$}
	\put(101.66,137.55){\fontsize{8.54}{10.24}\selectfont $\ga_3(k)$}
	\put(101.66,108.80){\fontsize{8.54}{10.24}\selectfont $\ga_4(k)$}
	\put(100.22,207.52){\fontsize{8.54}{10.24}\selectfont $\wf_1(k)$}
	\put(5.67,207.52){\fontsize{8.54}{10.24}\selectfont $\wf_0(k)$}
	\put(5.67,77.13){\fontsize{8.54}{10.24}\selectfont $\wf_2(k)$}
	\put(100.22,76.54){\fontsize{8.54}{10.24}\selectfont $\wf_3(k)$}
	\put(59.59,157.58){\fontsize{8.54}{10.24}\selectfont $f$}
	\put(90.56,127.79){\fontsize{8.54}{10.24}\selectfont $Y$}
	\put(59.59,128.59){\fontsize{8.54}{10.24}\selectfont $f$}
	\put(59.59,99.76){\fontsize{8.54}{10.24}\selectfont $f$}
	\put(225.02,137.55){\fontsize{8.54}{10.24}\selectfont $\ga_3(k)$}
	\put(225.02,108.80){\fontsize{8.54}{10.24}\selectfont $\ga_4(k)$}
	\put(213.92,127.79){\fontsize{8.54}{10.24}\selectfont $Y$}
	\put(182.95,128.59){\fontsize{8.54}{10.24}\selectfont $f$}
	\put(213.92,98.96){\fontsize{8.54}{10.24}\selectfont $Z$}
	\put(182.95,99.76){\fontsize{8.54}{10.24}\selectfont $f$}
	\end{picture}%
	\caption{\label{Fig:KAFw-64-rnd}%
		The \kafwsg cipher variants with notations (for the intermediate values) used in this paper. $f$ is a {\it public} round-function---either a random function $F$, or a random permutation $P$. (Left) 4-round \kafwsg; (Right) 6-round \kafwsg.}
\end{figure}

To be closer to the reality, we do not assume the components
\emph{independent}. Instead, we assume: (i) all the
round-functions $f_1,\ldots,f_t$ are \textbf{the same one}
denoted $f$, and (ii) each sub-key is derived from an $n$-bit
master-key $k$ via an efficiently computable $n$-to-$n$-bit
transformation, i.e. $k_i=\ga_i(k)$ for $i=1,\ldots,t$,
and $wk_j=\wf_j(k)$ for $j=1,2,3,4$.\footnote{While
$n$-bit master-keys may be uncommon in practice, it suffices for serving some insights (as will be seen). To address longer master-keys, the difficulty lies in modeling key-schedules: see the discussion in page \pageref{page:open-questions}.} Please see Fig.
\ref{Fig:KAFw-64-rnd} for the instances with 4 and 6 rounds. Denote by $(\wf,\ga)$ such a key-schedule function for $t$-rounds, $\wf=(\wf_0,\wf_1,\wf_2,\wf_3)$, $\ga=(\ga_1,\ldots,\ga_t)$; and denote by $\kafwsg$ the ``single-function'' \kafw~model with round-function $f$ and key-schedule $(\wf,\ga)$.

\arrangelittlespace

\noindent\textbf{On Other Models.}
We re-stress our model should be distinguished from the mentioned {\it Luby-Rackoff model} built upon a PRF $\underline{G_{k_i}(W_R)}$. In such a round-function the key
is ``embedded'' in a {\it non-obvious way}, and it thus overlooks many structural properties in practical Feistel ciphers.

We did not notice any previous work on our \kafw~model.\footnote{On the practical side, the cipher CLEFIA recommended by the
ISO/IEC standard~\cite{CLEFIAISO} is a 4-line generalization of \kafw.} However, by appropriately shuffling the key-schedule
$(\wf,\ga)=((\wf_0,\ldots,\wf_3),(\ga_1,\ldots,\ga_t))$,
\kafw~unifies existing famous theoretical models, and captures the structures of a large range of Feistel ciphers.
To see this, we first note that (as mentioned) by setting the
whitening keys to 0, we recover the \kaf~model, a.k.a. {\it Feistel-2 schemes} in the cryptanalytic community~\cite{IS13AttackFeistelAC13}, which has been deeply understood from the cryptanalysis point of
view~\cite{904500,complementFeistelFSE2013,IS13AttackFeistelAC13} and frequently used as instructive examples for illustrating new attacks~\cite{slideJoc17}. The \kaf~model
roughly captures the structures of DES~\cite{DESDesign}, GOST~\cite{GOSTDesign}, and Camellia variant without $FL/FL^{-1}$ functions~\cite{complementFeistelFSE2013}.

We then note that in the aforementioned Lucifer-like structure,
each round-key is xored {\it after} the corresponding
round-function, i.e. the state at round $i$ is updated according to
\begin{align}
W_L\|W_R\mapsto W_R\|W_L\xor \underline{f_i(W_R)\xor k_i}.
\label{eq:round-function-of-kafv}
\end{align}
This afterwards manner effectively eliminates the key interruption in the 1st round and in the last round and allows the analyst to analyze an equivalent two-round-reduced variant~\cite{attackLuciferJoC}, using the original 1st and last round-keys as whitening keys: $0\|k_1$ for pre-whitening, and $k_t\|0$ for post-whitening (we include a formal clarification in Appendix \ref{subsec:lucifer-to-kafv}). We denote by \kafv~the resulted whitening key-based
\kaf~Variant. Roughly, \kafv~or the Lucifer-like model
and their multi-line generalizations capture Blowfish~\cite{BlowfishDesign}, TEA~\cite{TEADesign}, XTEA~\cite{XTEADesign}, SIMON~\cite{simonspeckDesign}, Piccolo (multi-line \kafv)~\cite{PiccoloDesign}, and RC2~\cite{RC2Design}. Most importantly to us, \emph{each \kafv~instance is also captured by a \kafw~instance with a corresponding key-schedule} (a formal analysis is given in section \ref{subsec:kafv-results}). Therefore, our model \kafw~seems the most general.

By the above discussion, it seems the three models \kafw, \kaf,
and \kafv~are cryptographically equivalent modulo different key-schedules. But this contradicts existing understandings. For example, it was commented that the Lucifer-like structure blocks the complementation property, while in \kaf~the first and last rounds are more effective~\cite{attackLuciferJoC}; and that \kafv~seems stronger
against RKAs, which appears one of the motivations to use
it~\cite{complementingFSE2010}. And, assuming
\textbf{independent} random round-functions and
\textbf{identical} round-keys, the 21-round $\kafv$ variant is
indifferentiable from ideal ciphers~\cite{MRH04iffTCC},
while the \kaf~variant is {\it never}
indifferentiable~\cite{iffKAFTCC15} (even worse, such
\kaf~would collapse to a 1-round \kac~built on a keyless
multi-round Feistel permutation! see page \pageref{label:kaf-collapse}). As will be unveiled
in this paper, this distinction stems from the fact that
\emph{to achieve the same level of security, \kaf~and
\kafv~models require different properties from the involved
key-schedules}; and \emph{with common key-schedule designs,
\kafv~has a higher chance of being secure against RKAs than
\kaf}! (For details please see below.)

\arrangelittlespace

\noindent\textbf{Our Contributions.}
We first focus on the $\kafwsg$ model and prove general results, and then derive concrete implications on the more popular \kaf~and \kafv~models.

In detail, we analyze both the case of $(\wf,\ga)$ being (highly) non-linear (with respect to $\xor$) and the case of $(\wf,\ga)$ being purely affine. In each case, (as mentioned) $\kafwsg$ uses \textbf{identical round-functions} and sub-keys derived from \textbf{an $n$-bit master-key}. For the round-function $f$, we consider both $f=F$ a
random $n$-to-$n$-bit function (denoted $\kafwsf$) and $f=P$ a random $n$-bit permutation
(denoted $\kafwsp$---distinguished by the superscript). The consideration here is two-fold. First,
both choices have been adopted in practice, e.g. GOST uses a
32-bit permutation, while SIMON$2n/\kappa$ uses an
$n$-to-$n$-bit non-bijective function. Second, both choices
have advantages: random functions are theoretically attractive
since they have less structural properties than random
permutations, while the latter allow practical instantiations
using e.g. SHA-3 permutations~\cite{KeccakRefer} (will be discussed later).

In all, we analyzed four cases: $\kafwsf$ and $\kafwsp$ with
non-linear $(\wf,\ga)$, and $\kafwsf$ and \kafwsp with affine $(\wf,\ga)$. With non-linear $(\wf,\ga)$, our main result states sufficient conditions on the key-schedule so that the 4-round \kafwsf~and \kafwsp~ciphers are secure against $\xor$-RKAs up to $\widetilde{\mathcal{O}}(2^{n/2})$ queries, where the $\widetilde{\mathcal{O}}(\cdot)$ notation hides factors that depend on $(\wf,\ga)$. Such good key-schedules can be instantiated via field arithmetics. For example, with the following key-schedule, the 4-round \kafwsf and \kafwsp are secure up to $c\cdot 2^{n/2}$ queries for a small constant $c$ (which is given in section \ref{sect:towards-minimizing}):
\begin{itemize}
	\item $\wf_0(k)=\wf_3(k)=\ga_2(k)=\ga_3(k)=0$;
	\item $\wf_1(k)\xor\ga_1(k)=\mathbf{M}_1\otimes k\xor
	k^3$, and $\wf_2(k)\xor\ga_4(k)=\mathbf{M}_4\otimes k\xor k^3$, where $\mathbf{M}_1\neq\mathbf{M}_4$ are two non-zero constants chosen from $\Zn$, and $\otimes$ denotes multiplications taken over the finite field $\mathbb{F}_{2^n}$.
\end{itemize}
Interestingly, this means one could set $\ga_1(k)=\ga_4(k)=0$, i.e., the security of the 4-round Feistel cipher can be fully based on carefully chosen pre- and post-whitening keys.

For any cipher with an $n$-bit master-key, an RKA adversary could leverage collisions between secret related-keys and offline guesses for distinguishing with $2^{n/2}$ queries~\cite{rksecurity03}. Our birthday bound is thus tight. The 4 rounds are also tight, as otherwise a standard (i.e., non related-key), adaptive chosen-plaintext and ciphertext attack (CCA) is possible (see e.g.~\cite{patarin04}).

%For any cipher with an $n$-bit master-key, an RKA adversary could distinguish by detecting collisions between secret related-keys and offline guesses~\cite{rksecurity03}. Such a collision can be detected with $2^{n/2}$ queries. Therefore, the birthday-type bound is tight. The round complexity 4 is also tight, since with one less round a standard (i.e., non related-key), adaptive chosen-plaintext and ciphertext attack (CCA) is possible (see e.g.~\cite{patarin04}).

\arrangelittlespace

Without non-linearity, using a related-key boomerang~\cite{rkBoomerangEC2005}
distinguisher we break four rounds with any affine $(\wf,\ga)$, and further using the boomerang switch
trick~\cite{rkAESAsiacrypt09} we break five rounds under one more assumption on $(\wf,\ga)$. Our positive result states conditions on the key-schedule that suffice for $2^{n/2}$ security of 6-round \kafwsf~and \kafwsp. The (simple) conditions (roughly) prevent self-symmetry and complementation properties. An example, which also highlights the importance of the 1st and last round-keys, is as follows:
\begin{itemize}
	\item $\wf_0(k)=\wf_1(k)=\wf_2(k)=\wf_3(k)=0$, i.e., no whitening keys;
	\item $(\ga_1(k),\ga_2(k),\ga_3(k),\ga_4(k),\ga_5(k),\ga_6(k))=(k,0,0,0,0,$ $\pi(k))$, where $\pi(k_L\|k_R)=k_R\|k_L\xor k_R$.
\end{itemize}
Note that this $\pi$ is a linear orthomorphism, i.e., a permutation of $\Zn$ for which $x\mapsto x\xor\pi(x)$ is also a permutation. Orthomorphisms have been found helpful in establishing nice theoretical results, in particular minimal Luby-Rackoff models~\cite{minimizingFeistelEC92Josef} and 2-round \kac s~\cite{CLL14}. We remark that such a key-schedule seems rather weak. Yet, it suffices for our birthday provable security. Stronger key-schedules might help establish beyond-birthday security, which is left for future work.

\arrangelittlespace

\noindent\underline{\textsc{Implications on} \kaf~\textsc{and} \kafv}.
From the general results on \kafw~we can derive positive
results on 4- and 6-round \kaf~and \kafv, and that which
conditions on the key-schedules suffice for security.

For non-linear key derivation functions (KDFs) our results indicate they could increase the $\xor$-RKA security of \kaf. This confirms the theoretical soundness of designs with highly non-linear key-schedules, e.g. CAST-128~\cite{CAST128RFC}.\footnote{But in practice, this should be interpreted with caution. CLEFIA also employs a highly non-linear key-schedule, but suffers from weak-keys~\cite{complementCLEFIAAC14} in the RKA setting. Weak-keys couldn't be covered by these theoretical analyses.}

For affine KDFs the situation is a bit complicated (and more interesting). Roughly
speaking, for \kaf~(and also \kafw) ciphers, one should pay
additional attention on the interaction between the KDFs at the
odd rounds and even rounds respectively. On the other hand, for
\kafv~ciphers it (may) suffice to just focus on designing each
round-KDF, without considering the interactions between
different rounds. These explain the different behaviors of
\kaf~and \kafv~structures, and serve as theoretical evidence
that {\it with common ad hoc key-schedule designs, \kafv~variants
	do have a higher chance to achieve $\xor$-RKA security
	than \kaf~and \kafw}. This confirms the theoretical soundness of reverting
to \kafv~structures to improve RKA security, which---as mentioned,---seems a
folklore~\cite{attackLuciferJoC,complementingFSE2010}, and seems the idea underlying
many \kafv~ciphers mentioned before. For clearness, more discussion is deferred to Section \ref{section:to-kaf-and-kafv}, after we present the
concrete key-schedule conditions.

Aside from clarifying \kaf~and \kafv~models, our results also provide new insights into designing affine key-schedules for practical Feistel ciphers, which is a long-standing open problem hightlighted in e.g.~\cite{KHP12IT,KAprpEC12}. Note that affine key-schedules are usually preferred (e.g., DES, SIMON, etc.) due to their efficiency and compatibility with frequently rekeying.

\arrangelittlespace

%; and some \kac~instances such as AES and PRESENT~\cite{PRESENTDesign} also employed ``nearly'' affine schedules.}\footnote{In detail, their schedules contain a few calls to the S-boxes for some non-linearity, but are mostly affine.}         \\

%%%
%%%Our results indeed match the practice to some extent. For
%%%example, recall that every round-KDF in DES is a {\it
%%%bit-permutation}. For any such KDF $\ga$ it holds
%%%$\ga(0\text{xFF}\ldots\text{FF})=0\text{xFF}\ldots\text{FF}$.
%%%Therefore, key-schedules built upon such KDFs {\it do not}
%%%satisfy the conditions (more concretely, condition
%%%(\ref{condition:6-rnd-good-ks-13-46-interference}) in
%%%Definition
%%%\ref{definition:good-linear-key-schedule-6-rnd-KAFw}) we
%%%characterized for \kafsf~variants. This matches the fact that
%%%DES variants are vulnerable to complementing attacks.
%%%
%%%Both Corollary \ref{corollary:good-key-schedule-for-kaf} and
%%%Definition \ref{definition:good-linear-key-schedule-6-rnd-KAFw}
%%%requires $M_1\cdot\D\neq M_3\cdot\D$ and $M_4\cdot\D\neq
%%%M_6\cdot\D$. This means adding whitening keys to a \kaf~instance
%%%would not make it more secure against $\xor$-RKA.
%%%
%%%Moreover, note that the na\"{\i}ve round constants-based
%%%key-schedule $k\mapsto(k\xor C_1,k\xor C_2,\ldots,k\xor C_t)$
%%%does {\it not} satisfy the requirements on \kafsf~either.
%%%Actually, \kafsf~variants with this key-schedule are also
%%%vulnerable to complementing attacks.

%%
%%\reducelittlespace
%%
%%\paragraph{\underline{Reflections in Practice}.}

\noindent\underline{\textsc{Tweakable Feistel Ciphers}}.
By the general result of Bellare and Kohno~\cite{rksecurity03}, given a
$\xor$-RKA secure blockcipher $\e_k(M)$ with $n$-bit $k$, XORing the tweak
$t$ into the key, i.e. $\e_{k\xor t}(M)$, gives rise to a tweakable blockcipher (TBC) with $n$-bit tweaks and keys and provable security against $2^{n/2}$ queries. Therefore, efficient
tweakable Feistel ciphers with birthday security could be
obtained from our results. We stress that tweakable Feistel
ciphers obtained via our approach are in the Random Oracle
Model, i.e. with {\it public random} round-functions, which significantly deviates from the tweakable Luby-Rackoff ciphers~\cite{TweakingLubyRackoffAC07} built upon {\it secret random} functions.

\arrangelittlespace

\noindent\underline{\textsc{Modes for Permutations}}.
Alternatively, the variants \kafwsp,
\kafsp, and \kafvsp~can be viewed as modes for cryptographic permutations. With the appearance of reliable permutations such as the permutations underlying SHA-3~\cite{KeccakRefer} and the Simpira family~\cite{SimpiraDesignAC2016}, our results allow creating highly modular
wide-block ciphers with some level of provable $\xor$-RKA security support, or wide-block tweakable Feistel ciphers. These may find application in various settings, for example, instantiating provably secure robust authenticated encryption~\cite{DBLP:conf/eurocrypt/HoangKR15,SimpiraDesignAC2016}, Onion-AE~\cite{cryptoeprint:2018:126}, and disk encryption~\cite{5238748}.

For comparison, the \kac~results~\cite{FP15RKEM,seqiffSKA4EUROCRYPT15,XPXTweakableEMBartIC16,RKEMEprint16} also offered such permutation modes. But \kafwsp~achieves {\it domain extension} at the same time, i.e. it offers a provable TBC from ``smaller'' permutations. This may reduce implementation cost and increase security confidence.

Finally, we remind the reader that all of our results are derived in the Random Oracle Model. Once instantiated, arguments and security insurance turn heuristic~\cite{CGHROM98}.

\arrangelittlespace

%For example, if the 1600-bit permutation of SHA-3 is used, a 3200-bit blockcipher with 1600-bit key is built, which cannot be broken by $\xor$-RKA with complexity below (roughly) $2^{800}$ without the help of distinguishers on the SHA-3 permutation.

%For the latter purpose, one could
%(and actually, {\it should}) ``fill in the blanks'', e.g.
%mapping the master-key $k$ to $(\mathbf{M}_1\otimes k\xor
%k^3,k,k,\mathbf{M}_4\otimes k\xor k^3)$ for the 4-round
%\kafsp.

\noindent\textbf{Related Work and Comparison.}
As mentioned, Barbosa and Farshim (BF) have studied provable RKA-security of Luby-Rackoff models~\cite{RKAFeistelFSE2014}. Here we make a comprehensive comparison. In detail, BF proved the following 4-round Luby-Rackoff variant (see Eq. (\ref{eq:defn-luby-rackoff-round}) for the function $\Phi_{G_{k_i}}(X)$)
$$\textsf{LR}_{k_1,k_2}(M)=\Phi_{G_{k_2}}(\Phi_{G_{k_1}}(\Phi_{G_{k_2}}(\Phi_{G_{k_1}}(M))))$$
is CCA secure against RKAs, if $G$ is an RKA-secure PRF. BF's work has two advantages:
\begin{enumerate}
	\item Their results covered a much wider range of Related-Key Derivation (RKD) function set. Informally, this means $\textsf{LR}_{k_1,k_2}$ is secure even if the attacker queries $\textsf{LR}_{\psi(k_1,k_2)}$ for $\psi$ more complicated than $(k_1\|k_2)\xor\D$.
	\label{comparison-bf14-p1}
	\item Their round-functions are more ``generic'', and could be instantiated under complexity assumptions.
	\label{comparison-bf14-p2}
\end{enumerate}
For (\ref{comparison-bf14-p1}), as we argued, we aim at bridging theory and reality. The most widely-used attack model is $\xor$-RKA, and it's not clear whether the complicated RKD functions are indeed possible in reality. Moreover, for \kafw, RKA security against larger RKD sets isn't ``for-free'': since the sufficient key-schedule conditions heavily depend on the concrete RKD function (e.g. see Definition \ref{definition:good-non-linear-ks-4-rnd-KAFw}), more complicated key-schedules are likely required. Random oracle KDFs should be sufficient for all ``interesting'' RKD sets, but they fall short of providing insights into practical designs. In all, it seems questionable to spend a lot of complexity on the key-schedules to buy security against somewhat artificial RKD sets. These clarify why we concentrate on $\xor$-RKAs. Still, considering larger RKD sets is of theoretical interest, and is a possible future direction.     \label{label:argument-for-RKD-sets}

For (\ref{comparison-bf14-p2}), we argue switching from Luby-Rackoff to \kafw~is a significant step in cryptography along two axes.   \label{page:comparison-to-BF14}

First, viewing Feistel networks as abstract models of real-world blockciphers, we already argued that the Luby-Rackoff model $\textsf{LR}_{k_1,k_2}(M)$, though seems generic, is arguably too far from cryptographic reality in the RKA setting. Even in theory there remains imperfectness: the Luby-Rackoff model doesn't show how to concretely design {\it keyed} primitives from (conceptually) simpler {\it keyless} primitives; it just ``defers'' the task to designing {\it keyed} round-function $G_{k_i}$. In the RKA setting, this requires an RKA-secure PRF $G_{k_i}$ from keyless primitives, which is even harder.

In contrast, \kafw~results demonstrate how to construct blockciphers from keyless permutations or functions, which fitted into a hot topic (see the \kac~papers~\cite{KAprpEC12}), and has been recently re-emphasized by Diffie (in Leiden, March, 2018). This nicely fills in the gap left by Luby-Rackoff results.

Second, viewing Feistel networks as modes, this represents switching from {\it modes for PRFs/blockciphers} to {\it modes for keyless permutations}. Permutation-based modes not only offer more choices, but also reduce the burden of designers (they could focus on designing one permutation without considering RKA issues). Therefore, it has been a long trend, with prominent examples include the popular multi-purpose sponge functions~\cite{spongeFunction}, permutation-based hash functions~\cite{6085614,7874114}, and authenticated encryption modes~\cite{5571911,DBLP:journals/csr/AbedFL16}.

In summary, BF's work is more foundational, and shows how to build RKA secure PRPs from RKA secure PRFs, while our work tries to shed more light on the practical side. BF's Luby-Rackoff approach also gives rise to RKA-secure ciphers and TBCs, but it requires an RKA secure PRF, for which it may not be easy to find an efficient and reliable candidate (especially when a large block-size is desired).

\arrangelittlespace

A concurrent work of Cogliati et al. shows how to construct wide-block TBC from SPNs~\cite{DBLP:conf/crypto/CogliatiDKLSTZ18}. They focus on (better) beyond-birthday bounds, while we proved $\xor$-RKA security which may not be implied by tweakable pseudorandomness. They shed lights on SPNs, while we on Feistel (that could use non-invertible functions). In all, the two works are complementary.

\arrangelittlespace

Concentrating on Feistel ciphers in the ideal model, previous works only considered \kaf~and \kafv. In the provable setting, \kaf~has been analyzed by Lampe and Seurin~\cite{KAFFSE2014}. While they proved better bounds of $2^{\frac{tn}{t+1}}$ queries for $6t$ rounds, they assumed \textit{completely independent} round-functions \emph{and} \textit{independent} round-keys and they only considered the \textit{single-key security}. A recent improvement considered {\it correlated round-keys}, and proved {\it multi-user} security with birthday bounds $2^{n/2}$ at 4 rounds and beyond-birthday bounds $2^{2n/3}$ at 6 rounds~\cite{KAFAC16full}. The 4-round ``minimal'' \kaf~scheme given in~\cite{KAFAC16full} consumes a (linear) orthomorphism for the key-schedule, which is very similar to ours. Thus in some sense, our results indicate that stronger key-schedule assumptions (i.e., non-linearity) buy $\xor$-RKA security. We additionally considered {\it round permutation case}, and this gives rise to permutation-modes. Another (mentioned) work is the indifferentiability of \kafv~of~\cite{iffKAFTCC15}, the security bound of which was however too weak.

Initiated in~\cite{Lucks04RKA}, a series of papers established efficient generic approaches to obtain RKA secure blockciphers from PRPs~\cite{RKAFeistelFSE2014,StefanoAC15}, which are complementary to our ``concrete'' results. Generic transformations however fall short of deepening the understanding of widely-deployed structures.

Finally, in the ideal model, key-schedule conditions that suffice for some level of security have been characterized for single-key security of
Luby-Rackoff~\cite{NandiFeistel2010}, \kac s~\cite{CLL14}, and
SPNs~\cite{DBLP:conf/crypto/CogliatiDKLSTZ18}, for $\xor$-RKA security of \kac
s~\cite{seqiffSKA4EUROCRYPT15}, and for indifferentiability of
Luby-Rackoff~\cite{roicequalJoC2014} and \kac s, see~\cite{IffSKA5CRYTO17} and the reference therein. These results are complementary to ours. Since we identified {\it concrete} conditions, our work is closer to the series~\cite{CLL14,seqiffSKA4EUROCRYPT15,DBLP:conf/crypto/CogliatiDKLSTZ18}.

\arrangelittlespace

\noindent\textbf{Possible Future Works} include: investigating RKA security of \kafw~with respect to larger RKD function sets, posing beyond-birthday secure tweakable \kafw~variants, or studying key-schedules sufficient for chosen-key security~\cite{pubiffFeistel6tcc12}. The most attractive direction seems to prove beyond-birthday security for \kafw~models with $\ge2n$-bit master-keys. This is much closer to reality, but it requires modeling the combinatorial properties of ``non-trivial'' key-schedules for longer master-keys, which seems quite hard. For RKA security, some level of dependence has to be assumed between the round-keys~\cite{KAprpEC12}. The dependence should be both close to reality and enough for proofs. So which type of dependence is satisfying? A natural idea is to consider an alternating form of round-keys $\ga_1(k)$, $\ga_2(k')$, $\ga_3(k)$, $\ga_4(k'),\ldots$, where $k$ and $k'$ are the two halves of a $2n$-bit master-key. But this model seems too artificial. \label{page:open-questions}

\arrangelittlespace

\noindent\textbf{Organization.}
Section \ref{section:preliminary} presents notations, definitions, and tools. In Sections
\ref{section:KAFw-non-linear-KDF} and
\ref{section:KAFw-linear-KDF}, we analyze the $\xor$-RKA
security of \kafwsg with non-linear and affine
$(\wf,\ga)$ respectively. Then, from the \kafw~results we derive results on \kaf~and \kafv~in Section \ref{section:to-kaf-and-kafv}, and make discussion on theoretically best possible results in Section \ref{sect:towards-minimizing}. The complementing attacks are given in Appendix \ref{subsection:complementing-kaf-linear-kdf-any-rnd} to help understanding our proofs.

% ==================================================================

\section{Preliminaries}
\label{section:preliminary}

%\subsubsection{\textbf{General Notation}}
\noindent\textbf{General Notation.}
For integers $1\leq b\leq a$, we write $(a)_b=a(a-1)\ldots(a-b+1)$ and $(a)_0=1$ by convention. In all the following, we fix an integer $n\geq 1$ and denote $N=2^n$.
Further denote by $\functionset$ the set of all functions from $\{0,1\}^n$ to $\{0,1\}^n$, by $\permutationset$
the set of all permutations on $\{0,1\}^n$, and by
$\blockcipherset$ the set of all blockciphers with $2n$-bit
block size and $n$-bit keys. For a finite set $\mathcal{X}$, $X \sample \mathcal{X}$ means that an element $X$ is selected from $\mathcal{X}$ uniformly at random. For $X,Y\in\{0,1\}^n$, $X\|Y$ or simply $XY$ denotes their concatenation. Finally, throughout this paper, we denote $k\xor\D$ by $k_{\D}$ for simplicity.

\arrangelittlespace

\noindent\textbf{Non-linear and Affine Functions.} For a function $\ga:\{0,1\}^n\rightarrow\{0,1\}^n$, its non-linearity could be measured by        {\small
\begin{align}
%\delta=
{\max}_{a,b\in\{0,1\}^n,a\neq0}\Big|\{k\in\{0,1\}^n:\ga(k\xor a)\xor\ga(k)=b\}\Big|.
\label{eq:defn-non-linearity}
\end{align}
%A famous highly non-linear example is $\ga_(k)=\mathbf{M}\otimes k$, where $\mathbf{M}$ is a non-zero constants chosen from $\Zn$, and $\otimes$ denotes multiplications taken over the finite field $\mathbb{F}_{2^n}$ (as used in the Introduction). The $\delta$ quantity of this function is 2.
}%
Viewing the $n$-bit input $k$ as an $n$-dimensional vector over $\mathbb{F}_2$, an $n$-bit affine
function $\ga$ can be defined as
$$\ga(k)=M\cdot k\xor C$$
for a fixed $n\times n$ matrix over $\mathbb{F}_2$ and a fixed
$n$-dimensional vector $C$ over $\mathbb{F}_2$. By these, a
$t$-round affine key-schedule
$(\wf,\ga)=((\wf_0,\wf_1,\wf_2,\wf_3),(\ga_1,\ldots,\ga_t))$ (as mentioned in the Introduction) would be specified by $t+4$ fixed matrices $M_0^{(w)}$, $M_1^{(w)},M_2^{(w)},M_3^{(w)},M_1,\ldots,M_t$, and $t+4$ fixed vectors/$n$-bit constants $C_0^{(w)},C_1^{(w)},C_2^{(w)},C_3^{(w)},C_1,\ldots,C_t$:
\begin{align}
\wf_i(k)=M_i^{(w)}\cdot k\xor C_i^{(w)},\text{ }i=1,2,3,4,
\label{eq:defn-wf-affine-func}
\end{align}
and
\begin{align}
\ga_j(k)=M_j\cdot k\xor C_j,\text{ }j=1,\ldots,t.
\label{eq:defn-gamma-affine-func}
\end{align}
We stress that the multiplication $M\cdot k$ should be distinguished from the aforementioned field multiplication $\mathbf{M}\otimes k$.

%%%For an affine KDF $\ga$, a
%%%difference in the master-key would deterministically result in
%%%a predicable difference in the derived (round-)key. The use of
%%%affine key-schedules would thus ease the life of related-key
%%%adversaries. Yet, affine transformations are usually more efficient, and are thus widely used despite this shortage.   \\

\noindent\textbf{Uniform AXU Functions.} For conciseness, we characterize good non-linear key-schedules using standard notions of almost XOR-universality (AXU) and uniformity for keyed (hash) functions. To this end, we serve their definitions below. First, a keyed function $H_k(\cdot)$ from the domain $\mathcal{X}$ to $\{0,1\}^n$ is said to be {\it $\delta$-uniform}, if for any $x\in\mathcal{X}$ and $y\in\{0,1\}^n$,
$$\Pr[k\sample\mathcal{K}:H_k(x)=y]\leq\delta,$$
where $\mathcal{K}$ is its key space. $H$ is said {\it $\delta'$-almost XOR-universal ($\delta'$-AXU)} if for all distinct $x,x'\in\mathcal{X}$ and all $y\in\{0,1\}^n$,
$$\Pr[k\sample\mathcal{K}:H_k(x)\xor H_k(x')=y]\leq\delta'.$$

\noindent\textbf{\kafw~Ciphers.}  As mentioned in the Introduction, we focus on \kafwsg, the \kafw~variants with two features:
\begin{enumerate}
	\item the same function $f:\{0,1\}^n\rightarrow\{0,1\}^n$
	is used at each round, and
	\item the key-schedule is
	$(\wf,\ga)=((\wf_0,\wf_1,\wf_2,\wf_3),(\ga_1,\ldots,$ $\ga_t))$,
	i.e. the $i$-th whitening key $wk_i$ is derived from
	the $n$-bit master-key $k$ via $wk_i=\wf_i(k)$, and
	the $i$-th round-key $k_i$ is $k_i=\ga_i(k)$.
\end{enumerate}
For such variants, the $i$-th round transformation is defined as
\begin{align}
\Psi_{\ga_i(k)}^{f}(W_L\|W_R)=W_R\|W_L\oplus f(\ga_i(k)\oplus W_R),
\label{eq:kafw-1-round-transformation}
\end{align}
where $W_L$ and $W_R$ are respectively the left and right
$n$-bit halves of the input. Then the $t$-round \kafwsg~variant is defined as (cf. Fig. \ref{Fig:KAFw-64-rnd})    {\small
\begin{align*}
\kafwsg_k(W)  % \\
=  wk_{out}\xor\Psi_{\ga_t(k)}^{f}\circ\ldots\circ\Psi_{\ga_1(k)}^{f}\big(wk_{in}\xor W\big),
\end{align*}
}%
where $wk_{in}=\wf_0(k)\|\wf_1(k)$ and $wk_{out}=\wf_2(k)\|\wf_3(k)$. To make it more precise, we give formal descriptions for the 4- and 6-round \kafwsg~that will be studied later. For the 4-round $\kafwsg_k$, on the $2n$-bit input $W$ which is parsed into $L\|R$, the computation proceeds in 4 steps:
%\begin{small}
\begin{enumerate}
	\item $x_1\leftarrow\ga_1(k)\xor\wf_1(k)\xor R$, $y_1\leftarrow f(x_1)$, $X=\wf_0(k)\xor L\xor y_1$;
	\item $x_2\leftarrow\ga_2(k)\xor X$, $y_2\leftarrow f(x_2)$, $Y\leftarrow\wf_1(k)\xor R\xor y_2$;
	\item $x_3\leftarrow\ga_3(k)\xor Y$, $y_3\leftarrow f(x_3)$, $S\leftarrow X\xor y_3\xor\wf_2(k)$;
	\item $x_4\leftarrow\ga_4(k)\xor\wf_2(k)\xor S$, $y_4\leftarrow f(x_4)$, $T\leftarrow Y\xor y_4\xor\wf_3(k)$.
\end{enumerate}
%\end{small}
One could see Fig. \ref{Fig:KAFw-64-rnd} (left) for illustration.
For the 6-round $\kafwsg_k$, on input $W=L\|R$, the computation proceeds in 6 steps (as in Fig. \ref{Fig:KAFw-64-rnd} (right)):   % {\small
\begin{enumerate}
	\item $x_1\leftarrow\ga_1(k)\xor\wf_1(k)\xor R$, $y_1\leftarrow f(x_1)$, $X=\wf_0(k)\xor L\xor y_1$;
	\item $x_2\leftarrow\ga_2(k)\xor X$, $y_2\leftarrow f(x_2)$, $Y\leftarrow\wf_1(k)\xor R\xor y_2$;
	\item $x_3\leftarrow\ga_3(k)\xor Y$, $y_3\leftarrow f(x_3)$, $Z\leftarrow X\xor y_3$;		
	\item $x_4\leftarrow\ga_4(k)\xor Z$, $y_4\leftarrow f(x_4)$, $A\leftarrow Y\xor y_4$;		
	\item $x_5\leftarrow\ga_5(k)\xor A$, $y_5\leftarrow f(x_5)$, $S\leftarrow Z\xor y_5\xor\wf_2(k)$;
	\item $x_6\leftarrow\ga_6(k)\xor\wf_2(k)\xor S$, $y_6\leftarrow f(x_6)$, $T\leftarrow A\xor y_6\xor\wf_3(k)$.
\end{enumerate}   %     }

As noted in~\cite{DR07}, a \textsf{KAFw} cipher (even with independent round-functions) with an even number of rounds can be seen as a special case of a \kac. In detail, the $i$-th and $(i+1)$-th rounds with round-functions $f_i$ and $f_{i+1}$ and round-keys $k_i$ and $k_{i+1}$ can be rewritten as   \label{label:kaf-collapse}
$$\Psi_{k_{i+1}}^{f_{i+1}}\circ\Psi_{k_i}^{f_i}(W)=(k_{i+1}\|k_i)\oplus\Psi_{0}^{f_{i+1}}\circ\Psi_{0}^{f_i}((k_{i+1}\|k_i)\oplus W),$$
where $\Psi_{0}^{f_{i+1}}\circ\Psi_{0}^{f_i}$ is a keyless
2-round Feistel permutation. However, provable results on
\kafw~\textbf{cannot} be derived by black-box composition of
existing results on \kac s and keyless Feistel, since
\textbf{no} provable results can be seen on
$\Psi_{0}^{f_{i+1}}\circ\Psi_{0}^{f_i}$ (let alone the even weaker $\Psi_{0}^{f}\circ\Psi_{0}^{f}$).

%
%What's more, assuming
%\textbf{identical round-functions}, the corresponding 2-round
%permutation $\Psi_{0}^{F}\circ\Psi_{0}^{F}$ is almost of
%\textbf{no cryptographic strength} for black-box composition.

As a side remark, for a $2t$-round \kafw~cipher, if the $2t$ round-keys are identical $k'=\ga_1(k)=\ldots=\ga_{2t}(k)$, then it can be seen it's essentially a 1-round \kac, i.e. $(\wf_2(k)\xor k'\|\wf_3(k)\xor k')\xor\pi\big((\wf_0(k)\xor k'\|\wf_1(k)\xor k')\xor W\big)$, where $\pi=\Psi_{0}^{f_{2t}}\circ\ldots\circ\Psi_{0}^{f_1}$ is a keyless permutation. This is known to be insecure against RKAs~\cite{seqiffSKA4EUROCRYPT15}.

\arrangelittlespace

%%%%%%%%%%%
%%%%%%%%%%%\subsubsection{\kafvsf~Ciphers.}
%%%%%%%%%%%
%%%%%%%%%%%For the $t$-round \kafvsf~variant, we take $\ga_0$ and
%%%%%%%%%%%$\ga_{t+1}$ as the KDFs for the input and output whitening keys,
%%%%%%%%%%%and denote by $(\ga_0,\ga_1,\ldots,\ga_t,\ga_{t+1})$
%%%%%%%%%%%the entire key schedule, cf. Fig.
%%%%%%%%%%%\ref{Fig:KAF-64-rnd} (right). The $i$-th round
%%%%%%%%%%%transformation is defined as $\Psi
%%%%%%%%%%%v_{k_i}^{F}(W_L\|W_R)=(W_R\|W_L\oplus F(W_R)\xor k_i).$ The
%%%%%%%%%%%entire \kafvsf~variant is specified by a public round function
%%%%%%%%%%%$F$, a master key $k$, and a key-schedule $(\ga_0,\ldots,\ga_{t+1})$:
%%%%%%%%%%%$$\kafvsf_{k}^{F}(W_L\|W_R)=(\ga_{t+1}(k)\|0)\xor\Psi v_{\ga_t(k)}^{F}\circ\ldots\circ\Psi v_{\ga_1(k)}^{F}((W_L\|W_R)\xor(0\|\ga_0(k))).$$

%%%Such whitening-based structures are seen in Blowfish and
%%%Piccolo (the latter is a 4-line GFN, yet its corresponding
%%%reduced 2-line version is a \kafvsf~instance).
%%%
%%%
%%%In Lucifer, SIMON, and XTEA, there is no whitening key, cf.
%%%Fig. \ref{Fig:simon-KAFv} (left); however, since the first and
%%%last round functions can be evaluated by the adversary freely,
%%%one could remove them and move the first and last round key
%%%addition operations ``one round inside'', and this will yield
%%%\kafvsf, cf. Fig. \ref{Fig:simon-KAFv} (right). Therefore, such
%%%ciphers are also nicely modeled by \kafvsf.

\noindent\textbf{$\xor$-RKA Security.} We follow Cogliati and Seurin~\cite{seqiffSKA4EUROCRYPT15} to
formalize $\xor$-RKA security in the ideal model. In detail,
let $\e$ be a $(n,2n)$-blockcipher, and fix a key
$k\in\{0,1\}^{n}$. We define the $\xor$-restricted related-key
oracle $\textsf{RK}[\e_k]$, which takes as input an ``offset''
$\D\in\{0,1\}^{n}$ and a plaintext $LR\in\{0,1\}^{2n}$, and
returns $\textsf{RK}[\e_k](\D,LR):=\e_{k\xor\D}(LR)$. It allows inverse queries, which we denote
$\textsf{RK}[\e_k]^{-1}(\D,ST):=\e_{k\xor\D}(ST)$. Then, we
consider a $\xor$-restricted related-key adversary $D$ which
has access to a function oracle $f$ and a related-key
oracle, and must distinguish between two worlds as follows:
\begin{itemize}
	\item the ``real'' world, where it interacts with
	$(\textsf{RK}[\e_k],f)$, and $k$ is randomly drawn;
	\item the ``ideal'' world where it interacts with
	$(\textsf{RK}[\textsf{IC}_k],f)$, where \textsf{IC}
	is an ideal cipher independent from $f$, and $k$ is
	randomly drawn.
\end{itemize}
The distinguisher is adaptive, and can make two-sided queries
to the related-key oracle. Note that in the ideal world, the oracle $\textsf{RK}[\textsf{IC}_k]$ essentially
implements an independent random permutation for each offset
$\D\in\Zn$. Formally, when $f=F$ is a random function, $D$'s distinguishing advantage on
\kafwsf is defined as   % {\small
\begin{align*}
& \mathbf{Adv}^{\xor\text{-rka}}_{\kafwsf_k}(D)
     \\
& =  
\Big|{\Pr}_{\textsf{IC},k,F}[D^{\textsf{RK}[\textsf{IC}_k],F}=1] - {\Pr}_{k,F}[D^{\textsf{RK}[\kafwsf_k],F}=1]\Big|,
\end{align*}
where the former probability is taken over the random draw of $\textsf{IC}\xleftarrow{\$}\blockcipherset,k\xleftarrow{\$}\{0,1\}^{n}, F\xleftarrow{\$}\functionset$, and the latter probability is taken over $k\xleftarrow{\$}\{0,1\}^{n},F\xleftarrow{\$}\functionset$.

For $\mathbf{Adv}^{\xor\text{-rka}}_{\kafwsp_k}(D)$, $P$ is randomly picked from the set $\permutationset$, i.e. $P\xleftarrow{\$}\permutationset$. Here the superscripts help distinguish between random function- and permutation-based \kafw.

Furthermore, we consider computationally unbounded
distinguishers, and we assume without loss of generality (wlog) that
the distinguisher is deterministic and never makes redundant
queries. For non-negative integers $q_e$, $q_f$, we define the
insecurity of the \kafwsg cipher against
$\xor$-restricted related-key attacks as
$$\mathbf{Adv}^{\xor\text{-rka}}_{\kafwsg_k}(q_f,q_e)=\text{max}_{D}\mathbf{Adv}_{\kafwsg_k}^{\xor\text{-rka}}(D),$$
where the maximum is taken over all distinguishers $D$ making
exactly $q_f$ queries to the function oracle and in total $q_e$
queries to the related-key oracle (termed as
$(q_f,q_e)$-distinguishers).

\arrangelittlespace

\noindent\textbf{The H-Coefficients Technique.}
We employ the H-coefficient technique~\cite{PatCoeffHSAC08},
and follow the paradigm of Chen and
Steinberger~\cite{KAtightboundEC2014}. To this end, we
summarize the information gathered by the distinguisher in tuples $\mathcal{Q}_E$ and $\mathcal{Q}_f$. The tuple
$$\mathcal{Q}_E=((\D_1,L_1R_1,S_1T_1),\ldots,(\D_{q_e},L_{q_e}R_{q_e},S_{q_e}T_{q_e}))$$
summarizes the queries to the related-key oracle, and means
that the $j$-th query was either a forward query
$(\D_j,L_jR_j)$ with answer $S_jT_j$, or a backward query
$(\D_j,S_jT_j)$ with answer $L_jR_j$. Throughout the remaining, we'll use the bold letter $\bft$ as a simplified notation for a tuple $(\D,LR,ST)$ in $\mathcal{Q}_E$.

Similarly to $\mathcal{Q}_E$, the tuple
$$\mathcal{Q}_f=((x_1,y_1),\ldots,(x_{q_f},y_{q_f}))$$
summarizes the queries to the round-function $f$, and
\begin{itemize}
	\item when $f=P$ is an invertible permutation, it means
	the $j$-th query was either a forward query $x_j$
	with answer $y_j$ or a backward query $y_j$ with
	answer $x_j$;
	\item when $f=F$ is a non-invertible
	function, it means $F$ was queried on
	$x_1,\ldots,x_{q_f}$ and answered
	$y_1,\ldots,y_{q_f}$ correspondingly.
\end{itemize}

To simplify the arguments (in particular, the definition of ``bad transcripts''), we reveal to the distinguisher the key $k$ at the end of the interaction. This is wlog
since $D$ is free to ignore this additional information to
compute its output bit. Formally, we append $k$ to $(\mathcal{Q}_E,\mathcal{Q}_f)$ and obtain what we
call the {\it transcript} $\tau=(\mathcal{Q}_E,\mathcal{Q}_f,k)$ of the attack. With respect to some fixed distinguisher $D$, a
transcript $\tau$ is said {\it attainable} if there exists oracles $(\textsf{IC},f)$ such that the interaction of $D$ with the ideal world $(\textsf{RK}[\textsf{IC}_k],f)$ yields $\tau$. We denote $\mathcal{T}$ the set of attainable transcripts. In all the following, we denote
$T_{re}$, resp. $T_{id}$, the probability distribution of the
transcript $\tau$ induced by the real world, resp. the ideal
world (note that these two probability distributions depend on the distinguisher). By extension, we use the same notation for a random variable distributed according to each distribution. And we define ${\Pr}_{re}(\tau)=\Pr[T_{re}=\tau]$ and ${\Pr}_{id}(\tau)=\Pr[T_{id}=\tau]$.

%%\begin{itemize}
%%\item [] ${\Pr}_{re}(\tau)=\Pr[T_{re}=\tau]$, and ${\Pr}_{id}(\tau)=\Pr[T_{id}=\tau]$.
%%\end{itemize}

Given a tuple $\mathcal{Q}_f$ of function queries and a
function $f$, we say that $f$ {\it extends} $\mathcal{Q}_f$,
denoted $f\vdash\mathcal{Q}_f$, if $f(x)=y$ for all
$(x,y)\in\mathcal{Q}_f$. Similarly, given a related-key oracle
transcript $\mathcal{Q}_E$, a blockcipher $\e$,
and a key $k\in\Zn$, we say the related-key oracle
$\textsf{RK}[\e_k]$ {\it extends} $\mathcal{Q}_E$, denoted
$\textsf{RK}[\e_k]\vdash\mathcal{Q}_E$, if $\e_{k\xor\D}(LR)=ST$
for all $(\D,LR,ST)\in\mathcal{Q}_E$. It is easy to see that
for any attainable transcript
$\tau=(\mathcal{Q}_E,\mathcal{Q}_f,k)$, the interaction of the
distinguisher with oracles $(\textsf{RK}[\e_k],f)$ produces
$\tau$ if and only if $\textsf{RK}[\e_k]\vdash\mathcal{Q}_E$ and
$f\vdash\mathcal{Q}_f$.

With all the above definitions, the main lemma of
H-coefficients technique is as follows.

\begin{lemma} [Lemma 1 in~\cite{CLL14}]
	\label{lemma:h-coefficients-main-lemma}	
	Fix a distinguisher $D$. Let
	$\mathcal{T}=\mathcal{T}_{good}\cup\mathcal{T}_{bad}$ be a
	partition of the set of attainable transcripts $\mathcal{T}$.
	Assume that there exists $\varepsilon_1$ such that for any
	$\tau\in\mathcal{T}_{good}$, one has
	\begin{align*}
	\frac{{\Pr}_{re}(\tau)}{{\Pr}_{id}(\tau)}\geq1-\varepsilon_1,
	\end{align*}
	and that there exists $\varepsilon_2$ such that
	$\Pr[T_{id}\in\mathcal{T}_{bad}]\leq\varepsilon_2$. Then
	$\mathbf{Adv}(D)\leq\varepsilon_1+\varepsilon_2$.
\end{lemma}
A proof could be found in~\cite{CLL14}.

%Finally, we remark that for an ideal $(n,2n)$-blockcipher
%$\textsf{IC}$, for the $i$-th query
%$(\D_i,L_iR_i,S_iT_i)\in\mathcal{Q}_E$, conditioned on that
%$\textsf{IC}_{k\xor\D_j}(L_jR_j)=S_jT_j$ for any $j<i$,
%$\textsf{IC}_{k\xor\D_i}(L_iR_i)$ is uniform in {\it at least}
%$N^2-(i-1)\geq N^2-q_e$ values. Therefore,

Finally, it's not hard to see
\begin{align*}
{\Pr}_{id}(\tau)&=\Pr[f\vdash\mathcal{Q}_f]\cdot\Pr[\textsf{RK}[\textsf{IC}_k]\vdash\mathcal{Q}_E]    \\
&\leq\Pr[f\vdash\mathcal{Q}_f]\cdot\bigg(\frac{1}{N^2-q_e}\bigg)^{q_e}.
\end{align*}

% ==================================================================

\section{\textsf{KAFw} with Non-linear Key-Schedules}
\label{section:KAFw-non-linear-KDF}

It is well-known that 3-round Feistel networks are not CCA
secure even in the single-key setting. So we consider 4-round \kafw. First, in section
\ref{subsection:good-ks-non-linear-4-round}, we present key-schedule conditions that are sufficient for the
$\xor$-RKA security of the 4-round \kafwsp~(which also turn out
sufficient for 4-round \kafwsf). Then, we start from \kafwsp, analyze it in section \ref{subsection:security-proof-4-round}, and then discuss how to adapt the proof for the 4-round
\kafwsf~variant (by ``dropping'' some modules from the proof for \kafwsp) in section \ref{subsection:security-kaf-rf-variant-non-linear-kdf-4-rnd}.

\subsection{Conditions on the Key-Schedules}
\label{subsection:good-ks-non-linear-4-round}

4-round key-schedules defined as follows would suffice.

\begin{definition}[Good Key-Schedule for 4 Rounds]
	\label{definition:good-non-linear-ks-4-rnd-KAFw} Consider a 4-round key-schedule $(\wf,\ga)$, where
	$\wf=(\wf_0,\wf_1,\wf_2,$ $\wf_3)$ for
	$\wf_i:\{0,1\}^n\rightarrow\{0,1\}^n$, and
	$\gamma=(\gamma_1,\gamma_2,\gamma_3,\gamma_4)$ for
	$\gamma_i:\{0,1\}^n\rightarrow\{0,1\}^n$. Then $(\wf,\ga)$ is \emph{good}, if $\underline{\varphi_1(k)=\wf_1(k)\xor\ga_1(k)}$ and
	$\underline{\varphi_4(k)=\wf_2(k)\xor\ga_4(k)}$ satisfy two conditions as follows:
	\begin{enumerate}
		\item %\emph{$\delta_1$-uniformness and $\delta_2$-non-linearity}:
		for $i=1,4$, the function $H_k(\D):=\varphi_i(k\xor\D)$ is $\delta_1$-uniform and $\delta_2$-AXU;
		\item %\emph{$\delta_3$-uniformness}:
		the function $H_k(\D,\D'):=\varphi_1(k\xor\D)\xor\varphi_4(k\xor\D')$ is $\delta_3$-uniform.
%
%		for any
%		image $k_i\in\Zn$, the number of master-keys $k$ such
%		that $\varphi_i(k)=k_i$ is at most
%		$\delta_1=\text{poly}(n)$;
%
%		\item \emph{$\delta_2$-non-linearity}: for $i=1,4$, the function $H_k(\D):=\varphi_i(k\xor\D)$ is $\delta_1$-uniform;
%
%		the non-linearity
%		measure (Eq. (\ref{eq:defn-non-linearity}))
%		of $\varphi_1$ and $\varphi_4$ is
%		$\delta_2=\text{poly}(n)$;
%		
%		\item \emph{$\delta_3$-``mutual-uniformness''}: for any
%		$k^*$, the number of master-keys $k$ such that
%		$\varphi_1(k)\xor\varphi_4(k)=k^*$ is at most
%		$\delta_3=\text{poly}(n)$;
%		
%		\item \emph{$\delta_4$-``mutual-non-linearity''}: for any
%		pair $(\nabla,\D)\in(\Zn\backslash\{0\})\times\Zn$,
%		the number of master-keys $k$ such that
%		$\varphi_1(k)\xor\varphi_4(k\xor\nabla)=\D$ is at
%		most $\delta_4=\text{poly}(n)$.
	\end{enumerate}
\end{definition}
An example of good key-schedules with $\delta_1,\delta_2,\delta_3\leq3/N$ was exhibited in the Introduction, cf. {\it Our Contributions}.

Note that $\varphi_1(k)$ and $\varphi_4(k)$ effectively mask (and
protect) the inputs to the 1st and last round-functions
respectively. This protection would be ineffective if the $\delta_1$-uniformness is seriously compromised. An extreme example is $\varphi_1(k)=0$, for which an adversary could freely compute the 2nd-round intermediate value as $R\|L\xor F(R)$.

Further note that, $\varphi_i(k\xor\D)$ is $\delta_2$-AXU essentially means the non-linearity (see Eq. (\ref{eq:defn-non-linearity})) of $\varphi_i$ is $\delta_2N$. This condition is intended to reduce the probability of 1-round related-key differentials with non-zero master-key differences; see the argument for condition (B-2) in page \pageref{condition:4-rnd-B-2}.

Finally, the 2nd condition is intended to prevent the derived round-keys from harmful ``palindrome-like'' properties~\cite{NandiFeistel2010} in the RKA setting. For example, consider a key-schedule $(\wf,\ga)$ such that
$\wf(k)=(0,0,0,0)$ and it's easy to derive $\D$ for which $\ga(k)=(k'',0,0,k')$ and $\ga(k\xor\D)=(k',0,0,k''')$ for any master-key $k$, i.e., $\varphi_1(k\xor\D)\xor\varphi_4(k)=\ga_1(k\xor\D)\xor\ga_4(k)=0$. Then it can be distinguished by querying $\textsf{RK}[\e_k](0,LR)\rightarrow ST$, $\textsf{RK}[\e_k](\D,TS)\rightarrow R'L'$, and checking if $R=R'$.

Actually, it might be possible to prove security without the 2nd condition. But this requires $\ga_2$ and $\ga_3$ to fulfill
more involved conditions. Therefore, our Definition
\ref{definition:good-non-linear-ks-4-rnd-KAFw}, with no
requirement on $\gamma_2$ and $\gamma_3$ at the expense of
slightly more requirements on $\varphi_1$ and $\varphi_4$,
captures a ``minimal'' group of conditions to some extent.

%	As a side remark, Barbosa and Farshim conjectured that for a {\bf Luby-Rackoff} cipher, if the derived round-keys are palindrome-free in the RKA setting, then the cipher is RKA-secure~\cite{RKAFeistelFSE2014}. This means (informally) it's unlikely to find two related-key derivation functions for which the derived round-key vectors $(k_1,\ldots,k_t)$ and $(k_1',\ldots,k_t')$ are such that $k_i=k_{t-i+1}'$ for $i=1,\ldots,t$. Clearly, the two round-key vectors $(k'',0,0,k')$ and $(k',0,0,k''')$ appeared in our instructive example
%	are {\it not} palindrome. Yet, their occurrence still ruins the RKA security of \kafw. This comparison again highlights the gap between Luby-Rackoff and \kafw~models (and likely, the reality).

\subsection{Security for 4 Rounds with Good Key-Schedules and f=P}
\label{subsection:security-proof-4-round}

Instantiated with a good key-schedule, the 4-round \kafwsp is
secure against $\xor$-RKAs.

\begin{theorem}
	\label{theorem:KAFwsp-rka-security-4-round}
	
	When $q_f+2q_e\leq N/2$, for the 4-round, random
	permutation-based \kafwsp cipher with a good
	key-schedule $(\wf,\ga)$ as specified in Definition
	\ref{definition:good-non-linear-ks-4-rnd-KAFw}, it holds
	{\footnotesize $$\mathbf{Adv}^{\xor\text{-rka}}_{\kafwsp_k}(q_f,q_e)\leq2\delta_1q_eq_f+(\delta_2+\delta_3)q_e^2+\frac{8q_eq_f+27q_e^2+4q_e}{N}.$$}
\end{theorem}
\noindent\underline{\emph{Proof}}. We first introduce some notations that will ease the subsequent analysis. Let $\tau=(\mathcal{Q}_E,\mathcal{Q}_P,k)$ be an attainable transcript, with $|\mathcal{Q}_E|=q_e$ and $|\mathcal{Q}_P|=q_f$. For convenience, for the involved $\mathcal{Q}_P=((x_1,y_1),\ldots,(x_{q_f},y_{q_f}))$,
we define two sets
\begin{align*}
\dom\xlongequal{\text{def}}\{ x_1,\ldots,x_{q_f} \},  \text{ and }
\rng\xlongequal{\text{def}}\{ y_1,\ldots,y_{q_f} \}.
\end{align*}

For any tuple $\bft=(\D,LR,ST)$ in $\mathcal{Q}_E$ and any function $f$ ($f=P$ or $F$; the former is the focus of this subsection), we define 10 functions     \label{page:helper-functions-4-rounds}
\begin{align*}
x_1(\bft) & =\varphi_1(k\xor\D)\xor R,    \\
y_1(\bft,f) & =f(x_1(\bft)),    \\
X(\bft,f) & =L\xor\wf_0(k\xor\D)\xor y_1(\bft,f),    \\
x_2(\bft,f) & =\ga_2(k\xor\D)\xor X(\bft,f),    \\
y_2(\bft,f) & =R\xor\wf_1(k\xor\D)\xor Y(\bft,f),    \\
Y(\bft,f) & =T\xor\wf_3(k\xor\D)\xor y_4(\bft,f),    \\
x_3(\bft,f) & =\ga_3(k\xor\D)\xor Y(\bft,f),    \\
y_3(\bft,f) & =S\xor\wf_2(k\xor\D)\xor X(\bft,f),    \\
x_4(\bft) & =\varphi_4(k\xor\D)\xor S,    \\
y_4(\bft,f) & =f(x_4(\bft)).
\end{align*}
The suffix $f$ emphasizes that the functions depend on $f$. Note that these values are derived in an ``$LR\rightarrow X,Y\leftarrow ST$'' manner, rather than the ``$LR\rightarrow X\rightarrow Y\rightarrow ST$'' manner. Moreover, $x_1(\bft)$ and $x_4(\bft)$ only depend on $\tau$.

To ease understanding our proofs, below we serve an overview of our strategy.

\arrangelittlespace

\subsubsection{{\bf Proof Strategy}}
\label{sec:proof-strategy}

Following Lemma \ref{lemma:h-coefficients-main-lemma}, with respect to a fixed $(q_f,q_e)$-distinguisher $D$, below in section \ref{sec:bad-transcripts} we define bad transcripts, and upper bound their probability of occurring in the ideal world. This probability is computed over the random choice of the key, and thus we could leverage the properties of good key-schedules.

Later in section \ref{sec:ratio-4-rounds}, we lower bound ${\Pr}_{re}(\tau)$ (and thus the ratio ${\Pr}_{re}(\tau)/{\Pr}_{id}(\tau)$) for any good $\tau$. In this step we follow~\cite{DBLP:conf/crypto/CogliatiDKLSTZ18} and define a ``bad'' predicate $\badf(P)$ on $P$, such that collisions in the $2q_e$ inputs in the 2nd and 3rd rounds
\begin{align}
x_2(\bft_1,P),\ldots,x_2(\bft_{q_e},P),x_3(\bft_1,P),\ldots,x_3(\bft_{q_e},P)
\label{eq:the-2qe-inputs}
\end{align}
and collisions in the $2q_e$ corresponding outputs
\begin{align}
y_2(\bft_1,P),\ldots,y_2(\bft_{q_e},P),y_3(\bft_1,P),\ldots,y_3(\bft_{q_e},P)
\label{eq:the-2qe-outputs}
\end{align}
are classified as conditions of $\badf(P)$. These values are determined by $P$ and thus random. Consequently, $\Pr[\badf(P)]$ could be upper bounded. In addition, as long as $\badf(P)$ is not fulfilled, it is easy to transform the probability ${\Pr}_{re}(\tau)$ into the (easy-to-bound) probability that
$$\Pr[\forall i\in\{1,\ldots,q_e\},j=2,3:P(x_j(\bft_i,P))=y_j(\bft_i,P)],$$
i.e., $P$ is consistent with the inputs/outputs of the middle two rounds. These cinch the final bound.

%Below we elaborate each step in detail.

\arrangelittlespace

\subsubsection{{\bf Bad Transcripts}}
\label{sec:bad-transcripts}
defined as follows.

\begin{definition}[Bad Transcripts for 4-Round \kafwsp]
	\label{definition:bad-transcripts-non-linear-ks-4-rnd-KAFwsp}
	An attainable transcript
	$\tau=(\mathcal{Q}_E,\mathcal{Q}_P,k)$ is \emph{bad}, if at
	least one of the following conditions is fulfilled:
	\begin{itemize}
		\item(B-1) $\exists\bft\in\mathcal{Q}_E:x_1(\bft)\in\dom$ or $x_4(\bft)\in\dom$;
		
		\item(B-2) $\exists\bft=(\Delta,LR,ST)$ and $\bft'=(\Delta',L'R',S'T')$ in
		$\mathcal{Q}_E$ such that $\Delta\neq\Delta'$,
		and $x_1(\bft)=x_1(\bft')$ or $x_4(\bft)=x_4(\bft')$;
		\label{condition:4-rnd-B-2}
		\item(B-3) $\exists\bft=(\Delta,LR,ST)$ and $\bft'=(\Delta',L'R',S'T')$ in
		$\mathcal{Q}_E$ such that $x_1(\bft)=x_4(\bft')$ (it could be $\bft=\bft'$);
		\item(B-4) there exist two distinct queries $(\Delta,LR,ST)$ and
		$(\Delta',L'R',S'T')$ in $\mathcal{Q}_E$ such that
		$\Delta=\Delta'$, and
		\begin{itemize}
			\item $L\xor L'=S\xor S'$, or $R\xor R'=T\xor
			T'$.
		\end{itemize}
		\item(B-5) there exists $(\Delta,LR,ST)\in\mathcal{Q}_E$ such that
		\begin{itemize}
			\item $L\xor\wf_0(k\xor\D)=S\xor\wf_2(k\xor\D)$, or $R\xor \wf_1(k\xor\D)=T\xor\wf_3(k\xor\D)$.
		\end{itemize}
	\end{itemize}
	Otherwise we say $\tau$ is {\it good}. Denote by
	$\mathcal{T}_{bad}$ the set of bad transcripts.
\end{definition}
We analyze the conditions in turn, with (B-1) the first. For any $\bft=(\D,LR,ST)$ in $\mathcal{Q}_E$ and any $x$, as $H_k(\D)=\varphi_1(k\xor\D)$ is $\delta_1$-uniform (cf. Definition
\ref{definition:good-non-linear-ks-4-rnd-KAFw}), we immediately have
{\small
$$\Pr[x_1(\bft)\in\dom]=\Pr[\exists x\in\dom:\varphi_1(k_{\D})=R\xor
x]\leq\delta_1q_f.$$  }%
Similarly, $\Pr[x_4(\bft)\in\dom]\leq\delta_1q_f$.
Since there are $q_e$ choices for $\bft$, we have
$$\Pr[\text{(B-1)}]\leq2\delta_1q_eq_f.$$

For (B-2), since $H_k(\D)=\varphi_i(k\xor\D)$ is $\delta_2$-AXU for $i=1,4$, for each pair $(\bft,\bft')$ with $\bft=(\D,LR,ST)$ and $\bft'=(\Delta',L'R',S'T')$ we have
\begin{align*}
& \Pr[x_1(\bft)=x_1(\bft')\text{ or }x_4(\bft)=x_4(\bft')]    \\
=& \Pr[\varphi_1(k\xor\D)\oplus
R=\varphi_1(k\xor\D')\oplus R' \\
& \ \ \ \ \text{ or }
\varphi_4(k\xor\D)\oplus S=\varphi_4(k\xor\D')\oplus S']\leq2\delta_2.
\end{align*}
As we have at most ${q_e\choose 2}\leq\frac{q_e^2}{2}$ choices for $(\bft,\bft')$ it holds
$\Pr[\text{(B-2)}]\leq\delta_2q_e^2$.

For \text{(B-3)}, since $H_k(\D,\D')=\varphi_1(k\xor\D)\xor\varphi_4(k\xor\D')$ is $\delta_3$-uniform, for each pair $(\bft,\bft')$ we have
\begin{align*}
  & \Pr[x_1(\bft)=x_4(\bft')]    \\
= & \Pr[\varphi_1(k\xor\D)\xor\varphi_4(k\xor\D')=R\xor S']\leq\delta_3.
\end{align*}
Summing over the $q_e^2$ choices of $(\bft,\bft')$ yields $\Pr[\text{(B-3)}]\leq\delta_3q_e^2$.

For \text{(B-4)}, consider a pair $(\bft,\bft')$. Wlog assume that $\bft'$ comes after $\bft$. If $\bft'$ was forward $\textsf{RK}[\textsf{IC}_k](\D,L'R')\rightarrow S'T'$, then the obtained $S'T'$ is uniform in a set of size at least $N^2-q_e$, and since $q_e\leq N$ we have
$${\Pr}_{\textsf{IC}}[S'=L\xor L'\xor S]\leq\frac{N}{N^2-q_e}\leq\frac{1}{N-1}\leq\frac{2}{N}.$$
Similarly, ${\Pr}_{\textsf{IC}}[T'=R\xor R'\xor T]\leq\frac{2}{N}$. If $\bft'$ was backward $\textsf{RK}[\textsf{IC}_k]^{-1}(\D,S'T')\rightarrow L'R'$, then similarly
$${\Pr}_{\textsf{IC}}[L'=L\xor S\xor S']\leq\frac{2}{N},\text{ }
{\Pr}_{\textsf{IC}}[R'=R\xor T\xor T']\leq\frac{2}{N}.$$
Therefore, for each of the ${q_e\choose 2}\leq\frac{q_e^2}{2}$ pairs $(\bft,\bft')$, (B-4) is fulfilled with probability at most $4/N$. Thus $\Pr[\text{(B-4)}]\leq\frac{2q_e^2}{N}$.

Finally consider \text{(B-5)}. Fix a query $\bft=(\D,LR,ST)$. For $k\in\{0,1\}^n$, denote by $\mathcal{R}_1$ the set of possible values of $\wf_0(k_{\D})\xor\wf_2(k_{\D})$, and by $\mathcal{R}_2$ the set of values of $\wf_1(k_{\D})\xor\wf_3(k_{\D})$. If $\bft$ was forward, then the obtained $ST$ is uniform in $\geq N^2-q_e$ values, and (as argued) ${\Pr}_{\textsf{IC}}[L\xor S=v]\leq\frac{2}{N}$ for any fixed value $v\in\mathcal{R}_1$ and ${\Pr}_{\textsf{IC}}[R\xor T=v']\leq\frac{2}{N}$ for any $v'\in\mathcal{R}_2$. Therefore,
\begin{align*}
& {\Pr}_{\textsf{IC}}[L\xor\wf_0(k_{\D})=S\xor\wf_2(k_{\D})]  \\
= &  \sum_{v\in\mathcal{R}_1}{\Pr}_{\textsf{IC}}[L\xor S=v]\cdot{\Pr}_{k}[\wf_0(k_{\D})\xor\wf_2(k_{\D})=v]  \\
\leq & \frac{2}{N}\cdot \underbrace{\sum_{v\in\mathcal{R}_1}{\Pr}_k[\wf_0(k_{\D})\xor\wf_2(k_{\D})=v]}_{=1} \leq \frac{2}{N}.
\end{align*}
%%Similarly,
%%\begin{align*}	k\xleftarrow{\$}\{0,1\}^n
%%{\Pr}_{\textsf{IC}}[R\xor \wf_1(k\xor\D)=T\xor\wf_3(k\xor\D)] \leq \frac{1}{N-q_e}.
%%\end{align*}
Similarly, ${\Pr}_{\textsf{IC}}[R\xor \wf_1(k_{\D})=T\xor\wf_3(k_{\D})] \leq \frac{2}{N}$. When $\bft$ was backward, $LR$ is uniform, and similar bounds hold. Taking a union bound for the $q_e$ queries gives $\Pr[\text{(B-5)}]\leq \frac{4q_e}{N}$. Summing over the above yields
\begin{align}
\Pr[T_{id}\in\mathcal{T}_{bad}]\leq2\delta_1q_eq_f+(\delta_2+\delta_3)q_e^2+\frac{2q_e^2+4q_e}{N}.
\label{bad-transcript-4-rnd-probability}
\end{align}

\subsubsection{\bf Ratio ${\Pr}_{re}(\tau)/{\Pr}_{id}(\tau)$ for Good $\tau$}
\label{sec:ratio-4-rounds}

%Fix a good transcript $\tau$. We follow the ``predicate'' approach in~\cite{DBLP:conf/crypto/CogliatiDKLSTZ18}. We first define a ``bad'' predicate $\badf(P)$ on $P$ in paragraph \ref{para:proof-predicate-4-rounds}. Then, it's easy to see

Fix a good transcript $\tau$. As per our remark before, we define the bad predicate $\badf(P)$ in paragraph \ref{para:proof-predicate-4-rounds}. Then, it's easy to see      {\small
\begin{align}
{\Pr}_{re}(\tau)
&  ={\Pr}_P\big[\textsf{RK}[\kafwsp_k]\vdash\mathcal{Q}_E\wedge P\vdash\mathcal{Q}_P\big]   \notag   \\
&  \geq{\Pr}_P\big[\textsf{RK}[\kafwsp_k]\vdash\mathcal{Q}_E\wedge P\vdash\mathcal{Q}_P\wedge\neg\badf(P)\big]  \notag   \\
&  \ge\textsf{p} \cdot\Big(1-{\Pr}_P[\badf(P)\mid P\vdash\mathcal{Q}_P]\Big)\cdot{\Pr}_P[P\vdash\mathcal{Q}_P],    \label{eq:transforming-Pre-tau}
\end{align}
}%
where
\begin{align*}
\textsf{p} & ={\Pr}_P\big[\rk[\kafwsp_k]\vdash\mathcal{Q}_E\mid P\vdash\mathcal{Q}_P\wedge\neg\badf(P)\big].
\end{align*}
We next argue
\begin{align}
\textsf{p}\geq\frac{1}{N^{2q_e}}
\label{eq:bound-for-prob-p-4-rounds}
\end{align}
in paragraphs \ref{para:proof-manipulating-p-4-rounds} and \ref{para:proof-2qe-eq-6-rounds}. Gathering this and Eq. (\ref{eq:transforming-Pre-tau}) yields        {\small
\begin{align}
{\Pr}_{re}(\tau)\geq\frac{{\Pr}_P[P\vdash\mathcal{Q}_P]}{N^{2q_e}}\bigg(1-{\Pr}_P[\badf(P)\mid P\vdash\mathcal{Q}_P]\bigg),    \label{eq:lower-bound-for-Pre-tau}
\end{align}
}%
which allows us to conclude in paragraph \ref{para:final-counting-4-rounds}.

\arrangelittlespace

\paragraph{\underline{The Bad Predicate $\badf(P)$}}
\label{para:proof-predicate-4-rounds}

For any $P\vdash\mathcal{Q}_P$, the predicate $\badf(P)$ holds, if any of the following is fulfilled:
\begin{itemize}
	\item(C-1) $\exists\bft,\bft'\in\mathcal{Q}_{E}:x_1(\bft)\neq x_1(\bft')$, yet either $x_2(\bft,P)=x_2(\bft',P)$ or $y_3(\bft,P)=y_3(\bft',P)$.
	\item(C-2) $\exists\bft,\bft'\in\mathcal{Q}_{E}$ (could be $\bft=\bft'$):
	\begin{itemize}
		\item $x_2(\bft,P)\in\dom$ or $y_3(\bft,P)\in\rng$, or
		\item $x_2(\bft,P)=x_1(\bft')$ or $x_2(\bft,P)=x_4(\bft')$, or
		\item $y_3(\bft,P)=y_1(\bft',P)$ or $y_3(\bft,P)=y_4(\bft',P)$.
	\end{itemize}
	\item(C-3) $\exists\bft,\bft'\in\mathcal{Q}_{E}:x_4(\bft)\neq x_4(\bft')$, yet either $x_3(\bft,P)=x_3(\bft',P)$ or $y_2(\bft,P)=y_2(\bft',P)$.
	\item(C-4) $\exists\bft,\bft'\in\mathcal{Q}_{E}$ (could be $\bft=\bft'$):
	\begin{itemize}
		\item $x_3(\bft,P)\in\dom$ or $ y_2(\bft,P)\in\rng$, or
		\item $x_3(\bft,P)\in\big\{x_1(\bft'),x_2(\bft',P),x_4(\bft')\big\}$, or
		\item $y_2(\bft,P)\in\big\{y_1(\bft',P),y_3(\bft',P),y_4(\bft',P)\big\}$.
%%%%			\item $x_3(\bft,P)=x_1(\bft')$ or $ x_3(\bft,P)=x_2(\bft',P)$ or $x_3(\bft,P)=x_4(\bft')$, or
%%%%			\item $y_2(\bft,P)=y_1(\bft',P)$ or $y_2(\bft,P)=y_3(\bft',P)$ or $y_2(\bft,P)=y_4(\bft',P)$.
	\end{itemize}
\end{itemize}
\noindent{\underline{{\it Remark}}.} As per our discussion before, collisions in the $2q_e$ values in Eq. (\ref{eq:the-2qe-inputs}) and in the $2q_e$ values in Eq. (\ref{eq:the-2qe-outputs}) are captured by (C-1) and (C-3) resp. Moreover, there should be no ``conflict'' between these $4q_e$ values and the inputs/outputs in 1st and 4th rounds, as captured by (C-2) and (C-4). This is crucial, as the values of the forms $P(x_1(\bft))$ and $P(x_4(\bft))$ will be used for bounding $\Pr[\badf(P)]$, and it's unclear how this affects their distribution. Finally, note that $x_2(\bft,P)$ and $y_3(\bft,P)$ depends on the same random value $P(x_1(\bft))$ (and could be analyzed at the same time), while $x_3(\bft,P)$ and $y_2(\bft,P)$ depends on $P(x_4(\bft))$: this clarifies the order of the above bad conditions.

\arrangespace

We now analyze $\Pr[\badf(P)]$. Let $\bft=(\D,LR,ST)$. Consider (C-1) first. For each pair $(\bft,\bft')$, the event $x_2(\bft,P)=x_2(\bft',P)$ implies
\begin{align}
  & \ga_2(k_{\D})\xor L\xor\wf_0(k_{\D})\xor P(x_1(\bft))    \notag   \\
= & \ga_2(k_{\D'})\xor L\xor\wf_0(k_{\D'})\xor P(x_1(\bft')).
\label{eq:analyze-B1-4-rounds}
\end{align}
Define a set of function values $\mathcal{S}=\big\{P(x_i(\bft'))\mid\bft'\in\mathcal{Q}_E,i=1,4,x_i(\bft')\ne x_1(\bft)\big\}$. Then $|\mathcal{S}|\leq 2q_e$, and $P(x_1(\bft'))\in\mathcal{S}$ since $x_1(\bft)\neq x_1(\bft')$. Furthermore, by $\neg$(B-1) we have $x_1(\bft)\notin\dom$. Thus conditioned on $P\vdash\mathcal{Q}_P$ and further the function values in $\mathcal{S}$, $P(x_1(\bft))$ is uniform in a set of size {\it at least} $N-q_f-2q_e$. This means the left hand side of Eq. (\ref{eq:analyze-B1-4-rounds}) is random conditioned on the right hand side, thus $\Pr[x_2(\bft,P)=x_2(\bft',P)]\leq\frac{1}{N-q_f-2q_e}$. Similarly, $\Pr[y_3(\bft,P)=y_3(\bft',P)]\leq\frac{1}{N-q_f-2q_e}$. As we have ${q_e\choose 2}\leq\frac{q_e^2}{2}$ pairs $(\bft,\bft')$, it holds $\Pr[\text{(C-1)}]\leq\frac{q_e^2}{N-q_f-2q_e}$. A symmetrical analysis yields $\Pr[\text{(C-3)}]\leq{q_e\choose 2}\cdot\frac{2}{N-q_f-2q_e}\leq\frac{q_e^2}{N-q_f-2q_e}$.

We next consider (C-2). As argued, for any $\bft$, $X(\bft,P)$ is uniform in $\ge N-q_f-2q_e$ possibilities. On the other hand, all the values in $\dom$ are fixed by $\tau$ and thus independent from the function values of $P$. Therefore,
\begin{align}
\Pr[x_2(\bft,P)\in\dom]=  & 
\Pr[\gamma_2(k\oplus\D)\oplus X(\bft,P)\in\dom]    \notag  \\
\leq&\frac{q_f}{N-q_f-2q_e}.
\label{eq:argument-C2-4-rounds-subeve-1}
\end{align}
Similarly,
\begin{align}
& \Pr[\exists\bft':x_2(\bft,P)=x_1(\bft')\text{ or }
x_2(\bft,P)=x_4(\bft')]   \notag \\
\leq &  \frac{2q_e}{N-q_f-2q_e}, \label{eq:argument-C2-4-rounds-subeve-2} \\
& \Pr[y_3(\bft,P)\in\rng]\leq\frac{q_f}{N-q_f-2q_e}.
\label{eq:argument-C2-4-rounds-subeve-3}
\end{align}
Now consider $\Pr[\exists\bft':y_3(\bft,P)=y_4(\bft',P)]$. If this event happens, then
$$L\oplus\wf_0(k_{\D})\xor P(x_1(\bft))\xor\wf_2(k_{\D})\xor S=P(x_4(\bft')).$$
By $\neg$(B-3) we have $x_1(\bft)\ne x_4(\bft')$, so $P(x_4(\bft'))$ is random conditioned on the left hand side. Therefore,
\begin{align}
\Pr[\exists\bft':y_3(\bft,P)=y_4(\bft',P)]\leq\frac{q_e}{N-q_f-2q_e}.
\label{eq:argument-C2-4-rounds-subeve-4}
\end{align}
Finally consider $\Pr[\exists\bft':y_3(\bft,P)=y_1(\bft',P)]$. If this event happens, then for $\bft$ there exists $\bft'\in\mathcal{Q}_{E}$ such that
\begin{align}
L\oplus\wf_0(k_{\D})\xor P(x_1(\bft))\xor\wf_2(k_{\D})\xor S=P(x_1(\bft')).
\label{eq:argument-C2-IND-discussion-4-rounds}
\end{align}
We distinguish two cases:
\begin{enumerate}
	\item Case 1: $x_1(\bft)\ne x_1(\bft')$. Then $P(x_1(\bft'))$ is random conditioned on $P(x_1(\bft))$, and $\Pr[\text{Eq. }(\ref{eq:argument-C2-IND-discussion-4-rounds})]\leq\frac{1}{N-q_f-2q_e}$;
	\item Case 2: $x_1(\bft)=x_1(\bft')$. Then for this tuple $\bft$ we have $L\oplus\wf_0(k_{\D})=\wf_2(k_{\D})\xor S$, which contradicts $\neg$(B-5) (Definition \ref{definition:bad-transcripts-non-linear-ks-4-rnd-KAFwsp}).  \label{label:use-of-B5-4-round}
\end{enumerate}
As we have $q_e$ choices for $\bft'$ we obtain
\begin{align}
\Pr[\exists\bft':y_3(\bft,P)=y_1(\bft',P)]\leq\frac{q_e}{N-q_f-2q_e}.
\label{eq:argument-C2-4-rounds-subeve-5}
\end{align}
Summing over (\ref{eq:argument-C2-4-rounds-subeve-1}), (\ref{eq:argument-C2-4-rounds-subeve-2}), (\ref{eq:argument-C2-4-rounds-subeve-3}), (\ref{eq:argument-C2-4-rounds-subeve-4}), and (\ref{eq:argument-C2-4-rounds-subeve-5}), and taking union bound on the $q_e$ choices of $\bft$, we obtain    {\small
\begin{align}
\Pr[\text{(C-2)}] & \leq\frac{q_e(q_f+2q_e+q_f+q_e+q_e)}{N-q_f-2q_e} \leq\frac{2q_e(q_f+2q_e)}{N-q_f-2q_e}.
%
%+\frac{q_eq_f}{N-q_f-2q_e}+\frac{2q_e^2}{N-q_f-2q_e}  \notag  \\  & \leq\frac{2q_e(q_f+2q_e)}{N-q_f-2q_e}.   \notag
\end{align}    }%

\arrangespace

The analysis for (C-4) is similar by symmetry: for each $\bft=(\D,LR,ST)\in\mathcal{Q}_E$, $P(x_4(\bft))$ and further $Y(\bft,P)=T\oplus\wf_3(k_{\D})\xor P(x_4(\bft))$, $x_3(\bft,P)$, and $y_2(\bft,P)$ are uniform. By this, for $\bft$,      {\small
\begin{align}
& \Pr[x_3(\bft,P)\in\dom\text{ or }y_2(\bft,P)\in\rng]\leq\frac{2q_f}{N-q_f-2q_e}, \label{eq:argument-C4-4-rounds-subeve-1} \\
& \Pr[\exists\bft':x_3(\bft,P)=x_1(\bft')\text{ or }
x_3(\bft,P)=x_4(\bft')]  \notag  \\
 & \leq \frac{2q_e}{N-q_f-2q_e}, \label{eq:argument-C4-4-rounds-subeve-2}
\end{align} }%

Now consider $\Pr[\exists\bft':x_3(\bft,P)=x_2(\bft',P)]$. If it happens, then for $\bft$ there exists $\bft'=(\D',L'R',S'T')$ such that
\begin{align}
& \ga_3(k_{\D})\oplus T\oplus\wf_3(k_{\D})\xor P(x_4(\bft))   \notag   \\
= & \ga_2(k_{\D'})\xor L'\oplus\wf_0(k_{\D'})\xor P(x_1(\bft')).   \label{eq:analysis-C4-EX2-cap-EX3}
\end{align}
By $\neg$(B-3) we have $x_4(\bft)\neq x_1(\bft')$, so the right hand side of (\ref{eq:analysis-C4-EX2-cap-EX3}) is random conditioned on $P(x_4(\bft))$. Thus we have $\Pr[\text{Eq. }(\ref{eq:analysis-C4-EX2-cap-EX3})]\leq\frac{1}{N-q_f-2q_e}$, and further
\begin{align}
\Pr[\exists\bft':x_3(\bft,P)=x_2(\bft',P)]\leq\frac{q_e}{N-q_f-2q_e}. \label{eq:argument-C4-4-rounds-subeve-3}
\end{align}
By similar arguments, it can be shown
\begin{align}
& \Pr[\exists\bft':y_2(\bft,P)=y_1(\bft')\text{ or }y_2(\bft,P)=y_3(\bft',P)]    \notag  \\
\leq & \frac{2q_e}{N-q_f-2q_e}. \label{eq:argument-C4-4-rounds-subeve-4}
\end{align}

Finally consider $\Pr[\exists\bft':y_2(\bft,P)=y_4(\bft',P)]$. For $\bft$ if it happens then there exists $\bft'=(\D',L'R',S'T')$ such that
\begin{align}
R\xor\wf_1(k_{\D})\xor T\oplus\wf_3(k_{\D})\xor P(x_4(\bft))
=P(x_4(\bft')).  \label{eq:analysis-C4-EY2-cap-EY4}
\end{align}
If $x_4(\bft)\ne x_4(\bft')$ then the right hand side of (\ref{eq:analysis-C4-EY2-cap-EY4}) is random conditioned on $P(x_4(\bft))$ and thus $\Pr[\text{Eq. }(\ref{eq:analysis-C4-EY2-cap-EY4})]\leq\frac{1}{N-q_f-2q_e}$; otherwise i.e. $x_4(\bft)=x_4(\bft')$, then it implies $R\xor\wf_1(k_{\D})=T\oplus\wf_3(k_{\D})$, contradicting $\neg$(B-5). So
\begin{align}
\Pr[\exists\bft':y_2(\bft,P)=y_4(\bft',P)]\leq\frac{q_e}{N-q_f-2q_e}. \label{eq:argument-C4-4-rounds-subeve-5}
\end{align}
Summing over (\ref{eq:argument-C4-4-rounds-subeve-1}), (\ref{eq:argument-C4-4-rounds-subeve-2}), (\ref{eq:argument-C4-4-rounds-subeve-3}), (\ref{eq:argument-C4-4-rounds-subeve-4}), and (\ref{eq:argument-C4-4-rounds-subeve-5}), and taking union over $q_e$ yield
\begin{align}
\Pr[\text{(C-4)}] & \leq\frac{q_e(2q_f+2q_e+q_e+2q_e+q_e)}{N-q_f-2q_e} \leq\frac{2q_e(q_f+3q_e)}{N-q_f-2q_e}.   \notag
\end{align}

\arrangespace

Finally, summing over the four conditions yields
\begin{align}
{\Pr}[P\xleftarrow{\$}\permutationset:\badf(P)\mid P\vdash\mathcal{Q}_P]\leq\frac{4q_eq_f+12q_e^2}{N-q_f-2q_e}.
\label{kaf-4-rnd-bad-f-prob}
\end{align}

\arrangelittlespace

\arrangelittlespace

\paragraph{\underline{The Probability \textsf{p}}}
\label{para:proof-manipulating-p-4-rounds}

For any $P^*\vdash\mathcal{Q}_P$ such that $\badf(P^*)$ doesn't hold, we define an ``extended transcript''
\begin{align*}
\mathcal{Q}^{out}(P^*)=\big\{\big(x_1(\bft),y_1(\bft,P^*)\big),\big(x_4(\bft),y_4(\bft,P^*)\big)\big\}_{\bft\in\mathcal{Q}_E}.
\end{align*}
We further define $\mathcal{T}^{out}$ as the set of all such extended transcripts, i.e.,
\begin{align*} \mathcal{T}^{out}=\big\{\mathcal{Q}^{out}(P)\big\}_{P\in\permutationset},
\end{align*}
and a set of ``good'' extended transcripts based on permutations that don't fulfill the bad predicate, i.e.,
\begin{align*}
& \mathcal{T}_{good}^{out}=\big\{\mathcal{Q}^{out}(P^*)\big\}_{P^*\vdash\mathcal{Q}_P,\neg\badf(P^*)}.
\end{align*}
Next, for any instance $\mathcal{Q}^{out}\in\mathcal{T}^{out}$, we define another extended transcript $\mathcal{Q}^{mid}(\mathcal{Q}^{out})$. Formally, let $P^*$ be a permutation such that $\mathcal{Q}^{out}(P^*)=\mathcal{Q}^{out}$, then
\begin{align*}
\mathcal{Q}^{mid}(\mathcal{Q}^{out})=\big\{\big(x_2(\bft),y_2(\bft,P^*)\big),\big(x_3(\bft),y_3(\bft,P^*)\big)\big\}_{\bft\in\mathcal{Q}_E}.
\end{align*}
It's easy to see that such a choice of $P^*$ may not be unique, but for all $P^*$ with $\mathcal{Q}^{out}(P^*)=\mathcal{Q}^{out}$, the transcripts $\mathcal{Q}^{mid}(\mathcal{Q}^{out})$ defined as above are the same since the condition $\mathcal{Q}^{out}(P^*)=\mathcal{Q}^{out}$ ensures that $P^*$ is consistent with the input-output relations defined in $\mathcal{Q}^{out}$ which will fully characterize $\mathcal{Q}^{mid}(\mathcal{Q}^{out})$.

With these, by the definitions of $\kafw$ we have        {\small
\begin{align*}
\textsf{p}
&  = \sum_{\mathcal{Q}^{out}\in\mathcal{T}^{out}}\Pr[P\vdash\mathcal{Q}^{out}\mid P\vdash\mathcal{Q}_P\wedge\neg\badf(P)]   \\
& \ \ \ \ \ \ \ \ \ \ \ \ \ \ \ \ \cdot \Pr[P\vdash\mathcal{Q}^{mid}(\mathcal{Q}^{out})\mid P\vdash(\mathcal{Q}^{out}\cup\mathcal{Q}_P)\wedge\neg\badf(P)]       \\ 
&  \geq  \underbrace{\sum_{\mathcal{Q}^{out}\in\mathcal{T}_{good}^{out}}\Pr[P\vdash\mathcal{Q}^{out}\mid P\vdash\mathcal{Q}_P\wedge\neg\badf(P)]}_{=1}    \\
& \ \ \ \ \ \ \ \ \ \ \ \ \ \ \ \ \cdot \Pr[P\vdash\mathcal{Q}^{mid}(\mathcal{Q}^{out})\mid P\vdash(\mathcal{Q}^{out}\cup\mathcal{Q}_P)\wedge\neg\badf(P)] %\label{eq:transforming-Pre-tau}.
\end{align*}
}%
For any $\mathcal{Q}^{out}\in\mathcal{T}_{good}^{out}$, the conditions $\neg$(C-2) and $\neg$(C-4) ensure that
\begin{align*}
& \{x\text{ }\big|\text{ }\exists y:(x,y)\in\mathcal{Q}^{mid}(\mathcal{Q}^{out})\}     \\
& \ \ \ \ \ \ \ \ \ \ \bigcap\{x'\text{ }\big|\text{ }\exists y':(x',y')\in(\mathcal{Q}^{out}\cup\mathcal{Q}_P)\}=\emptyset,    \\
& \{y\text{ }\big|\text{ }\exists x:(x,y)\in\mathcal{Q}^{mid}(\mathcal{Q}^{out})\}     \\
& \ \ \ \ \ \ \ \ \ \ \bigcap\{y'\text{ }\big|\text{ }\exists x':(x',y')\in(\mathcal{Q}^{out}\cup\mathcal{Q}_P)\}=\emptyset.
\end{align*}
Thus      {\small
$$\Pr[P\vdash\mathcal{Q}^{mid}(\mathcal{Q}^{out})\mid P\vdash(\mathcal{Q}^{out}\cup\mathcal{Q}_P)\wedge\neg\badf(P)]\geq\frac{1}{N^{|\mathcal{Q}^{mid}(\mathcal{Q}^{out})|}}.$$
}%
In the next paragraph, we show $|\mathcal{Q}^{mid}(\mathcal{Q}^{out})|=2q_e$ to complete the proof of Eq. (\ref{eq:bound-for-prob-p-4-rounds}) and further (\ref{eq:lower-bound-for-Pre-tau}).

\arrangelittlespace

\paragraph{\underline{$2q_e$ Relations for Good $P$}}
\label{para:proof-2qe-eq-6-rounds}
By the definitions, for any $\mathcal{Q}^{out}\in\mathcal{T}_{good}^{out}$, there exists $P\vdash\mathcal{Q}_P$ such that $\badf(P)$ doesn't hold, and $\mathcal{Q}^{out}(P)=\mathcal{Q}^{out}$. Now we can write
$$\mathcal{Q}^{mid}(\mathcal{Q}^{out})=\{(x_2(\bft,P),y_2(\bft,P)),(x_3(\bft,P),y_3(\bft,P))\}.$$
We show $\big|\{x_2(\bft,P),x_3(\bft,P)\text{ }\big|\text{ }\bft\in\mathcal{Q}_E\}\big|=2q_e$ and $\big|\{y_2(\bft,P),y_3(\bft,P)\text{ }\big|\text{ }\bft\in\mathcal{Q}_E\}\big|=2q_e$. First, by $\neg\badf(P)$ (i.e., $\neg$(C-4)), for any pair $(\bft,\bft')$, it holds $x_2(\bft,P)\neq x_3(\bft',P)$ and $y_2(\bft,P)\neq y_3(\bft',P)$. It remains to show
\begin{itemize}
	\item $x_2(\bft,P)\neq x_2(\bft',P)$, $y_2(\bft,P)\neq y_2(\bft',P)$, and
	\item $x_3(\bft,P)\neq x_3(\bft',P)$, $y_3(\bft,P)\neq y_3(\bft',P)$.
\end{itemize}
Consider $(x_2(\bft,P),x_2(\bft',P))$ and $(y_3(\bft,P),y_3(\bft',P))$ first: their proof flows are similar. In detail, let $\bft=(\D,LR,ST)$ and $\bft'=(\D',L'R',S'T')$, then we exclude possibility of $x_2(\bft,P)=x_2(\bft',P)$ or $y_3(\bft,P)=y_3(\bft',P)$ for each case:
\begin{enumerate}
	\item Case 1: $\D\neq\D'$. Then $x_1(\bft,P)\neq x_1(\bft',P)$ by $\neg$(B-2)
	(see Definition
	\ref{definition:bad-transcripts-non-linear-ks-4-rnd-KAFwsp}), and further $x_2(\bft,P)\ne x_2(\bft',P)$ and $y_3(\bft,P)\ne y_3(\bft',P)$ by $\neg$(C-1);
	\item Case 2: $\D=\D'$, yet $R\neq R'$. Then still $x_1(\bft,P)\neq x_1(\bft',P)$, thus further $x_2(\bft,P)\ne x_2(\bft',P)$ and $y_3(\bft,P)\ne y_3(\bft',P)$;
	\item Case 3: $\D=\D'$ and $R=R'$. Then it has to be
	$L\neq L'$ since $\bft'\ne\bft'$. Now:
	\begin{itemize}
		\item On one hand, $L\neq L'$ immediately implies $x_2(\bft,P)=L\xor\wf_0(k_{\D})\xor y_1(\bft,P)\xor\ga_2(k_{\D})$ and
		$x_2(\bft',P)=L'\xor\wf_0(k_{\D})\xor y_1(\bft,P)\xor\ga_2(k_{\D})$ are distinct;
		\item On the other hand, $\D=\D'$ and $R=R'$ imply $X(\bft,P)\xor X(\bft',P)=L\xor L'$. By this,
		$y_3(\bft,P)=X(\bft,P)\xor\wf_2(k_{\D})\xor
		S=y_3(\bft',P)=X(\bft',P)\xor\wf_2(k_{\D})\xor S'$ would imply $L\xor L'=S\xor S'$, contradicting $\neg$(B-4).
	\end{itemize}
	\label{label:the-place-of-using-B4}
\end{enumerate}
By the above, it does hold $x_2(\bft,P)\neq x_2(\bft',P)$ and $y_3(\bft,P)\ne y_3(\bft',P)$ for any $\bft'\ne\bft'$. A symmetrical argument could establish $x_3(\bft,P)\neq x_3(\bft',P)$ and $y_2(\bft,P)\ne y_2(\bft',P)$ for any $\bft'\ne\bft'$.

\arrangelittlespace

%\noindent{\it \underline{The Final Counting}.}
\paragraph{\underline{The Final Counting}}
\label{para:final-counting-4-rounds}
By the above discussion and (\ref{eq:lower-bound-for-Pre-tau}) and (\ref{kaf-4-rnd-bad-f-prob}), when $q_f+2q_e\leq N/2$, for any
$\tau\in\mathcal{T}_{good}$ we have      {\small
\begin{align*}
\frac{{\Pr}_{re}(\tau)}{{\Pr}_{id}(\tau)}
\geq  &   \frac{{\Pr}_P[P\vdash\mathcal{Q}_P]}{N^{2q_e}}\bigg(1-{\Pr}_P[\badf(P)]\bigg)\Bigg/\frac{{\Pr}_P[P\vdash\mathcal{Q}_P]}{(N^2-q_e)^{q_e}}          \\
%	\geq  &   \frac{\Pr[F\vdash\mathcal{Q}_F]}{N^{2q_e}}  \bigg( 1-  \frac{4q_eq_f+12q_e^2}{N-q_f-2q_e} \bigg)\Bigg/\Pr[F\vdash\mathcal{Q}_F]\cdot\bigg(   \frac{1}{N^2-q_e}    \bigg)^{q_e}      \\
\geq  &
\left(1-\frac{4q_eq_f+12q_e^2}{N-q_f-2q_e}\right)
\left(\frac{N^2-q_e}{N^2}\right)^{q_e}  \\
\geq  & \left(1-\frac{8q_eq_f+24q_e^2}{N}\right)  \left(1-\frac{q_e^2}{N^2}\right)   \\
\geq  &  1 - \frac{8q_eq_f+25q_e^2}{N}.
\end{align*}   }%
Gathering this, (\ref{bad-transcript-4-rnd-probability}), and Lemma \ref{lemma:h-coefficients-main-lemma} yields Theorem
\ref{theorem:KAFwsp-rka-security-4-round}.
%(when $q_f+2q_e\leq N/2$, we also have $\frac{q_e^2+2q_e}{N-q_e}\leq\frac{2q_e^2+4q_e}{N}$).
% save a math
%\begin{align*}
%\frac{{\Pr}_{re}(\tau)}{{\Pr}_{id}(\tau)}
%\geq  &   \frac{\frac{1}{(N)_{q_f}\cdot N^{2q_e}}  \bigg( 1-  \frac{4q_eq_f+12q_e^2}{N-q_f-2q_e} \bigg)}{\frac{1}{(N)_{q_f}}\bigg(   \frac{1}{N^2-q_e}    \bigg)^{q_e}}      \\
%\geq  &
%\left(1-\frac{4q_eq_f+12q_e^2}{N-q_f-2q_e}\right)
%\left(\frac{N^2-q_e}{N^2}\right)^{q_e}  \\
%\geq  & \left(1-\frac{8q_eq_f+24q_e^2}{N}\right)  \left(1-\frac{q_e^2}{N^2}\right)   \\
%\geq  &  1 - \frac{8q_eq_f+25q_e^2}{N}.
%\end{align*}

\subsection{When f=F is a Random Function}
\label{subsection:security-kaf-rf-variant-non-linear-kdf-4-rnd}

With a good key-schedule specified in Definition
\ref{definition:good-non-linear-ks-4-rnd-KAFw}, the $\xor$-RKA
security claim still holds when we use a random function $F$ for
$f$. For the proof, we make some moderate modifications to the
previous proof for \kafwsp. First, (of course) the helper functions $y_1(\bft,P),X(\bft,P),\ldots$ here are defined on $F$ instead of $P$, i.e. $y_1(\bft,F),X(\bft,F),\ldots$

Then, note that since $F$ is a random function, for the to-be-derived $2q_e$
equalities
\begin{align*}
\big\{F(x_2(\bft,F))=y_2(\bft,F),F(x_3(\bft,F))=y_3(\bft,F)\text{ }\big|\text{ }\bft\in\mathcal{Q}_E\big\},
\end{align*}
collisions within the image set
$\{y_2(\bft,F),y_3(\bft,F)\text{ }\big|\text{ }\bft\in\mathcal{Q}_E\}$ would not be troublesome. Therefore, the main task is to drop definitions and arguments concerning these image values.

In detail, we recall
that in the definition of bad transcripts (Definition
\ref{definition:bad-transcripts-non-linear-ks-4-rnd-KAFwsp}),
	\begin{itemize}
		\item the condition (B-4) is only used for proving $\big|\{y_2(\bft,F)\text{ }\big|\text{ }\bft\in\mathcal{Q}_E\}\big|=\big|\{y_3(\bft,F)\text{ }\big|\text{ }\bft\in\mathcal{Q}_E\}\big|=q_e$ in the subsequent analysis, cf. the Case 3 in page
		\pageref{label:the-place-of-using-B4}, and
		\item (B-5) is only used for bounding $\Pr[\exists\bft,\bft':y_2(\bft,F)=y_4(\bft',F)]$ and $\Pr[\exists\bft,\bft':y_3(\bft,F)=y_1(\bft',F)]$, cf. Eq. (\ref{eq:argument-C2-IND-discussion-4-rounds}) and (\ref{eq:analysis-C4-EY2-cap-EY4}) in page \pageref{label:use-of-B5-4-round}.
	\end{itemize}
	So both (B-4) and (B-5) could be dropped. On the other
hand, (B-1), (B-2), and (B-3) and their probabilities remain unchanged. Subtracting the corresponding terms from
(\ref{bad-transcript-4-rnd-probability}) yields the following
bound for 4-round \kafwsf
\begin{align}
\Pr[T_{id}\in\mathcal{T}_{bad}]\leq2\delta_1q_eq_f+(\delta_2+\delta_3)q_e^2.
\label{kafwsf-4-rnd-bad-transcript-probability}
\end{align}

%%%For clearness, we present the modified definition of bad
%%%transcripts as follows.
%%%
%%%\begin{definition}[Bad Transcripts for 4-Round \textsf{KAFSF}, Non-Linear KS Case]
%%%\label{definition:bad-transcripts-non-linear-ks-4-rnd-KAFsf} An
%%%attainable transcript $\tau=(\mathcal{Q}_E,\mathcal{Q}_{F},k)$
%%%is \emph{bad}, if at least one of the following conditions is
%%%fulfilled:
%%%\begin{itemize}
%%%  \item(B-1) there exists $(\D,LR,ST)\in\mathcal{Q}_E$ such
%%%      that $\gamma_1(k\oplus\Delta)\oplus R\in Dom$ or
%%%      $\gamma_4(k\oplus\Delta)\oplus S\in Dom$;
%%%%      \label{bad-transcript-4-rnd-collision-in-qeqf}
%%%  \item(B-2) there exists two distinct $(\Delta,LR,ST)$ and
%%%      $(\Delta',L'R',S'T')$ in $\mathcal{Q}_E$ such that
%%%      $\Delta\neq\Delta'$, and
%%%      \begin{enumerate}
%%%        \item $\gamma_1(k\oplus\Delta)\oplus
%%%            R=\gamma_1(k\oplus\Delta')\oplus R'$, or
%%%        \item $\gamma_4(k\oplus\Delta)\oplus
%%%            S=\gamma_4(k\oplus\Delta')\oplus S'$.
%%%      \end{enumerate}
%%%%      \label{bad-transcript-4-rnd-collision-in-mere-qe}
%%%  \item(B-3) there exists two (not necessarily distinct)
%%%      $(\Delta,LR,ST)$ and $(\Delta',L'R',S'T')$ in
%%%      $\mathcal{Q}_E$ such that
%%%      $\gamma_1(k\oplus\Delta)\oplus
%%%      R=\gamma_4(k\oplus\Delta')\oplus S'$.
%%%\end{itemize}
%%%Otherwise we say $\tau$ is {\it good}. Denote by
%%%$\mathcal{T}_{bad}$ the set of bad keys.
%%%\end{definition}

We then modify the definition of $\badf(P)$ into $\badf(F)$. We
remark that for any value $x$ such that $F(x)$ remains
unknown, the function value $F(x)$ is uniform in
$\Zn$, which slightly deviates from the permutation case. Then, following the idea as before, we make the following
modifications:
\begin{enumerate}
	\item Dropping $y_3(\bft,F)=y_3(\bft',F)$ in (C-1). This
	decreases $\Pr[\text{(C-1)}]$ to $\frac{q_e^2}{2N}$ (with the above remark in mind);
	\item Dropping the condition(s) $\exists\bft,\bft':y_3(\bft,F)\in\rng\vee y_3(\bft,F)=y_1(\bft',F)\vee y_3(\bft,F)=y_4(\bft',F)$ in (C-2). This decreases $\Pr[\text{(C-2)}]$ to
	$\frac{q_e(q_f+2q_e)}{N}$;
	\item Dropping $y_2(\bft,F)=y_2(\bft',F)$ in (C-3). This
	decreases $\Pr[\text{(C-3)}]$ to $\frac{q_e^2}{2N}$;
	\item Dropping the condition(s) $\exists\bft,\bft':y_2(\bft,F)\in\rng\vee y_2(\bft,F)=y_1(\bft',F)\vee y_2(\bft,F)=y_3(\bft',F)\vee y_2(\bft,F)=y_4(\bft',F)$ in (C-4). This decreases $\Pr[\text{(C-2)}]$ to
	$\frac{q_e(q_f+3q_e)}{N}$.
\end{enumerate}
In total we have
\begin{align*}
{\Pr}[F\xleftarrow{\$}\functionset:\badf(F)\mid F\vdash\mathcal{Q}_F]\leq\frac{2q_eq_f+6q_e^2}{N}.
%\label{kaf-4-rnd-bad-f-prob}
\end{align*}

Finally,
\begin{align}
& \Pr[F\xleftarrow{\$}\functionset:\textsf{RK}[\kafwsf_k]\vdash\mathcal{Q}_E\mid F\vdash\mathcal{Q}_F\wedge\neg\badf(F)]       \notag   \\
 &  \geq {\Pr}_F\big[\forall\bft\in\mathcal{Q}_E:F(x_2(\bft,F))=y_2(\bft,F)     \notag   \\
& \ \ \ \ \ \ \ \ \ \ \ \ \ \ \ \ \wedge F(x_3(\bft,F))=y_3(\bft,F)\big]=\frac{1}{N^{2q_e}}.
\notag
\end{align}
Therefore,
\begin{align}
\frac{{\Pr}_{re}(\tau)}{{\Pr}_{id}(\tau)}\geq1-\frac{2q_eq_f+6q_e^2}{N}-\frac{q_e^2}{N^2}.
\label{kafwsf-4-rnd-bad-f-probability}
\end{align}

Gathering (\ref{kafwsf-4-rnd-bad-transcript-probability}) and
(\ref{kafwsf-4-rnd-bad-f-probability}) yields

\begin{theorem}
\label{theorem:KAFw-rf-variant-rka-security-4-round}

For the 4-round, random function-based \kafwsf~cipher with a
good key-schedule $(\wf,\ga)$ as specified in Definition
\ref{definition:good-non-linear-ks-4-rnd-KAFw}, it holds    {\small
$$\mathbf{Adv}^{\xor\text{-rka}}_{\kafwsf_k}(q_f,q_e)\leq2\delta_1q_eq_f+(\delta_2+\delta_3)q_e^2+\frac{2q_eq_f+7q_e^2}{N}.$$}
\end{theorem}

% ==================================================================

\section{\textsf{KAFw} with Affine Key-Schedules}
\label{section:KAFw-linear-KDF}

This section provides a comprehensive analysis of \kafw~with affine key-schedules. First, in section \ref{subsection:attack-kaf-linear-kdf-4-and-5-rnd}, we describe attacks against 4- and 5-round \kafw. These attacks can be easily adapted to \kaf~(of more general interest for attacks). Then, we prove security for 6-round \kafwsp~and \kafwsf~in sections \ref{subsect:app-proof-kafwsp-6-rounds} and \ref{subsect:app-proof-sketch-kafwsf-6-rounds} respectively.

\subsection{Insecurity for 4 and 5 Rounds}
\label{subsection:attack-kaf-linear-kdf-4-and-5-rnd}

We stress that, for attacks we consider $\kafw$ built upon {\it any} round-functions, and thus notations used in this subsection have slightly different meanings than those from section \ref{section:preliminary}. In detail, let $(\wf,\ga)$ be a $t$-round key-schedule, and $\overrightarrow{f}=(f_1,\ldots,f_t)$ be any $t$ functions. Then we define a $t$-round \kafw~variant      {\small
\begin{align*}
\kafw_k^{\overrightarrow{f},(\wf,\ga)}(W)
=   wk_{out}\xor\Psi_{\ga_t(k)}^{f_t}\circ\ldots\circ\Psi_{\ga_1(k)}^{f_1}\big(wk_{in}\xor W\big),
\end{align*}
}%
where $wk_{in}=\wf_0(k)\|\wf_1(k)$ and $wk_{out}=\wf_2(k)\|\wf_3(k)$. And for any distinguisher $D$, we define
\begin{align*}
& \mathbf{Adv}^{\xor\text{-rka}}_{\kafw_k^{\overrightarrow{f},(\wf,\ga)}}(D)
    \\
& = 
\Big|{\Pr}_{\textsf{IC},k}[D^{\textsf{RK}[\textsf{IC}_k],\overrightarrow{f}}=1] - {\Pr}_{k}[D^{\textsf{RK}[\kafw_k^{\overrightarrow{f},(\wf,\ga)}],\overrightarrow{f}}=1]\Big|.
\end{align*}
With these notations, subsubsections \ref{subsection:attack-kaf-linear-kdf-4-rnd} and \ref{subsection:discussion-kaf-linear-kdf-5-rnd} below present negative results on 4 and 5 rounds respectively.

%\subsection{Insecurity for 4 Rounds with Any Affine Key-Schedules}
\subsubsection{{\bf Insecurity for 4 Rounds with Any Affine Key-Schedules}}
\label{subsection:attack-kaf-linear-kdf-4-rnd}

\renewcommand\theenumi{\arabic{enumi}}
\renewcommand\labelenumi{(\theenumi)}

From a cryptanalytic point of view, note that for \kafw~with affine key-schedules, we have 2-round related-key differential characteristics with probability 1: see Eq. (\ref{eq:differential-1-2-rnd}) and (\ref{eq:differential-3-4-rnd}) below. Concatenating them would yield a 4-round related-key boomerang distinguisher~\cite{rkBoomerangEC2005} that consumes only four related-key oracle queries. Formally, we have

\begin{theorem}
	\label{theorem:insecurity-4-rounds}
	
	There exists a $(0,4)$-distinguisher $D$ such that, for any 4 functions $\overrightarrow{f}=(f_1,f_2,f_3,f_4)$ and any 4-round affine key-schedule $(\wf,\ga)$ where $\wf$ and $\ga$ are as defined in Eq. (\ref{eq:defn-wf-affine-func}) and (\ref{eq:defn-gamma-affine-func}), it holds
	$$\mathbf{Adv}^{\xor\text{-rka}}_{\kafw_k^{\overrightarrow{f},(\wf,\ga)}}(D)\ge1-\frac{1}{N^2-1}.$$
\end{theorem}
%This insecurity claim {\it does not require identical round-functions}. The same for the attack in the next subsection.
\begin{IEEEproof}
	We denote generically $(\rk[\e_k],\overrightarrow{f})$ the oracles to which the adversary has access, where $\e$ is either $\kafw^{\overrightarrow{f},(\wf,\ga)}$ or \textsf{IC}. The distinguisher $D$ proceeds as:
	\begin{enumerate}
		\item choose arbitrary values $L,R,\D\in\Zn$, $\D\neq0$, let $\nabla_1=(M_0^{(w)}\xor
		M_2)\cdot\D$, $\nabla_2=(M_1^{(w)}\xor
		M_1)\cdot\D$, $\nabla_3=(M_4\xor
		M_2^{(w)})\cdot\D$, and $\nabla_4=(M_3\xor
		M_3^{(w)})\cdot\D$. Make two queries $\textsf{RK}[\e_k](0,L\|R)\rightarrow S\|T$
		and $\textsf{RK}[\e_k](\D,L\xor\nabla_1\|R\xor\nabla_2)\rightarrow S'\|T'$;
		\item make two decryption queries $\textsf{RK}[\e_k]^{-1}(\D,S''\|T'')\rightarrow L''\|R''$ and
		$\textsf{RK}[\e_k]^{-1}(0,S'''\|T''')\rightarrow L'''\|R'''$, for $S''\|T''=S\xor\nabla_3\|T\xor\nabla_4$ and $S'''\|T'''=S'\xor\nabla_3\|T'\xor\nabla_4$;
		\item if $(L''\|R'')\xor(L'''\|R''')=\nabla_1\|\nabla_2$ then output
		1 to indicate $\e$ is $\kafw^{\overrightarrow{f},(\wf,\ga)}$, and
		otherwise 0: $\e$ is \textsf{IC}.
	\end{enumerate}
	We show the output is always 1 when $\e$ is $\kafw^{\overrightarrow{f},(\wf,\ga)}$. It's not hard to see for any $i$ and any $V,\D\in\Zn$, it holds
	$$\ga_i(k)\xor V=\ga_i(k_{\D})\xor V\xor M_i\cdot\D$$
	for $\kd=k\xor\D$. By this, for any $\D$, it holds
	\begin{align}
 &\Pr_{k,V,W}\big[\Psi_{\ga_i(k)}^{f_i}(V\|W)\xor
		\Psi_{\ga_i(k)}^{f_i}(V\xor M_{i+1}\cdot\D\|W\xor M_i\cdot\D)   \notag   \\
	&\ \ \ \ \ \ \ \ \ \ \ \ \ \ \ \ 	=M_i\cdot\D\|M_{i+1}\cdot\D\big]=1.\notag
	\end{align}
	This is essentially a 1-round related-key differential with probability 1. To ease exposition, we follow the notation in~\cite{KHP12IT} and denote this phenomena by		
	\begin{align}
	\Pr\Big(M_{i+1}\cdot\D\|M_i\cdot\D\xrightarrow[\D]{\Psi_{\ga_i(k)}^{f_i}}M_i\cdot\D\|M_{i+1}\cdot\D\Big)=1.   \notag
	\end{align}
	Concatenating two such differentials gives rise to two 2-round related-key differentials with probability 1 as follows
	
	{\footnotesize
	\begin{align}
	\Pr\Big(\nabla_1\|\nabla_2\xrightarrow[\D]{\Psi_{\ga_2(k)}^{f_2}\circ\Psi_{\ga_1(k)}^{f_1}\circ\xornotation_{wk_{in}}}M_2\cdot\D\|M_1\cdot\D\Big)=1, \label{eq:differential-1-2-rnd}     \\
	\Pr\Big(M_4\cdot\D\|M_3\cdot\D\xrightarrow[\D]{\xornotation_{wk_{out}}\circ\Psi_{\ga_4(k)}^{f_4}\circ\Psi_{\ga_3(k)}^{f_3}}\nabla_3\|\nabla_4\Big)=1, \label{eq:differential-3-4-rnd}
	\end{align}   }%
	where $wk_{in}=\wf_0(k)\|\wf_1(k)$, $wk_{out}=\wf_2(k)\|\wf_3(k)$, and $\xornotation_{wk}(W)=wk\xor W$.
	
	Therefore, for the two forward queries, if we assume
	\begin{align}
	\big(\Psi_{\ga_2(k)}^{f_2}\circ\Psi_{\ga_1(k)}^{f_1}\big)
	\big(wk_{in}\xor(L\|R)\big)=X\|Y,
	\label{eq:rkquery1-intermediate-value-after-2-rounds}
	\end{align}
	then by (\ref{eq:differential-1-2-rnd}) it holds    {\footnotesize
	\begin{align}
	& \big(\Psi_{\ga_2(\kd)}^{f_2}\circ\Psi_{\ga_1(\kd)}^{f_1}\big)\big(wk_{in}^{\D}\xor(L\xor\nabla_1\|R\xor\nabla_2)\big)   \notag  \\
	= & X\xor M_2\cdot\D\|Y\xor M_1\cdot\D
	\label{eq:rkquery2-intermediate-value-after-2-rounds}
	\end{align}    }%
	for $wk_{in}^{\D}=\wf_0(\kd)\|\wf_1(\kd)$. Eq. (\ref{eq:rkquery1-intermediate-value-after-2-rounds}) and (\ref{eq:rkquery2-intermediate-value-after-2-rounds}) also mean
	\begin{align*}
	&   \big(\Psi_{\ga_4(k)}^{f_4}\circ\Psi_{\ga_3(k)}^{f_3}\big)^{-1}\big(wk_{out}\xor(S\|T)\big)=X\|Y,       \\
	&   \big(\Psi_{\ga_4(\kd)}^{f_4}\circ\Psi_{\ga_3(\kd)}^{f_3}\big)^{-1}\big(wk_{out}^{\D}\xor(S'\|T')\big)   \\
	= & X\xor M_2\cdot\D\|Y\xor M_1\cdot\D,
	\end{align*}
	where $wk_{out}^{\D}=\wf_2(\kd)\|\wf_3(\kd)$, and $S,T,S',T'$ are the values appeared during the attack. Consider the two backward queries, and assume that
	\begin{align*}
	&  X''\|Y''=\big(\Psi_{\ga_4(\kd)}^{f_4}\circ\Psi_{\ga_3(\kd)}^{f_3}\big)^{-1}\big(wk_{out}^{\D}\xor(S''\|T'')\big),   \\
	&  X'''\|Y'''=\big(\Psi_{\ga_4(k)}^{f_4}\circ\Psi_{\ga_3(k)}^{f_3}\big)^{-1}\big(wk_{out}\xor(S'''\|T''')\big).
	\end{align*}
%%		Since $(S\|T)\xor(S''\|T'')=(S'\|T')\xor(S'''\|T''')=\nabla_3\|\nabla_4$, 
	By (\ref{eq:differential-3-4-rnd}) we have
	\begin{align*}
	&   X''\|Y''=  X\xor M_4\cdot\D\|Y\xor M_3\cdot\D,\text{ }   \\
	&   X'''\|Y'''=X\xor M_2\cdot\D\xor M_4\cdot\D\|Y\xor M_1\cdot\D\xor M_3\cdot\D,
	\end{align*}
	thus $(X''\|Y'')\xor(X'''\|Y''')=M_2\cdot\D\|M_1\cdot\D$. By this and (\ref{eq:differential-1-2-rnd}) it can be seen $(L''\|R'')\xor (L'''\|R''')=\nabla_1\|\nabla_2$.
	
	On the other hand, when interacting with
	$\textsf{RK}[\textsf{IC}_k]$, the last response $L'''\|R'''$ is uniform in $\{0,1\}^{2n}\backslash\{LR\}$. So ${\Pr}_{\textsf{IC}}[(L''\|R'')\xor(L'''\|R''')=(\nabla_1\|\nabla_2)]=\frac{1}{N^2-1}$, which is also the probability that the distinguisher outputs 1 in the ideal world. Thus the claimed bound.
\end{IEEEproof}

%\subsection{(In)security for 5 Rounds}
\subsubsection{{\bf (In)security for 5 Rounds}}
\label{subsection:discussion-kaf-linear-kdf-5-rnd}

We first exhibit an attack with only one additional assumption on the key-schedule: it's easy to derive $\D\neq0$ such that $M_1\cdot\D=M_5\cdot\D$. This is possible: e.g., if $\ga_1$ and $\ga_5$ are	bit-permutations, then for $\D=0\text{xFF}\ldots\text{FF}$ it holds $M_1\cdot\D=M_5\cdot\D=0\text{xFF}\ldots\text{FF}$.

From a cryptanalytic point of view, the core trick is: in the boomerang attack setting, under some conditions, Feistel schemes allow a {\it Feistel boomerang switch} trick~\cite{rkAESAsiacrypt09}, which enables penetrating one more round. Applying this trick to the 4-round related-key boomerang mentioned before yields a 5-round related-key boomerang distinguisher. Formally,
\begin{theorem}
	\label{theorem:insecurity-5-rounds}
	
	There exists a $(0,4)$-distinguisher $D$ such that, for any 5 functions $\overrightarrow{f}=(f_1,f_2,f_3,f_4,f_5)$ and any 5-round affine key-schedule $(\wf,\ga)$ where $\wf$ and $\ga$ are as defined in Eq. (\ref{eq:defn-wf-affine-func}) and (\ref{eq:defn-gamma-affine-func}) and satisfy that it's easy to derive $\D\neq0$ such that $M_1\cdot\D=M_5\cdot\D$, it holds
	$$\mathbf{Adv}^{\xor\text{-rka}}_{\kafw_{k}^{\overrightarrow{f},(\wf,\ga)}}(D)\ge1-\frac{1}{N^2-1}.$$
\end{theorem}
\begin{IEEEproof}
	The distinguisher $D$ proceeds as:
	\begin{enumerate}
		\item derive a difference $\D\neq0$ such
		that $M_1\cdot\D=M_5\cdot\D$;
		
		\item choose two arbitrary values $L,R\in\Zn$, and let $\nabla_1=(M_0^{(w)}\xor
		M_2)\cdot\D$, $\nabla_2=(M_1^{(w)}\xor
		M_1)\cdot\D$, $\nabla_3=(M_5\xor
		M_2^{(w)})\cdot\D$, and $\nabla_4=(M_4\xor
		M_3^{(w)})\cdot\D$. Make two queries
		$\textsf{RK}[\e_k](0,L\|R)\rightarrow S\|T$ and
		$\textsf{RK}[\e_k](\D,L\xor\nabla_1\|R\xor\nabla_2)\rightarrow S'\|T'$;
		\item query $\textsf{RK}[\e_k]^{-1}(\D,S\xor\nabla_3\|T\xor\nabla_4)\rightarrow L''\|R''$ and
		$\textsf{RK}[\e_k]^{-1}(0,S'\xor\nabla_3\|T'\xor\nabla_4)\rightarrow L'''\|R'''$;
		\item if $(L''\|R'')\xor(L'''\|R''')=\nabla_1\|\nabla_2$ then output
		1 to indicate $\e$ is $\kafw^{\overrightarrow{f},(\wf,\ga)}$, and otherwise 0: $\e$ is \textsf{IC}.
	\end{enumerate}
	We show the output is always 1 when $\e$ is $\kafw^{\overrightarrow{f},(\wf,\ga)}$. Assume that $wk_{in}=\wf_0(k)\|\wf_1(k)$, and
	\begin{align*}
	\big(\Psi_{\ga_2(k)}^{f_2}\circ\Psi_{\ga_1(k)}^{f_1}\big)
	\big(wk_{in}\xor(L\|R)\big)=X\|Y,
	\end{align*}
	then by (\ref{eq:differential-1-2-rnd}) we have
	\begin{align*}
 & \big(\Psi_{\ga_2(\kd)}^{f_2}\circ\Psi_{\ga_1(\kd)}^{f_1}\big)\big(wk_{in}^{\D}\xor(L\xor\nabla_1\|R\xor\nabla_2)\big)   \\
	= & X\xor M_2\cdot\D\|Y\xor M_1\cdot\D
	\end{align*}
	for $k_{\D}=k\xor\D$ and $wk_{in}^{\D}=\wf_0(\kd)\|\wf_1(\kd)$. Computing one more round, we obtain
	\begin{align}
 & \big(\Psi_{\ga_3(k)}^{f_3}\circ\Psi_{\ga_2(k)}^{f_2}\circ\Psi_{\ga_1(k)}^{f_1}\big)
	\big(wk_{in}\xor(L\|R)\big)   \notag   \\
	= & Y\|X\xor f_3(\ga_3(k)\xor Y)
	\label{eq:rkquery1-intermediate-value-after-3-rounds-at-5-rnd}
	\end{align}
	and
	\begin{align}
	& \big(\Psi_{\ga_3(\kd)}^{f_3}\circ\Psi_{\ga_2(\kd)}^{f_2}\circ\Psi_{\ga_1(\kd)}^{f_1}\big)\big(wk_{in}^{\D}\xor(L\xor\nabla_1\|R\xor\nabla_2)\big)   \notag   \\
	= & Y\xor M_1\cdot\D\|X\xor M_2\cdot\D   \notag  \\ 
	&\codeindent\codeindent\codeindent  \xor f_3(\ga_3(k)\xor M_3\cdot\D\xor Y\xor M_1\cdot\D).
	\label{eq:rkquery2-intermediate-value-after-3-rounds-at-5-rnd}
	\end{align}

	The differential Eq. (\ref{eq:differential-3-4-rnd}) should be adapted to 5 rounds:    {\footnotesize
	\begin{align}
	\Pr\Big(M_5\cdot\D\|M_4\cdot\D\xrightarrow[\D]{\xornotation_{wk_{out}}\circ\Psi_{\ga_5(k)}^{f_5}\circ\Psi_{\ga_4(k)}^{f_4}}\nabla_3\|\nabla_4\Big)=1,   \label{eq:differential-4-5-rnd}
	\end{align}    }%
	where $wk_{out}=\wf_2(k)\|\wf_3(k)$. By this and Eq. (\ref{eq:rkquery1-intermediate-value-after-3-rounds-at-5-rnd}) and (\ref{eq:rkquery2-intermediate-value-after-3-rounds-at-5-rnd}), if we assume $wk_{out}^{\D}=\wf_2(\kd)\|\wf_3(\kd)$,    {\small
	\begin{align*}
	Y''\|Z''=&\big(\Psi_{\ga_5(\kd)}^{f_5}\circ\Psi_{\ga_4(\kd)}^{f_4}\big)^{-1}\big(wk_{out}^{\D}\xor(S\xor\nabla_3\|T\xor\nabla_4)\big),   \\
	X''\|Y''=&\big(\Psi_{\ga_3(\kd)}^{f_3}\big)^{-1}\big(Y''\|Z''\big),  \\ Y'''\|Z'''=&\big(\Psi_{\ga_5(k)}^{f_5}\circ\Psi_{\ga_4(k)}^{f_4}\big)^{-1}\big(wk_{out}\xor(S'\xor\nabla_3\|T'\xor\nabla_4)\big),    \\
	X'''\|Y'''=&\big(\Psi_{\ga_3(k)}^{f_3}\big)^{-1}\big(Y'''\|Z'''\big),
	\end{align*}
}%
	then it necessarily holds
	\begin{align*}
	&  Y''\|Z''=Y\xor M_5\cdot\D\|X\xor f_3(\ga_3(k)\xor Y)\xor M_4\cdot\D,   \\
	& X''=X\xor f_3(\ga_3(k)\xor Y)\xor M_4\cdot\D   \\
	&\codeindent\codeindent\codeindent\xor f_3(\ga_3(k)\xor M_3\cdot\D\xor Y\xor M_5\cdot\D),  \\
	&  Y'''\|Z'''=\underbrace{Y\xor
		M_1\cdot\D\xor M_5\cdot\D}_{=Y,\text{ since }M_1\cdot\D=M_5\cdot\D}\|X\xor M_2\cdot\D    \\
	& \codeindent\codeindent\codeindent\xor f_3(\ga_3(k)\xor M_3\cdot\D\xor Y\xor M_1\cdot\D)\xor M_4\cdot\D,    \\
	&  X'''=X\xor M_2\cdot\D\xor f_3(\ga_3(k)\xor M_3\cdot\D\xor Y\xor M_1\cdot\D)  \\
	&\codeindent\codeindent\codeindent\xor M_4\cdot\D\xor f_3(\ga_3(k)\xor Y).
	\end{align*}
	Now since $M_5\cdot\D=M_1\cdot\D$, it can be seen
	$$(X''\|Y'')\xor(X'''\|Y''')=M_2\cdot\D\|M_5\cdot\D=M_2\cdot\D\|M_1\cdot\D,$$
	which further indicates $(L''\|R'')\xor
	(L'''\|R''')=\nabla_1\|\nabla_2$
	by Eq. (\ref{eq:differential-1-2-rnd}).
	
	We've proved that the probability of outputting 1 in the ideal world is $1/(N^2-1)$ in the proof of Theorem \ref{theorem:insecurity-4-rounds}. Thus the claim.
\end{IEEEproof}

Note that we did not assume $\exists\D\neq0$ such that $M_1\cdot\D\neq M_3\cdot\D$ or $M_3\cdot\D\neq M_5\cdot\D$. In this case, the scheme suffers from simpler complementation-based attacks, see Appendix \ref{subsection:complementing-kaf-linear-kdf-any-rnd}. On the other hand, if there is no $\D\neq0$ such that $M_1\cdot\D=M_5\cdot\D$, then the above attack is not effective. In fact, we conjecture security in this (latter) case, but the proof would be a significantly different from those in this paper. Moreover, it's inferior in the sense that it requires
\emph{additional assumptions} on the key-schedule (i.e. $\forall\D\neq0,M_1\cdot\D\neq M_5\cdot\D$). We thereby leave it for future, and revert to 6 rounds.

\subsection{Security for 6 Rounds when f=P}
\label{subsect:app-proof-kafwsp-6-rounds}

We first present the conditions on the
key-schedule $(\wf,\ga)$ that are sufficient for security proof
for 6-round \kafwsg.

\begin{definition}[Good Affine Key-Schedule for 6 Rounds]
	\label{definition:good-linear-key-schedule-6-rnd-KAFw} We say
	that a 6-round key-schedule $(\wf,\ga)$, for which
	$\wf=(\wf_0,\wf_1,\wf_2,\wf_3)$, $\wf_i(k)=M_i^{(w)}\cdot k\xor
	C_i^{(w)}$, $\ga=(\ga_1,\ga_2,\ga_3,\ga_4,\ga_5,\ga_6)$, and
	$\ga_i(k)=M_i\cdot k\xor C_i$, is \emph{good}, if it satisfies
	the following conditions:
	\begin{enumerate}
		\item $\varphi_1$, $\varphi_6$, and
		$\varphi_1\oplus\varphi_6$ are bijective maps of
		$\Zn$, where $\underline{\varphi_1(k)=\wf_1(k)\xor\ga_1(k)}$, $\underline{\varphi_6(k)=\wf_2(k)\xor\ga_6(k)}$; \label{condition:6-rnd-good-ks-1-6-bijective}
		\item for any $\D\neq0$, $M_1\cdot\D\neq M_3\cdot\D$,
		$M_4\cdot\D\neq M_6\cdot\D$.
		\label{condition:6-rnd-good-ks-13-46-interference}
	\end{enumerate}
\end{definition}
The 1st condition resembles those in Definition
\ref{definition:good-non-linear-ks-4-rnd-KAFw}. On the other
hand, the 2nd condition prevents the complementing attacks. One could see Appendix
\ref{subsection:complementing-kaf-linear-kdf-any-rnd} for further insights.

%	With such a good key-schedule, the 6-round \kafwsp is secure against $\xor$-RKA.

\begin{theorem}
	\label{theorem:KAFw-rka-security-6-round}
	
	When $q_f+4q_e\leq N/2$, for the 6-round, random
	permutation-based \kafwsp~cipher with a good key-schedule $(\wf,\ga)$ as
	specified in Definition
	\ref{definition:good-linear-key-schedule-6-rnd-KAFw}, it holds
	$$\mathbf{Adv}^{\xor\text{-rka}}_{\kafwsp_k}(q_f,q_e)\leq\frac{14q_eq_f+57q_e^2+4q_e}{N}.$$
\end{theorem}
\noindent\underline{{\it Proof}.} The proof strategy is similar to that described in section \ref{sec:proof-strategy}. For any function transcript
$\mathcal{Q}_f=((x_1,y_1),\ldots,$ $(x_{q_f},y_{q_f}))$, we
define $\dom$ and $\rng$ as the sets
$\{x_1,\ldots,x_{q_f}\}$ and $\{y_1,\ldots,y_{q_f}\}$. We also define 16 functions for any tuple $\bft=(\D,LR,ST)$ in $\mathcal{Q}_E$ and any function $f$ ($f=P\in\permutationset$ in this subsection):
\begin{itemize}
	\item $x_1(\bft) =\varphi_1(k\xor\D)\xor R$,
	\item $y_1(\bft,f) =f(x_1(\bft))$,
	\item $X(\bft,f) =L\xor\wf_0(k\xor\D)\xor y_1(\bft,f)$,
	\item $x_2(\bft,f)=\ga_2(k\xor\D)\xor X(\bft,f)$,
	\item $y_2(\bft,f)=f(x_2(\bft,f))$,
	\item $Y(\bft,f) =R\xor\wf_1(k\xor\D)\xor y_2(\bft,f)$,
	\item $x_3(\bft,f)=\ga_3(k\xor\D)\xor Y(\bft,f)$,
	\item $y_3(\bft,f)=X(\bft,f)\xor Z(\bft,f)$,
	\item $Z(\bft,f)=S\xor\wf_2(k\xor\D)\xor y_5(\bft,f)$,
	\item $x_4(\bft,f)=\ga_4(k\xor\D)\xor Z(\bft,f)$,
	\item $y_4(\bft,f)=Y(\bft,f)\xor A(\bft,f)$,
	\item $A(\bft,f) =T\xor\wf_3(k\xor\D)\xor y_6(\bft,f)$,
	\item $x_5(\bft,f)=\ga_5(k\xor\D)\xor A(\bft,f)$,
	\item $y_5(\bft,f)=f(x_5(\bft,f))$,
	\item $x_6(\bft) =\varphi_6(k\xor\D)\xor S$,
	\item $y_6(\bft,f) =f(x_6(\bft))$.
\end{itemize}

\arrangespace
	
\noindent{\bf Bad Transcripts\ } are then defined as follows.

\begin{definition}[Bad Transcripts for 6-Round \kafwsp]
	\label{definition:bad-transcripts-linear-ks-6-rnd-KAFwsp} An
	attainable transcript $\tau=(\mathcal{Q}_E,\mathcal{Q}_P,k)$
	is \emph{bad}, if at least one of the following conditions is
	fulfilled:
	\begin{itemize}
		\item(B-1) $\exists\bft\in\mathcal{Q}_E:x_1(\bft)\in
		\dom$ or $x_6(\bft)\in\dom$;
		\item(B-2) $\exists\bft,\bft'\in\mathcal{Q}_E:x_1(\bft)=x_6(\bft')$;
		\item(B-3) there exist two queries $\bft=(\Delta,LR,ST)$ and
		$\bft'=(\Delta',L'R',S'T')$ in $\mathcal{Q}_E$ such that
		$\D\neq\D'$, and $R\xor R'=(M_1^{(w)}\xor
		M_1)\cdot(\D\xor\D')$ and $S\xor S'=(M_6\xor
		M_2^{(w)})\cdot(\D\xor\D')$.
	\end{itemize}
	Otherwise we say $\tau$ is {\it good}. Denote by
	$\mathcal{T}_{bad}$ the set of bad transcripts.
\end{definition}
Recall that    {\small
\begin{align*}
&\varphi_1(k)=\wf_1(k)\xor\ga_1(k)=M_1^{(w)}\cdot k\xor C_1^{(w)}\xor M_1\cdot k\xor C_1,\text{ and}  \\
&\varphi_6(k)=\wf_2(k)\xor\ga_6(k)=M_2^{(w)}\cdot k\xor C_2^{(w)}\xor M_6\cdot k\xor C_6.
\end{align*}   }%
Since both $\varphi_1$ and $\varphi_6$ are bijective maps of
$\mathbb{F}_2^n$, $\Pr[\text{(B-1)}]\leq\frac{2q_eq_f}{N}$ is obvious. On the other hand, since
$\varphi_1\oplus\varphi_6$ is also bijective, for each choice
of $\bft=(\Delta,LR,ST)$ and $\bft'=(\Delta',L'R',S'T')$ it holds    {\small
\begin{align*}
& \Pr[x_1(\bft)=x_6(\bft')]   \\
=& \Pr\big[(M_1^{(w)}\xor M_1)\cdot(k_{\D})\xor C_1^{(w)}\xor C_1\xor R   \\
&  \textcolor{white}{\Pr[]}=(M_2^{(w)}\xor M_6)\cdot(k_{\D})\xor (M_2^{(w)}\xor M_6)\cdot(\D\xor\D')      \\
& \codeindent\codeindent\codeindent \xor C_2^{(w)}\xor C_6\xor S'\big]   \\
=& \Pr\big[(\varphi_1\oplus\varphi_6)(k_{\D})=(M_2^{(w)}\xor M_6)\cdot(\D\xor\D')\xor R\oplus S'\big]    \\
=&\frac{1}{N}.
\end{align*}    }%
Therefore, $\Pr[\text{(B-2)}]\leq\frac{q_e^2}{N}$.
Ultimately, for (B-3), for any such two queries
$(\Delta,LR,ST)$ and $(\Delta',L'R',S'T')$, following an analysis similar to (B-4)
in Definition
\ref{definition:bad-transcripts-non-linear-ks-4-rnd-KAFwsp}, the probability
that $R\xor R'=(M_1^{(w)}\xor M_1)\cdot(\D\xor\D')$ and
$S\xor S'=(M_6\xor M_2^{(w)})\cdot(\D\xor\D')$ are both fulfilled is at most $2/N$ when $q_e\leq N$. Thus $\Pr[\text{(B-3)}]\leq\frac{q_e^2}{N}$. In all,
\begin{align}
\Pr[T_{id}\in\mathcal{T}_{bad}]\leq\frac{2q_eq_f+2q_e^2}{N}.
\label{bad-transcripts-6-rnd-probability-kafsp}
\end{align}

\arrangespace

\noindent{\bf Ratio ${\Pr}_{re}(\tau)/{\Pr}_{id}(\tau)$ for Good $\tau$.}
We define two bad predicates on $P$ in turn. Then, using an argument similar to subsection \ref{sec:ratio-4-rounds}, we show that if neither of the two predicates holds, then $\Pr[\kafwsp_k\vdash\mathcal{Q}_E]\geq\frac{1}{N^{2q_e}}$. These cinch the bounds.

\arrangespace

\noindent{\it \underline{First Bad Predicate}.}
For any $P\vdash\mathcal{Q}_P$, the predicate
$\badfi(P)$ holds, if any of the following
conditions is fulfilled:
\begin{itemize}
	\item(C-11) there exist $\bft=(\D,LR,ST)$ and $\bft'=(\D',L'R',S'T')$ in $\mathcal{Q}_{E}$ such that $x_1(\bft)\neq x_1(\bft')$, yet $x_2(\bft,P)=x_2(\bft',P)$ or $X(\bft,P)\xor
	X(\bft',P)=M_6\cdot(\D\xor\D')$;
	\item(C-12) $\exists\bft,\bft'\in\mathcal{Q}_{E}$ (could be $\bft=\bft'$): $x_2(\bft,P)\in\dom$, or $x_2(\bft,P)=x_1(\bft')$, or $x_2(\bft,P)=x_6(\bft')$;
	\label{condition:6-rnd-badF1-coll-Ext2-other}
	\item(C-13) there exist $\bft=(\D,LR,ST)$ and $\bft'=(\D',L'R',S'T')$ in $\mathcal{Q}_{E}$ such that $x_6(\bft)\neq x_6(\bft')$, yet $x_5(\bft,P)=x_5(\bft',P)$ or $A(\bft,P)\xor
	A(\bft',P)=M_1\cdot(\D\xor\D')$;
	\label{condition:6-rnd-badF1-coll-x5}
	\item(C-14) $\exists\bft,\bft'\in\mathcal{Q}_{E}$ (could be $\bft=\bft'$):	$x_5(\bft,P)\in\dom$, or $x_5(\bft,P)\in\big\{x_1(\bft'),x_2(\bft',P),x_6(\bft')\big\}$;
	\label{condition:6-rnd-badF1-coll-Ext5-other}
	\item(C-15) there exists a query
	$\bft=(\D,LR,ST)$ in $\mathcal{Q}_{E}$ such that
	\begin{itemize}
		\item $L\xor\wf_0(k\xor\D)\xor S\xor\wf_2(k\xor\D)=P(x_1(\bft))$, or
		\item $R\xor\wf_1(k\xor\D)\xor T\xor\wf_3(k\xor\D)=P(x_6(\bft)).$
	\end{itemize}
\end{itemize}
For (C-11), for each pair $(\bft,\bft')$ with $\bft=(\D,LR,ST)$ and $\bft'=(\D',L'R',S'T')$, the event $x_2(\bft,P)=x_2(\bft',P)$ is equivalent to $X(\bft,P)\xor
X(\bft',P)=M_2\cdot(\D\xor\D')$, which is further equivalent to
\begin{align}
& L\xor\wf_0(k_{\D})\oplus P(x_1(\bft))  \notag  \\
=& L'\xor\wf_0(k_{\D'})\oplus P(x_1(\bft'))\xor M_2\cdot(\D\xor\D').
\label{eq:argument-C11-6-rounds}
\end{align}
Since $\tau$ is good, it holds $x_1(\bft)\notin\dom$. Conditioned on $P\vdash\mathcal{Q}_P$ and the $\le2q_e$ function values $\big\{P(x_i(\bft'))\mid\bft'\in\mathcal{Q}_{E},i=1,6,x_i(\bft')\ne x_1(\bft)\big\}$ (which includes $P(x_1(\bft'))$ since $x_1(\bft)\neq x_1(\bft')$), $P(x_1(\bft))$ is uniform in at least $N-q_f-2q_e$ possibilities. Therefore, for each pair $(\bft,\bft')$, $\Pr[x_2(\bft,P)=x_2(\bft',P)]=\Pr[\text{Eq. } (\ref{eq:argument-C11-6-rounds})]\le\frac{1}{N-q_f-2q_e}$. For the same reason, $\Pr[X(\bft,P)\xor
X(\bft',P)=M_6\cdot(\D\xor\D')]\le\frac{1}{N-q_f-2q_e}$. Thus
$$\Pr[\text{(C-11)}]\le{q_e\choose 2}\cdot\frac{2}{N-q_f-2q_e}\le\frac{q_e^2}{N-q_f-2q_e}.$$

Then, the value $x_2(\bft,P)$ relies on $P(x_1(\bft))$, and is thus uniform. Since the values in $\dom$ and the values of the form $x_1(\bft')$ and $x_6(\bft')$ are all independent from $P(x_1(\bft))$, it holds $$\Pr[\text{(C-12)}]\leq\frac{q_e q_f}{N-q_f-2q_e}+\frac{q_e\cdot 2q_e}{N-q_f-2q_e}=\frac{q_e(q_f+2q_e)}{N-q_f-2q_e}.$$

For (C-13)
the analysis is similar to (C-11) by symmetry, yielding the same bound
$$\Pr[\text{(C-13)}]\le{q_e\choose 2}\cdot\frac{2}{N-q_f-2q_e}\le\frac{q_e^2}{N-q_f-2q_e}.$$
Similarly, the main claim in (C-14) can be bounded:     {\footnotesize
	\begin{align*}
	&\Pr[\exists\bft,\bft':x_5(\bft,P)\in\dom\text{ or }x_5(\bft,P)=x_1(\bft')\text{ or }x_5(\bft,P)=x_6(\bft')]   \\
	\leq&\frac{q_e(q_f+2q_e)}{N-q_f-2q_e}
	\end{align*}
}%
The remaining subevent of (C-14), i.e. $\exists\bft,\bft':x_5(\bft,P)=x_2(\bft',P)$, is equivalent to
\begin{align}
& \ga_5(k\xor\D)\xor T\xor\wf_3(k\xor\D)\xor P(x_6(\bft))  \notag  \\
=& \ga_2(k\xor\D')\xor L'\xor\wf_0(k\xor\D')\xor P(x_1(\bft')).
\label{eq:argument-C14-6-rounds}
\end{align}
By $\neg$(B-2), $x_1(\bft')\ne x_6(\bft)$, thus $P(x_1(\bft'))$---as well as the entire right hand side---is random conditioned on $P(x_6(\bft))$. Thus $\Pr[\exists\bft,\bft':\text{Eq. }(\ref{eq:argument-C14-6-rounds})\text{ holds}]\leq\frac{q_e^2}{N-q_f-2q_e}$, and       {\small
$$\Pr[\text{(C-14)}]\leq\frac{q_e(q_f+2q_e)}{N-q_f-2q_e}+\frac{q_e^2}{N-q_f-2q_e}\leq\frac{q_e(q_f+3q_e)}{N-q_f-2q_e}.$$
}%
Finally, since both $P(x_1(\bft))$ and $P(x_6(\bft))$ are uniform for each
$\bft$, we immediately obtain $\Pr[\text{(C-15)}]\leq\frac{2q_e}{N-q_f-2q_e}$.
Summing over $\Pr[\text{(C-11)}]$ to $\Pr[\text{(C-15)}]$, we reach     {\small
\begin{align}
{\Pr}[P\xleftarrow{\$}\permutationset:\badfi(P)\mid
P\vdash\mathcal{Q}_P]\leq\frac{2q_eq_f+7q_e^2+2q_e}{N-q_f-2q_e}.
\label{kaf-6-rnd-bad-f-1-prob}
\end{align}
}%

%%
%%We finally remark that as $F$ is a permutation, the events $\neg$(C-12) and $\neg$(C-14) imply
%%\begin{align*}
%%& \extrngii\cap(\rng\cup \extrngi\cup
%%\extrngvi)=\emptyset,\text{ and}   %\\
%%%&
%%\text{ }\extrngv\cap(\rng\cup \extrngi\cup
%%\extrngvi\cup \extrngii)=\emptyset.      \\
%%\end{align*}

%$\extdomii\cap(\dom\cup \extdomi\cup\extdomvi)=\emptyset$ and $\extdomv\cap(\dom\cup \extdomi\cup\extdomvi\cup \extdomii)=\emptyset$ imply $$\extrngii\cap(\rng\cup \extrngi\cup\extrngvi)=\emptyset$$ and $$\extrngv\cap(\rng\cup \extrngi\cup\extrngvi\cup \extrngii)=\emptyset.$$

\arrangespace

\noindent{\it \underline{Second Bad Predicate}.}
We then consider a random permutation $P$ such that
$P\vdash\mathcal{Q}_P$ and $\neg\badfi(P)$. For this $P$, the predicate $\badfii(P)$ holds if any of the following conditions is fulfilled:
\begin{itemize}
	\item(C-21) $\exists\bft,\bft'\in\mathcal{Q}_{E}:x_2(\bft,P)\neq x_2(\bft',P)$, yet either $x_3(\bft,P)=x_3(\bft',P)$ or
	$y_4(\bft,P)=y_4(\bft',P)$;
	\label{condition:6-rnd-badF1-coll-x3}
	\item(C-22) $\exists\bft,\bft'\in\mathcal{Q}_{E}$ (could be $\bft=\bft'$):
	\begin{itemize}
		\item $x_3(\bft,P)\in\dom$ or $y_4(\bft,P)\in\rng$, or
		\item $x_3(\bft,P)\in\big\{x_1(\bft'),x_2(\bft',P),x_5(\bft',P),x_6(\bft')\big\}$, or
		\item $y_4(\bft,P)\in\big\{y_1(\bft',P),y_2(\bft',P),y_5(\bft',P),y_6(\bft',P)\big\}$.
%%%%%			\item $x_3(\bft,P)=x_1(\bft')$, or $x_3(\bft,P)=x_2(\bft',P)$, or $x_3(\bft,P)=x_5(\bft',P)$, or $x_3(\bft,P)=x_6(\bft')$, or
%%%%%			\item $y_4(\bft,P)=y_1(\bft',P)$, or $y_4(\bft,P)=y_2(\bft',P)$, or $y_4(\bft,P)=y_5(\bft',P)$, or $y_4(\bft,P)=y_6(\bft',P)$.
	\end{itemize}
	\label{condition:6-rnd-badF1-coll-Ext3-other}
	\item(C-23)  $\exists\bft,\bft'\in\mathcal{Q}_{E}:x_5(\bft,P)\neq x_5(\bft',P)$, yet either $x_4(\bft,P)=x_4(\bft',P)$ or
	$y_3(\bft,P)=y_3(\bft',P)$;
	\label{condition:6-rnd-badF1-coll-x4}
	\item(C-24) $\exists\bft,\bft'\in\mathcal{Q}_{E}$ (could be $\bft=\bft'$):
	\begin{itemize}
		\item $x_4(\bft,P)\in\dom$ or $y_3(\bft,P)\in\rng$, or
		\item $x_4(\bft,P)\in\big\{x_1(\bft'),x_2(\bft',P),x_3(\bft',P),x_5(\bft',P),$ $x_6(\bft')\big\}$, or
		\item $y_3(\bft,P)\in\big\{y_1(\bft',P),y_2(\bft',P),y_4(\bft',P),y_5(\bft',P),$ $y_6(\bft',P)\big\}$.
%%%%%			\item $x_4(\bft,P)=x_1(\bft')$, or $x_4(\bft,P)=x_2(\bft',P)$, or $x_4(\bft,P)=x_3(\bft',P)$, or $x_4(\bft,P)=x_5(\bft',P)$, or $x_4(\bft,P)=x_6(\bft')$, or
%%%%%			\item $y_3(\bft,P)=y_1(\bft',P)$, or $y_3(\bft,P)=y_2(\bft',P)$, or $y_3(\bft,P)=y_4(\bft',P)$, or $y_3(\bft,P)=y_5(\bft',P)$, or $y_3(\bft,P)=y_6(\bft',P)$.
	\end{itemize}
	\label{condition:6-rnd-badF1-coll-Ext4-other}
\end{itemize}
First, for each $\bft=(\D,LR,ST)$, conditioned on $P\vdash\mathcal{Q}_P$ and the $\le4q_e$ values
\begin{align*}
\{P(x_i(\bft')),
P(x_j(\bft',P)),\text{ }\big|\text{ }&\bft'\in\mathcal{Q}_{E},i=1,6,j=2,5,    \\
&x_j(\bft',P)\neq x_2(\bft,P)\},
\end{align*}
the value $y_2(\bft,P)=P(x_2(\bft,P))$
%	\begin{align*}
%	y_2(\bft,F)  &  =F(x_2(\bft,F))   \\
%	& \markgreen{  =F\big(\ga_2(k\xor\D)\xor L\xor\wf_0(k\xor\D)\oplus F(x_1(\bft))\big)  }
%	\end{align*}
remains uniform in at least $N-q_f-4q_e$ possibilities. So $Y(\bft,P)$, $x_3(\bft,P)$, and $y_4(\bft,P)$ derived from $y_2(\bft,P)$ are all uniform. These show:  

{\small
\begin{align}
&  \Pr[\text{(C-21)}]\leq{q_e\choose 2}\cdot\frac{2}{N-q_f-4q_e}\leq \frac{q_e^2}{N-q_f-4q_e},  \notag     \\
&  \Pr[\exists\bft:x_3(\bft,P)\in\dom]\leq\frac{q_eq_f}{N-q_f-4q_e},   \label{eq:bound-C22-group-2}    \\
&   \Pr[\exists\bft,\bft':x_3(\bft,P)\in\big\{x_1(\bft'),x_6(\bft')\big\}]\leq\frac{q_e\cdot 2q_e}{N-q_f-4q_e},  \label{eq:bound-C22-group-3}     \\
&   \Pr[\exists\bft:y_4(\bft,P)\in\rng]\leq\frac{q_eq_f}{N-q_f-4q_e}.   \label{eq:bound-C22-group-4}
\end{align}    }

Second, for cleanness let $k_{\D}=k\xor\D$ and $k_{\D'}=k\xor\D'$, then
\begin{align*}
& \Pr[\exists\bft,\bft':x_3(\bft,P)=x_2(\bft',P)]   \notag  \\
=& \Pr[\exists\bft,\bft':\underbrace{\ga_3(k_{\D})\xor R\xor\wf_1(k_{\D})}_{\con_1\text{, will be used below}}\xor\underline{P(x_2(\bft,P))}   \notag  \\
&\codeindent\codeindent\codeindent\codeindent=\ga_2(k_{\D'})\xor L'\xor\wf_0(k_{\D'})\xor \underline{P(x_1(\bft'))}].
\end{align*}
By $\neg$(C-12), $x_2(\bft,P)\ne x_1(\bft')$, $x_2(\bft,P)\ne x_6(\bft')$, $x_2(\bft,P)\ne x_6(\bft)$, so for the involved equality the right hand side is random conditioned on the left hand side. Therefore,
\begin{align}
\Pr[\exists\bft,\bft':x_3(\bft,P)=x_2(\bft',P)]
\le \frac{q_e^2}{N-q_f-4q_e}.
\label{eq:bound-C22-group-5}
\end{align}
For similar reasons,     {\small
\begin{align}
& \Pr[\exists\bft,\bft':x_3(\bft,P)=x_5(\bft',P)]   \notag  \\
=& \Pr[\exists\bft,\bft':\con_1\xor\underline{P(x_2(\bft,P))}  \notag \\
& \codeindent\codeindent\codeindent\codeindent\codeindent=\ga_5(k_{\D'})\xor T'\xor\wf_3(k_{\D'})\xor \underline{P(x_6(\bft'))}]     \notag   \\
\le& \frac{q_e^2}{N-q_f-4q_e},
\label{eq:bound-C22-group-6}       \\
& \Pr\big[\exists\bft,\bft':y_4(\bft,P)=y_1(\bft',P)\big]   \notag  \\
=& \Pr\big[\exists\bft,\bft':\big(R\xor\wf_1(k_{\D})\xor P(x_2(\bft,P))\big)    \notag  \\
&\codeindent\codeindent\codeindent \codeindent\codeindent \xor\big(T\xor\wf_3(k_{\D})\xor P(x_6(\bft))\big)=P(x_1(\bft'))\big]   \notag  \\
=& \Pr\big[\exists\bft,\bft':P(x_2(\bft,P))=R\xor\wf_1(k_{\D})\xor T     \notag  \\
&\codeindent\codeindent\codeindent\codeindent\codeindent   \xor\wf_3(k_{\D})\xor P(x_6(\bft))\xor P(x_1(\bft'))\big]   \notag  \\
\le & \frac{q_e^2}{N-q_f-4q_e}.
\label{eq:bound-C22-group-7}       \\
& \Pr\big[\exists\bft,\bft':y_4(\bft,P)=y_6(\bft',P)\big]   \notag  \\
=& \Pr\big[\exists\bft,\bft':P(x_2(\bft,P))=R\xor\wf_1(k_{\D})\xor T    \notag   \\
& \codeindent\codeindent\codeindent\codeindent\codeindent \xor\wf_3(k_{\D})\xor P(x_6(\bft))\xor P(x_6(\bft'))\big]   \notag  \\
\le & \frac{q_e^2}{N-q_f-4q_e}.
\label{eq:bound-C22-group-8}
\end{align}    }

Furthermore, by $\neg$(C-14), $\forall\bft,\bft',x_2(\bft,P)\ne x_5(\bft',P)$. So
\begin{align}
& \Pr[\exists\bft,\bft':y_4(\bft,P)=y_5(\bft',P)]   \notag  \\
=& \Pr[\exists\bft,\bft':R\xor\wf_1(k_{\D})\xor \underline{P(x_2(\bft,P))}\xor T\xor\wf_3(k_{\D})   \notag \\
&\codeindent\codeindent\codeindent\codeindent\xor P(x_6(\bft))=\underline{P(x_5(\bft',P))}]   \notag \\
\le&\frac{q_e^2}{N-q_f-4q_e}.
\label{eq:bound-C22-group-9}
\end{align}

Finally, for a pair $(\bft,\bft')$, $y_4(\bft,P)=y_2(\bft',P)$ would imply
\begin{align}
& R\xor\wf_1(k_{\D})\oplus \underline{P(x_2(\bft,P))}\xor T\xor\wf_3(k_{\D})\xor P(x_6(\bft))   \notag   \\
= & \underline{P(x_2(\bft',P))}.
\label{eq:analysis-for-6-round-eq-F-block-1}
\end{align}
Then,
\begin{enumerate}
	\item If $x_2(\bft,P)=x_2(\bft',P)$, then for $\bft=(\D,LR,ST)$ it holds
	$$R\xor\wf_1(k\xor\D)\xor T\xor\wf_3(k\xor\D)=P(x_6(\bft)),$$
	contradicting $\neg$(C-15); \label{label:use-of-C15-6-rounds}
	\item Otherwise, $P(x_2(\bft',P))$ is random conditioned on the left hand side of (\ref{eq:analysis-for-6-round-eq-F-block-1}), thus $\Pr[\text{Eq. }(\ref{eq:analysis-for-6-round-eq-F-block-1})]\leq\frac{1}{N-q_f-4q_e}$.
\end{enumerate}
As the number of pairs $(\bft,\bft')$ is at most $q_e^2$,
\begin{align}
\Pr[\exists\bft,\bft':y_4(\bft,P)=y_2(\bft',P)]
\le \frac{q_e^2}{N-q_f-4q_e}.
\label{eq:bound-C22-group-10}
\end{align}

Summing over (\ref{eq:bound-C22-group-2})-(\ref{eq:bound-C22-group-10}), we obtain $$\Pr[\text{(C-22)}]\leq\frac{2q_e(q_f+4q_e)}{N-q_f-4q_e}.$$

%	\begin{align*}
%	y_5(\bft,F)  &  =F(x_5(\bft,F))   \\
%	& \markgreen{ =F\big(\ga_5(k\xor\D)\xor T\xor\wf_3(k\xor\D)\oplus F(x_6(\bft))\big)  }
%	\end{align*}

Third, symmetrically, for each $\bft=(\D,LR,ST)\in\mathcal{Q}_{E}$, the value $y_5(\bft,P)=P(x_5(\bft,P))$ remains random. So $Z(\bft,P)$, $x_4(\bft,P)$, and $y_3(\bft,P)$ are all uniform. Therefore, $\Pr[\text{(C-23)}]\leq\frac{q_e^2}{N-q_f-4q_e}$. In addition, in a similar vein to the analysis of (C-22), we have
\begin{itemize}
	\item $\Pr[\exists\bft:x_4(\bft,P)\in\dom\text{ or }y_3(\bft,P)\in\rng]\leq\frac{2q_eq_f}{N-q_f-4q_e}$;
	\item $\Pr[\exists\bft,\bft':x_4(\bft,P)=x_1(\bft')\text{ or }x_4(\bft,P)=x_6(\bft')]\leq\frac{2q_e^2}{N-q_f-4q_e}$.
\end{itemize}
By $\neg$(C-14), $\forall\bft,\bft',x_5(\bft,P)\ne x_1(\bft')$. So: ($k_{\D}=k\xor\D,$ $k_{\D'}=k\xor\D'$)
\begin{align*}
& \Pr[\exists\bft,\bft':x_4(\bft,P)=x_2(\bft',P)] \\
= & \Pr[\underbrace{\ga_4(k_{\D})\xor S\xor\wf_2(k_{\D})}_{\con_2\text{, will be used below}}\xor \underline{P(x_5(\bft,P))}    \\
	& \codeindent\codeindent\codeindent=\ga_2(k_{\D'})\xor L'\xor\wf_0(k_{\D'})\xor \underline{P(x_1(\bft'))}]   \\
\leq & \frac{q_e^2}{N-q_f-4q_e}, \text{ and }   \\
& \Pr[\exists\bft,\bft':y_3(\bft,P)=y_1(\bft',P)]    \\
= &\Pr[\big(L\xor\wf_0(k_{\D})\xor P(x_1(\bft))\big)
\xor \big(S\xor\wf_2(k_{\D})    \\
&\codeindent\codeindent\codeindent\codeindent\codeindent \xor P(x_5(\bft,P))\big)=P(x_1(\bft'))]    \\
= &\Pr\big[P(x_5(\bft,P))=\underbrace{L\xor\wf_0(k_{\D})\xor S\xor\wf_2(k_{\D})}_{\con_3\text{, will be used later}}    \\
& \codeindent\codeindent\codeindent\codeindent\codeindent\xor P(x_1(\bft))\xor \underline{P(x_1(\bft'))}\big]   \\
\leq & \frac{q_e^2}{N-q_f-4q_e}.
\end{align*}
By $\neg$(C-14), $\forall\bft,\bft',x_5(\bft,P)\ne x_2(\bft',P)$. So
\begin{align*}
& \Pr[\exists\bft,\bft':x_4(\bft,P)=x_3(\bft',P)]  \\
= & \Pr[\con_2\xor \underline{P(x_5(\bft,P))}    \\
&\codeindent\codeindent\codeindent\codeindent=\ga_3(k_{\D'})\xor R'\xor\wf_1(k_{\D'})\xor \underline{P(x_2(\bft',P))}]   \\
\leq & \frac{q_e^2}{N-q_f-4q_e}, \text{ and }   \\
& \Pr[\exists\bft,\bft':y_3(\bft,P)=y_2(\bft',P)]    \\
%%	= & \Pr[X(\bft,P)\xor S\xor\wf_2(k\xor\D)\xor P(x_5(\bft,P))=P(x_2(\bft',P))]    \\
= & \Pr[P(x_5(\bft,P))=\con_3\xor P(x_1(\bft))\xor \underline{P(x_2(\bft',P))}]   \\
\leq & \frac{q_e^2}{N-q_f-4q_e}.
\end{align*}
By $\neg$(C-14), $\forall\bft,\bft',x_5(\bft,P)\ne x_6(\bft')$. So
\begin{align*}
&  \Pr[\exists\bft,\bft':x_4(\bft,P)=x_5(\bft',P)]  \\
= & \Pr[\con_2\xor\underline{P(x_5(\bft,P))}    \\
& \codeindent\codeindent\codeindent\codeindent =\ga_5(k_{\D'})\xor T'\xor\wf_3(k_{\D'})\xor\underline{P(x_6(\bft'))}]   \\
\leq & \frac{q_e^2}{N-q_f-4q_e}, \text{ and }   \\
&  \Pr[\exists\bft,\bft':y_3(\bft,P)=y_6(\bft',P)]    \\
%	= &  \Pr[X(\bft,P)\xor S\xor\wf_2(k\xor\D)\xor P(x_5(\bft,P))=P(x_6(\bft'))]    \\
= & \Pr[P(x_5(\bft,P))=\con_3\xor P(x_1(\bft))\xor \underline{P(x_6(\bft'))}]   \\
\leq & \frac{q_e^2}{N-q_f-4q_e}.
\end{align*}
By $\neg$(C-14), $\forall\bft,\bft',x_5(\bft,P)\ne x_2(\bft',P)$ and $x_5(\bft,P)\ne x_6(\bft')$. So
\begin{align*}
&  \Pr[\exists\bft,\bft':y_3(\bft,P)=y_4(\bft',P)]   \\
%	= & \Pr[X(\bft,P)\xor S\xor\wf_2(k\xor\D)\xor P(x_5(\bft,F))    \\
%	 & \text{\ \ \ \ }=R'\xor\wf_1(k\xor\D')\xor P(x_2(\bft',P))\xor T'\xor\wf_3(k\xor\D')\xor P(x_6(\bft'))]    \\
= & \Pr[P(x_5(\bft,F))=\con_3\xor P(x_1(\bft))\xor \big(R'\xor\wf_1(k_{\D'})    \\
& \text{\ \ \ \ \ \ \ \ \ \ \  }\xor\underline{P(x_2(\bft',P))}\big)\xor \big(T'\xor\wf_3(k_{\D'})\xor\underline{P(x_6(\bft'))}\big)]    \\
\leq  & \frac{q_e^2}{N-q_f-4q_e}
\end{align*}

Finally consider $\Pr[\exists\bft,\bft':y_3(\bft,P)=y_5(\bft',P)]$. If it happens then we have
\begin{align}
& L\xor\wf_0(k_{\D})\xor P(x_1(\bft))\xor S\xor\wf_2(k_{\D})\xor \underline{P(x_5(\bft,P))}    \notag   \\
= & \underline{P(x_5(\bft',P))}.  \label{eq:analysis-C24-EY3-cap-EY5-6-rnd}
\end{align}
If $x_5(\bft,P)\ne x_5(\bft',P)$ then the right hand side of (\ref{eq:analysis-C24-EY3-cap-EY5-6-rnd}) is random given $P(x_5(\bft,P))$ and $\Pr[\text{Eq. }(\ref{eq:analysis-C24-EY3-cap-EY5-6-rnd})]\leq\frac{1}{N-q_f-2q_e}$; otherwise we reach $L\xor\wf_0(k\xor\D)\xor S\xor\wf_2(k\xor\D)=P(x_1(\bft))$, contradicting $\neg$(C-15). So
\begin{align}
\Pr[\exists\bft,\bft':y_3(\bft,P)=y_5(\bft',P)]\leq\frac{q_e^2}{N-q_f-2q_e}. \label{eq:argument-C24-6-rounds-subeve-5}
\end{align}

In all, we have
$$\Pr[\text{(C-24)}]\leq\frac{2q_e(q_f+5q_e)}{N-q_f-4q_e},$$
and further        {\footnotesize
\begin{align}
{\Pr}[P\xleftarrow{\$}\permutationset:\badfii(P)\mid P\vdash\mathcal{Q}_P\wedge\neg\badfi(P)]\leq\frac{4q_eq_f+20q_e^2}{N-q_f-4q_e}.
\label{kaf-6-rnd-bad-f-2-prob}
\end{align}
}%
Define $\badf(P)=\badfi(P)\vee\badfii(P)$.
Then Eq. (\ref{kaf-6-rnd-bad-f-1-prob}) and (\ref{kaf-6-rnd-bad-f-2-prob}) yield	{\small
\begin{align}
{\Pr}[P\xleftarrow{\$}\permutationset:\badf(P)\mid P\vdash\mathcal{Q}_P]\leq\frac{6q_eq_f+27q_e^2+2q_e}{N-q_f-4q_e}.   \label{kaf-6-rnd-bad-f-total-prob}
\end{align}
}

\noindent{\it \underline{$2q_e$ Equations}.}
Similarly to the 4-round case, we show
\begin{align}
& {\Pr}_P[\textsf{RK}[\kafwsp_k]\vdash\mathcal{Q}_E\mid P\vdash\mathcal{Q}_P\wedge\neg\badf(P)]\geq\frac{1}{N^{2q_e}}.
\notag
\end{align}
Here $\neg\badf(P)$ indicates
\begin{itemize}
	\item $\forall\bft\in\mathcal{Q}_E,i=3,4$, $x_i(\bft,P)\notin\dom$, $y_i(\bft,P)\notin\rng$, and
	\item $\{x_i(\bft,P)\text{ }\big|\text{ }i=3,4,\bft\in\mathcal{Q}_E\}\cap\{x_j(\bft,P)\text{ }\big|\text{ }j=1,2,5,6,\bft\in\mathcal{Q}_E\}=\emptyset$, and
	\item $\{y_i(\bft,P)\text{ }\big|\text{ }i=3,4,\bft\in\mathcal{Q}_E\}\cap\{y_j(\bft,P)\text{ }\big|\text{ }j=1,2,5,6,\bft\in\mathcal{Q}_E\}=\emptyset$, and
	\item $\forall\bft,\bft'\in\mathcal{Q}_E,x_3(\bft,P)\ne x_4(\bft',P),y_3(\bft,P)\ne y_4(\bft',P)$.
\end{itemize}
By an analysis similar to subsection \ref{sec:ratio-4-rounds}, we only need to show
$$\Big|\big\{x_i(\bft,P)\text{ }\big|\text{ }\bft\in\mathcal{Q}_E\big\}\Big|=\Big|\big\{y_i(\bft,P)\text{ }\big|\text{ }\bft\in\mathcal{Q}_E\big\}\Big|=q_e$$
for $i=3,4$. For this, we argue $\bft\ne\bft'\Rightarrow x_3(\bft,P)\neq x_3(\bft',P)\text{ and }y_4(\bft,P)\neq y_4(\bft',P)$ for any $\bft=(\D,LR,ST)$ and
$\bft'=(\D',L'R',S'T')$:

\arrangespace

\noindent \underline{Case 1:} $\bft$ and $\bft'$ are such that
$R\xor R'=(M_1^{(w)}\xor M_1)\cdot(\D\xor\D')$ and $L\xor
L'=(M_0^{(w)}\xor M_2)\cdot(\D\xor\D')$. By the definition of
$\varphi_0,\varphi_1,\ga_1$, and $\ga_2$, the former implies
$\varphi_1(k\xor\D)\xor R=\varphi_1(k\xor\D')\xor R',$
i.e. $x_1(\bft)=x_1(\bft')$; and the latter further
implies
\begin{align*}
& \ga_2(k\xor\D)\xor L\xor\wf_0(k\xor\D)\xor
P(x_1(\bft))    \\
= & \ga_2(k\xor\D')\xor L'\xor\wf_0(k\xor\D')\xor
P(x_1(\bft)),
\end{align*}
i.e. $x_2(\bft,P)=x_2(\bft',P)$. And it
necessarily be $\D\neq\D'$, otherwise $\D\xor\D'=0$ and thus
$R=R'$ and $L=L'$ and $\bft=\bft'$, a contradiction. Then:
\begin{itemize}
	\item $x_3(\bft,P)\ne x_3(\bft',P)$, otherwise it
	implies   {%\small
		\begin{align*}
	 & \ga_3(k\xor\D)\xor
		R\xor\wf_1(k\xor\D)\xor
		P(x_2(\bft,P))    \\
		= & \ga_3(k\xor\D')\xor
		R'\xor\wf_1(k\xor\D')\xor P(x_2(\bft',P)).
		\end{align*}
	}Then we have
	\begin{align*}
	\underbrace{\ga_3(k\xor\D)\xor\ga_3(k\xor\D')}_{=M_3\cdot\D\xor M_3\cdot\D'}&=R\xor R'\xor
	M_1^{(w)}\cdot(\D\xor\D')    \\
	&=M_1\cdot(\D\xor\D'),
	\end{align*}
	thus
	$M_3\cdot(\D\xor\D')=M_1\cdot(\D\xor\D')$,
	contradicting condition
	(\ref{condition:6-rnd-good-ks-13-46-interference}) in
	Definition
	\ref{definition:good-linear-key-schedule-6-rnd-KAFw}
	({\it good key-schedule for 6 rounds});
	\item $y_4(\bft,P)\ne y_4(\bft',P)$. Because the
	assumption on $R\xor R'$ implies $S\xor S'\neq
	(M_6\xor M_2^{(w)})\cdot(\D\xor\D')$ by $\neg$(B-3). By $\neg$(C-13) we further have $A(\bft,P)\xor
	A(\bft',P)\neq M_1\cdot(\D\xor\D')$. However, in this case, it necessarily be
	\begin{align*}
	Y(\bft,P)\xor Y(\bft',P) & =R\xor R'\xor\wf_1(k_{\D})\xor \wf_1(k_{\D'})      \\
	& =M_1\cdot(\D\xor\D').
	\end{align*}
	Therefore, we must have
	$Y(\bft,P)\xor Y(\bft',P)\neq A(\bft,P)\xor A(\bft',P)$, i.e. $y_4(\bft,P)\ne y_4(\bft',P)$;
	\label{label:the-place-of-using-B3-6-rnd}
\end{itemize}

\arrangespace

\noindent \underline{Case 2:} for $(\bft,\bft')$, $x_1(\bft)=x_1(\bft')$, yet $x_2(\bft,P)\ne x_2(\bft',P)$. Then by $\neg$(C-21) we immediately have $x_3(\bft,P)\ne x_3(\bft',P)$ and $y_4(\bft,P)\ne y_4(\bft',P)$;

\arrangespace

\noindent \underline{Case 3:} for $(\bft,\bft')$, $x_1(\bft)\ne x_1(\bft')$. This implies $x_2(\bft,P)\ne x_2(\bft',P)$ by $\neg$(C-11), and further $x_3(\bft,P)\ne x_3(\bft',P)$ and $y_4(\bft,P)\ne y_4(\bft',P)$ by $\neg$(C-21).

\arrangespace

So
$\big|\big\{x_3(\bft,P)\text{ }\big|\text{ }\bft\in\mathcal{Q}_E\big\}\big|=\big|\big\{y_4(\bft,P)\text{ }\big|\text{ }\bft\in\mathcal{Q}_E\big\}\big|=q_e$. The argument
for $x_4(\bft,P)$ and $y_3(\bft,P)$ is similar
by symmetry (utilizing the property $M_4\cdot\D\neq
M_6\cdot\D$ for $\D\neq0$ given in Definition
\ref{definition:good-linear-key-schedule-6-rnd-KAFw} and the
condition $\neg$(C-13)). By all the above discussion and
(\ref{kaf-6-rnd-bad-f-total-prob}), for any good $\tau$, when
$q_f+4q_e\leq N/2$, via a counting similar to that in the
previous section we reach

{\footnotesize
	\begin{align*}
	\frac{{\Pr}_{re}(\tau)}{{\Pr}_{id}(\tau)}
	\geq  &  	\frac{\Pr[P\vdash\mathcal{Q}_P]}{N^{2q_e}}  \bigg( 1-
	\frac{6q_eq_f+27q_e^2+2q_e}{N-q_f-4q_e} \bigg)  \Bigg/ \frac{\Pr[P\vdash\mathcal{Q}_P]}{(N^2-q_e)^{q_e}}         \\
	\geq  &   \left(1-\frac{12q_eq_f+54q_e^2+4q_e}{N}\right)
	\left(\frac{N^2-q_e}{N^2}\right)^{q_e}   \\
	\geq  &  1 -
	\frac{12q_eq_f+55q_e^2+4q_e}{N}.
	\end{align*}
}Gathering this and
(\ref{bad-transcripts-6-rnd-probability-kafsp}) and Lemma
\ref{lemma:h-coefficients-main-lemma} yields Theorem \ref{theorem:KAFw-rka-security-6-round}.
% Save a math below
%  {\small
%	\begin{align*}
%	\frac{{\Pr}_{re}(\tau)}{{\Pr}_{id}(\tau)}
%	\geq  &  \frac{	\frac{1}{(N)_{q_f}\cdot N^{2q_e}}  \bigg( 1-
%		\frac{6q_eq_f+27q_e^2\revision{+2q_e}}{N-q_f-4q_e} \bigg)}{\frac{1}{(N)_{q_f}} \bigg(   \frac{1}{N^2-q_e}    \bigg)^{q_e}  }    \\
%	\geq  &   \left(1-\frac{12q_eq_f+54q_e^2\revision{+4q_e}}{N}\right)
%	\left(\frac{N^2-q_e}{N^2}\right)^{q_e}   \\
%	\geq  &  1 -
%	\frac{12q_eq_f+55q_e^2\revision{+4q_e}}{N}.
%	\end{align*}
%}

\subsection{When f=F is a Random Function}
\label{subsect:app-proof-sketch-kafwsf-6-rounds}

For the proof, we need the following modifications on the proof for 6-round \kafwsp:
\begin{enumerate}
	\item in Definition
	\ref{definition:bad-transcripts-linear-ks-6-rnd-KAFwsp}
	(bad transcripts), (B-3) is only used
	for proving $\forall(\bft,\bft'):y_3(\bft,F)\ne y_3(\bft',F),y_4(\bft,F)\ne y_4(\bft',F)$, cf.
	page \pageref{label:the-place-of-using-B3-6-rnd}. We
	thus drop it and obtain
	\begin{align}
	\Pr[T_{id}\in\mathcal{T}_{bad}]\leq\frac{2q_eq_f+q_e^2}{N};  \label{bad-tau-prob-6-round-kafwsf}
	\end{align}
	
	\item in the definition of $\badfi(F)$, the two
	conditions $X(\bft,F)\xor X(\bft',F)=M_6\cdot(\D\xor\D')$ in (C-11)
	and $A(\bft,F)\xor A(\bft',F)=M_1\cdot(\D\xor\D')$ in (C-13) are only
	used for proving $\forall(\bft,\bft'):y_3(\bft,F)\ne y_3(\bft',F),y_4(\bft,F)\ne y_4(\bft',F)$, cf.
	page \pageref{label:the-place-of-using-B3-6-rnd}. In addition, (C-15) is only used for proving $\forall(\bft,\bft'):y_2(\bft,F)\ne y_4(\bft',F),y_3(\bft,F)\ne y_5(\bft',F)$, cf. page \pageref{label:use-of-C15-6-rounds}. We
	thus drop them, which decreases
	$\Pr[\badfi(F)]$ to
	$\frac{2q_eq_f+6q_e^2}{N}$;
	
	\item in the definition of $\badfii(F)$, we drop
	\begin{itemize}
		\item $y_4(\bft,F)=y_4(\bft',F)$ in (C-21),
		and
		\item $\exists\bft,\bft':y_4(\bft,F)\in\rng\text{ or }y_4(\bft,F)\in\{y_1(\bft',F),$ $y_2(\bft',F),y_5(\bft',F),y_6(\bft',F)\}$ in
		(C-22), and
		\item $y_3(\bft,F)=y_3(\bft',F)$ in (C-23),
		and
		\item $\exists\bft,\bft':y_3(\bft,F)\in\rng\text{ or }y_3(\bft,F)\in\{y_1(\bft',F),$ $y_2(\bft',F),y_4(\bft',F),y_5(\bft',F),y_6(\bft',F)\}$ in
		(C-24).
	\end{itemize}
	These decrease $\Pr[\badfii(F)]$ to
	$\frac{2q_eq_f+10q_e^2}{N}$.
\end{enumerate}
Therefore,
\begin{align}
\frac{{\Pr}_{re}(\tau)}{{\Pr}_{id}(\tau)}\geq1-\frac{4q_eq_f+16q_e^2}{N}-\frac{q_e^2}{N^2}.  \label{ratio-6-round-kafwsf}
\end{align}
Gathering (\ref{bad-tau-prob-6-round-kafwsf}) and (\ref{ratio-6-round-kafwsf}) gives rise to the following Theorem.

\begin{theorem}
	\label{theorem:KAFw-rf-variant-rka-security-6-round}
	
	For the 6-round, random function-based \kafwsf cipher
	with a good key-schedule $(\wf,\ga)$ as specified in Definition
	\ref{definition:good-linear-key-schedule-6-rnd-KAFw}, it holds
	$$\mathbf{Adv}^{\xor\text{-rka}}_{\kafwsf_k}(q_f,q_e)\leq\frac{6q_eq_f+18q_e^2}{N}.$$
\end{theorem}

% ==================================================================

\section{Deriving Results on \kaf~and \kafv~Ciphers}
\label{section:to-kaf-and-kafv}

\subsection{Results on \kaf}

Since \kaf~ciphers are \kafw~ciphers with no whitening keys,
results on the latter can be immediately transposed to the
former. In detail, denote by $\ga=(\ga_1,\ldots,\ga_t)$ a
$t$-round key-schedule of \kaf, then with the function $\Psi_{k}^{f}$ defined by Eq. (\ref{eq:kafw-1-round-transformation}) in section \ref{section:preliminary}, the
$t$-round \kafsg~cipher is defined as
$$\kafsg_k(W)=\Psi_{\ga_t(k)}^{f}\circ\ldots\circ\Psi_{\ga_1(k)}^{f}(W).$$
Setting $\varphi_i=\ga_i$ for $i=1,4,6$ in Definitions
\ref{definition:good-non-linear-ks-4-rnd-KAFw} and
\ref{definition:good-linear-key-schedule-6-rnd-KAFw} (good
key-schedules for \kafwsg) yields the corollary.

\begin{corollary}
	\label{corollary:good-key-schedule-for-kaf}
	
	A 4-round non-linear key-schedule
	$\ga=(\ga_1,\ga_2,\ga_3,\ga_4)$ is good for \kafsg,
	if $\varphi_1=\ga_1$ and $\varphi_4=\ga_4$ satisfy the uniformness and AXU conditions defined in
	Definition \ref{definition:good-non-linear-ks-4-rnd-KAFw}.
	
	A 6-round affine key-schedule $\ga=(\ga_1,\ldots,\ga_6)$,
	where $\ga_i(k)=M_i\cdot k\xor C_i$, is
	good for \kafsg, if:
	\begin{enumerate}
		\item $\ga_1$, $\ga_6$, and
		$\ga_1\oplus\ga_6$ are bijective maps of $\Zn$, and
		\item for
		any $\D\neq0$, \underline{$M_1\cdot\D\neq M_3\cdot\D$, $M_4\cdot\D\neq M_6\cdot\D$}.
	\end{enumerate}
	With such good key-schedules, 4- and 6-round idealized \kafsp~and \kafsf~ensure the same security bounds as described in Theorems \ref{theorem:KAFwsp-rka-security-4-round}, \ref{theorem:KAFw-rf-variant-rka-security-4-round}, \ref{theorem:KAFw-rka-security-6-round}, and \ref{theorem:KAFw-rf-variant-rka-security-6-round}.
\end{corollary}
For affine schedules, both Corollary \ref{corollary:good-key-schedule-for-kaf} and
Definition \ref{definition:good-linear-key-schedule-6-rnd-KAFw}
require the ``inner'' KDFs $\ga_3$ and $\ga_4$ to fulfill some conditions, i.e.
$M_1\cdot\D\neq M_3\cdot\D$ and $M_4\cdot\D\neq M_6\cdot\D$.
This means for \kaf~instances that suffer from $\xor$-RKAs,
adding whitening keys derived by affine KDFs would probably
{\it not} be beneficial for RKA security (since the ``inner''
KDFs remain ``bad''). For example, consider the attempt to prevent DES from the complementation property via using a DESX-like~\cite{FXJoC01} structure $\text{DESX}_{k}^*(M)=k\xor\text{DES}_{k'}(k\xor M)$, where the 56-bit DES key $k'$ are 56 bits chosen from the 64-bit master-key $k$. It can be seen while $\text{DESX}_{\overline{k}}^*(\overline{M})=\overline{\text{DESX}_{k}^*(M)}$ does not necessarily hold, the $\text{DESX}_{k}^*(M)$ still suffers from a (less trivial) complementation-based property $\text{DESX}_{\overline{k}}^*(M)=\text{DESX}_{k}^*(M)$.

\subsection{Results on \kafv}
\label{subsec:kafv-results}

The transition to \kafv~is a bit more complicated. Formally, \kafv~relies on the following round transformation
\begin{align}
\widetilde{\Psi}_{k}^{f}(W_L\|W_R)=W_R\|W_L\oplus\underline{f(W_R)\xor k}.
\label{eq:round-transformation-for-kafv}
\end{align}
With this, a $t$-round \kafv~needs $t+2$ sub-keys. To make a clear distinction from the notations for \kaf, we denote by
$\ga^*=(\ga^*_0,\ga^*_1,\ldots,\ga^*_t,\ga^*_{t+1})$ a
$t$-round key-schedule for \kafv: $\ga^*_1,\ldots,\ga^*_t$ for the $t$ round-keys, while $\ga^*_0$ and $\ga^*_{t+1}$ for the
two whitening keys. Then the entire \kafvsg~variant is  

{\footnotesize
\begin{align*}
\kafvsg_k(W)
=(\ga_{t+1}^*(k)\|0)\xor\widetilde{\Psi}_{\ga_t^*(k)}^{f}\circ\ldots\circ\widetilde{\Psi}_{\ga_1^*(k)}^{f}\big((0\|\ga_0^*(k))\xor W\big).
\end{align*}     }%
%\begin{align*}
%&\kafvsf_{k}^{F}(W_L\|W_R)  \\
%=&(\ga_{t+1}^*(k)\|0)\xor\widetilde{\Psi}_{\ga_t^*(k)}^{F}\circ\ldots\circ\widetilde{\Psi}_{\ga_1^*(k)}^{F}((W_L\|W_R)\xor(0\|\ga_0^*(k))).
%\end{align*}

For \kafvsg~we have
%%%%$$\kafvsf_{k,\ga^*}^{F}(W)=\kafwsf_{k,(\wf,\ga)}^{F}(W).$$

\begin{corollary}
	\label{corollary:good-key-schedule-for-kafv}
	
	A 4-round non-linear key-schedule
	$\ga^*=(\ga_0^*,\ldots,\ga_5^*)$ is good for \kafvsg, if $\varphi_1=\ga_0^*$ and $\varphi_4=\ga_5^*$ satisfy the two conditions defined in Definition
	\ref{definition:good-non-linear-ks-4-rnd-KAFw}.

	A 6-round affine key-schedule $\ga^*=(\ga_0^*,\ldots,\ga_7^*)$,
	where $\ga_i^*(k)=M_i^*\cdot k\xor C_i^*$, is good for
	\kafvsg, if:
	\begin{enumerate}
		\item $\ga_0^*$, $\ga_7^*$, and $\ga_0^*\oplus\ga_7^*$
		are bijective maps of $\Zn$, and
		\item for any $\D\neq0$, \underline{$M_2^*\cdot\D\neq0$,
			$M_5^*\cdot\D\neq0$}.
	\end{enumerate}
	With such good key-schedules, 4- and 6-round idealized \kafvsp~and \kafvsf~ensure the same security bounds as described in Theorems \ref{theorem:KAFwsp-rka-security-4-round}, \ref{theorem:KAFw-rf-variant-rka-security-4-round}, \ref{theorem:KAFw-rka-security-6-round}, and \ref{theorem:KAFw-rf-variant-rka-security-6-round}.
\end{corollary}
\begin{IEEEproof}
	For a $t$-round \kafv~key-schedule $\ga^*$, define a
	$t$-round \kafw~schedule $(\wf,\ga)$ as follows:
	\begin{itemize}
		\item $\ga_{2\ell+1}=\bigoplus_{i=0}^{\ell}\ga_{2i}^*$, where
		$\ell=0,\ldots,\lfloor\frac{t-1}{2}\rfloor$, and
		\item $\ga_{2\ell+2}=\bigoplus_{i=0}^{\ell}\ga_{2i+1}^*$, where
		$\ell=0,\ldots,\lfloor\frac{t-2}{2}\rfloor$, and
		\item $\wf_2=\ga_t\xor\ga_{t+1}^*$,
		$\wf_3=\ga_{t-1}\xor\ga_t^*$, while
		$\wf_0(k)=\wf_1(k)=0$.
		%%%  \item $\wf_2(k)=$ and $\wf_3(k)=0$ when $t$ is odd, and
		%%%      $\wf_2(k)=$ and $\wf_3(k)=0$ when $t$ is even.
	\end{itemize}
	Then it can be seen a $t$-round \kafv~with the key-schedule
	$\ga^*$ is a \kafw~instance with $(\wf,\ga)$, i.e.
	\begin{align}
	\kafvsg_k(W)=\kafwsg_k(W).
	\label{eq:kafv-to-kafw-transform}
	\end{align}
	Concretely, the 4-round \kafvsg~schedule
	$(\ga_0^*,\ldots,\ga_5^*)$ corresponds to the 4-round
	\kafwsg~schedule $(\wf,\ga)$ with $\wf_0(k)=\wf_1(k)=0$,
	$\ga_1=\ga_0^*$ (thus $\varphi_1=\wf_1\xor\ga_1=\ga_0^*$), and
	$\varphi_4=\ga_4\xor\wf_2=\ga_5^*$. The first half of the corollary thus
	follows from Definition
	\ref{definition:good-non-linear-ks-4-rnd-KAFw}.

	The 6-round affine key-schedule
	$\ga^*=(\ga_0^*,\ldots,\ga_7^*)$ corresponds to the 6-round
	affine schedule $(\wf,\ga)$, in which
	$\varphi_1=\wf_1\xor\ga_1=\ga_0^*$,
	$\varphi_6=\ga_6\xor\wf_2=\ga_7^*$, and:
	\begin{enumerate}
		\item $\ga_1\xor\ga_3=\ga_2^*$, and thus
		$M_2^*=M_1\xor M_3$;
		\item $\ga_4\xor\ga_6=\ga_5^*$, and thus
		$M_5^*=M_4\xor M_6$.
		\label{condition:6-rnd-good-ks-2-5-non-zero-kafv}
	\end{enumerate}
	Therefore, the second half follows from Definition
	\ref{definition:good-linear-key-schedule-6-rnd-KAFw}.
\end{IEEEproof}

We believe the requirements on schedules of \kafvsg~are more
relaxed than those required by \kafsg~(Corollary \ref{corollary:good-key-schedule-for-kaf}), since its condition
(\ref{condition:6-rnd-good-ks-2-5-non-zero-kafv}) only requires
to carefully design $\ga_2$ and $\ga_5$, without considering
the more complicated interactions between different round-KDFs
(comparing with the second condition in Definition \ref{definition:good-linear-key-schedule-6-rnd-KAFw}). In
particular, when designing affine key-schedules in practice,
one tends to choose {\it invertible	matrices} for $M_0,\ldots,M_{t+1}$ in order to ensure the largest possible amount of entropy in the round-keys, e.g. the bit-permutation-based key-schedules in DES. In this case, condition
(\ref{condition:6-rnd-good-ks-2-5-non-zero-kafv}) is naturally
satisfied, yet the second condition in Definition
\ref{definition:good-linear-key-schedule-6-rnd-KAFw} may not be
satisfied! (And when $M_1\cdot k$ and $M_3\cdot k$ define two
bit-permutations, the latter condition is indeed violated since $M_1\cdot\D=M_3\cdot\D=\D$ for $\D=\text{0xFF}\ldots\text{FF}$. This matches that DES is vulnerable to complementing attacks.)

Finally, we remark that whitening keys play a crucial role in the transformation Eq. (\ref{eq:kafv-to-kafw-transform}). This means \kafv---as well as the Lucifer-like model---cannot
be precisely captured by \kaf, the variant of \kafw~without whitening keys.

%
%For example, a very natural choice suffices for
%\kafvsf, i.e. taking $\ga_i(k)=k\xor C_i$ for 7 round constants
%$C_0,\ldots,C_6$, and $\ga_7=\pi$ for a linear orthomorphism
%$\pi$. On the other hand, its analogues for 6-round \kafsf,
%e.g. $(k\xor C_1,\cdot,k\xor C_3,\cdot,\cdot,\pi(k)\xor C_6)$,
%or $(k\xor C_1,\cdot,\cdot,\pi(k)\xor C_4,\cdot,\pi(k)\xor
%C_6)$ (where $\cdot$ could be any affine KDF), are {\it not}
%good, and indeed allow complementing attacks.

\section{Towards Minimalism}
\label{sect:towards-minimizing}

To maximize the efficiency of the resulted permutation modes, we derive theoretically ``minimal'' constructions. We focus on \kafsp~as it's of the most general interest, and it's wlog, since minimal \kafwsp~and \kafvsp~schemes can be easily derived similarly.

First, for the 4-round \kafsp, $\ga_1(k)=\mathbf{M}_1\otimes k\xor k^3$, $\ga_2(k)=\ga_3(k)=0$, and $\ga_4(k)=\mathbf{M}_4\otimes k\xor k^3$ is a group of good choices, where $\mathbf{M}_1\neq\mathbf{M}_4$ are two non-zero constants in $\Zn$, and $\otimes$ denotes multiplications taken over the finite field $\mathbb{F}_{2^n}$. With this choice, it can be seen the three parameters mentioned in Definition \ref{definition:good-non-linear-ks-4-rnd-KAFw} are such that $\delta_1\leq3/N$, $\delta_2\leq2/N$, and
$\delta_3\leq2/N$, and the concrete advantage bound is a
classical birthday one $\frac{14q_eq_f+31q_e^2+4q_e}{N}$.

Our choice of $\ga_1$ and $\ga_4$ is motivated by~\cite{RKAXUFSE16}. On the other hand, since no requirement is placed on $\ga_2$ and $\ga_3$ (see Corollary \ref{corollary:good-key-schedule-for-kaf} or Definition \ref{definition:good-non-linear-ks-4-rnd-KAFw}), they are completely absent: this matches the existing result that the two middle round-functions of 4-round Feistel do {\it not} need to be secret/``protected'' by round-keys~\cite{RR00feistelPub}. This \kafsp~variant seems ``minimal'' in the sense that removing any component harms security: reducing rounds ruins CCA security, choosing
$\mathbf{M}_1=\mathbf{M}_4$ introduces the weakness
$\kafsp_k(LR)=ST\Leftrightarrow\kafsp_k(TS)=RL$ and allows
trivially distinguishing, while reducing the non-linearity of KDFs would introduce related-key differentials with higher probability and compromise the concrete security.

Second, for 6-round \kafsp, using a linear orthomorphism $\pi$, the key-schedule $k\mapsto(k,0,0,0,0,\pi(k))$ is sufficient. It may be quite hard to believe many carefully designed sophisticated key-schedules (e.g. DES) are insufficient to prevent complementing attacks, while such an exotic design should be good. The reason is that the absence of the 3rd and 4th round-keys incidently prevents the complementation properties.

We stress that the key-schedule instances with many ``blanks'' mentioned here \textit{are for theoretically minimalism rather than for general purpose Feistel ciphers.} For the latter purpose, one could (actually, {\it should}) ``fill in the blanks''. For example, using $\pi(k_L\|k_R)=k_R\|k_L\xor k_R$ mentioned in the Introduction, it can be seen $k\mapsto(k,k,\pi(k),k,k,\pi(k))$ is a good key-schedule for 6-round \kafsp.

\subsubsection{{\bf A Tweakable \kac}}
\label{sec:TEM-variant}
%\noindent{\textsc{.}}
Finally, in 4-round \kafwsg, we can set $\wf_1(k)=\mathbf{M}_1\otimes k\xor k^3$ and $\wf_2(k)=\mathbf{M}_4\otimes k\xor k^3$, while omit all the other sub-keys. This results in a variant of the 1-round tweakable \kac~of~\cite{seqiffSKA4EUROCRYPT15}, with the permutation instantiated by a 4-round keyless Feistel network.

% ==================================================================

\section{Conclusion}

We've studied provable security of key-alternating Feistel/Feistel-2 variants against $\xor$-induced related-key attacks, which better model the reality of Feistel blockciphers. Assuming key-schedules being non-linear or purely affine, we identify (different) conditions on the key-schedules that are sufficient for a birthday-type security up to $2^{n/2}$ queries. The results and implications make a step towards understanding the behaviors of existing different Feistel cipher structures, and offer new insights.

\appendices

\section{Lucifer-like Model and \kafv}
\label{subsec:lucifer-to-kafv}

The Lucifer-like model \lucifer~also relies on the round transformation $\widetilde{\Psi}_{k}^{f}$ in Eq. (\ref{eq:round-transformation-for-kafv}). With this, a $t$-round \lucifer~model built upon $t$ round-functions $f_1,\ldots,f_t$ uses $t$ round-keys $k_1,\ldots,k_t$, and is
\begin{align}
\lucifer_{k_1,\ldots,k_t}^{f_1,\ldots,f_t}(W)
=\widetilde{\Psi}_{\ga_t^*(k)}^{f_t}\circ\ldots\circ\widetilde{\Psi}_{\ga_1^*(k)}^{f_1}(W).
\label{eq:lucifer-entire}
\end{align}
From section \ref{subsec:kafv-results} we know a $(t\text{--}2)$-round \kafv~uses $t-2$ round-functions $f_1,\ldots,f_{t-2}$ and $t$ sub-keys $k_1,\ldots,k_t$:     {\small
\begin{align*}
\kafv_{k_1,\ldots,k_t}^{f_1,\ldots,f_{t-2}}(W)
=(k_t\|0)\xor\widetilde{\Psi}_{k_{t-1}}^{f_{t-2}}\circ\ldots\circ\widetilde{\Psi}_{k_2}^{f_1}\big((0\|k_1)\xor W\big).
\end{align*}
}%
By these, it's not hard to see when $t\ge2$,
\begin{align*}
\lucifer_{k_1,\ldots,k_t}^{f_1,\ldots,f_t}(W)=\widetilde{\Psi}_{0}^{f_t}\circ\kafv_{k_1,\ldots,k_t}^{f_2,\ldots,f_{t-1}}\circ\widetilde{\Psi}_{0}^{f_1}(W),
\end{align*}
where $\widetilde{\Psi}_{0}^{f_1}$ and $\widetilde{\Psi}_{0}^{f_t}$ are two keyless permutations that can be freely evaluated by the adversary. It can be seen within a large range, any CCA attack $\mathcal{A}$ on $(t\text{--}2)$-round \kafv~can be turned into a CCA attack $\mathcal{A}'$ on $t$-round \lucifer: whenever $\mathcal{A}$ queries $\textsf{RK}[\kafv_k^{f_2,\ldots,f_{t-1}}](\D,LR)$, $\mathcal{A}'$ queries $\textsf{RK}[\lucifer_k^{f_1,\ldots,f_t}](\D,(\widetilde{\Psi}_{0}^{f_1})^{-1}(LR))$; whenever $\mathcal{A}$ queries $\textsf{RK}[\kafv_k^{f_2,\ldots,f_{t-1}}]^{-1}(\D,ST)$, $\mathcal{A}'$ queries $\textsf{RK}[\lucifer_k^{f_1,\ldots,f_t}]^{-1}(\D,\widetilde{\Psi}_{0}^{f_t}(ST))$. The formal characterization is out of the scope of this paper.

\section{Complementing Attacks}
\label{subsection:complementing-kaf-linear-kdf-any-rnd}

We don't claim novelty for these attacks, see~\cite{complementFeistelFSE2013}. We just include them to help understanding our provable results. We focus on \kafw~variants with key-schedules that do not
satisfy condition
(\ref{condition:6-rnd-good-ks-13-46-interference}) in
Definition
\ref{definition:good-linear-key-schedule-6-rnd-KAFw}. We first
brief how to break more than 4 rounds, then describe the attack
against any number of rounds for ``bad enough'' key-schedules.

\arrangelittlespace

\noindent{\bf On 5 Rounds.}	
Consider a 5-round key-schedule $(\wf,\ga)$, where
$\wf=(\wf_0,\wf_1,\wf_2,\wf_3)$ and
$\ga=(\ga_1,\ga_2,\ga_3,\ga_4,\ga_5)$, and $\ga$ is such that $M_1\cdot\D=M_3\cdot\D$ for a non-zero value $\D$. Then there exists a 1-round related-key differential for the 3rd round, i.e.
\begin{align}
\Pr\Big(M_2\cdot\D\|M_1\cdot\D\xrightarrow[\D]{\Psi_{\ga_3(k)}^{f_3}}M_1\cdot\D\|M_2\cdot\D\Big)=1.   \notag %\label{eq:differential-1-2-rnd}
\end{align}

Concatenating this differential with the mentioned 2-round related-key differential Eq. (\ref{eq:differential-1-2-rnd}) gives a 3-round differential:

{\footnotesize
\begin{align}	
\Pr\Big(\nabla_1\|\nabla_2\xrightarrow[\D]{\Psi_{\ga_3(k)}^{f_3}\circ\Psi_{\ga_2(k)}^{f_2}\circ\Psi_{\ga_1(k)}^{f_1}\circ\xornotation_{wk_{in}}}M_1\cdot\D\|M_2\cdot\D\Big)=1,     \notag %\label{eq:differential-1-2-rnd}
\end{align}
}%
where $\nabla_1=(M_0^{(w)}\xor M_2)\cdot\D$, $\nabla_2=(M_1^{(w)}\xor M_1)\cdot\D$, and $wk_{in}=\wf_0(k)\|\wf_1(k)$. Further concatenating this differential with the 2-round related-key differential Eq. (\ref{eq:differential-4-5-rnd}) yields a 5-round related-key boomerang distinguisher, which allows distinguishing 5 rounds with 4 queries.

\arrangelittlespace

\noindent{\bf On Any Rounds.}	
Consider a $2t$-round schedule $(\wf,\ga)$ with
$\wf=(\wf_0,\wf_1,\wf_2,\wf_3)$ and
$\ga=(\ga_1,\ga_2,\ldots,\ga_{2t})$, and $\ga$ satisfies: it's easy to derive $\D\neq0$ such that
\begin{itemize}
	\item
	$\D_1=M_1\cdot\D=M_3\cdot\D=M_5\cdot\D=\ldots=M_{2t-1}\cdot\D$,
	and
	\item
	$\D_2=M_2\cdot\D=M_4\cdot\D=M_6\cdot\D=\ldots=M_{2t}\cdot\D$.
\end{itemize}
Then it can be seen there exists related-key differentials with any number of rounds:
\begin{align*}
\Pr\Big(\nabla_1\|\nabla_2&\xrightarrow[\D]{\Psi_{\ga_1(k)}^{f_1}\circ\xornotation_{wk_{in}}}\D_1\|\D_2\xrightarrow[\D]{\Psi_{\ga_2(k)}^{f_2}}\D_2\|\D_1    \\
&\xrightarrow[\D]{\Psi_{\ga_3(k)}^{f_3}}\ldots\xrightarrow[\D]{\xornotation_{wk_{out}}\circ\Psi_{\ga_t(k)}^{f_t}}\delta\Big)=1,
\end{align*}
where $\nabla_1=(M_0^{(w)}\xor M_2)\cdot\D$, $\nabla_2=(M_1^{(w)}\xor M_1)\cdot\D$, $wk_{out}=\wf_2(k)\|\wf_3(k)$, the output difference $\delta=\D_2\xor M_2^{(w)}\cdot\D\|\D_1\xor M_3^{(w)}\cdot\D$ when $t$ is even, and $\delta=\D_1\xor M_3^{(w)}\cdot\D\|\D_2\xor M_2^{(w)}\cdot\D$ otherwise. This allows distinguishing any $t$ rounds with 2 queries. To save space we omit detailed
descriptions of these two (innovel) variants of complementing attacks.


\section*{Acknowledgements}
%%%%
I'd like to thank all the five anonymous reviewers of IEEE TIT and CRYPTO 2018 for carefully reading, identifying bugs and typos, supplying invaluable comments that significantly refine the presentations, and pointing the insights in section \ref{sec:TEM-variant} to me. As a post-doc paid by Fran\c{c}ois-Xavier Standaert by the ERC project SWORD (724725), I sincerely appreciate him for allowing (and encouraging) to complete this work.

\bibliographystyle{IEEEtran}
%	\begin{thebibliography}{1}

%		\bibitem{IEEEhowto:kopka}
%		H.~Kopka and P.~W. Daly, \emph{A Guide to \LaTeX}, 3rd~ed.\hskip 1em plus 0.5em minus 0.4em\relax Harlow, England: Addison-Wesley, 1999.

%	\end{thebibliography}

%\input{ref.tex}



%	\begin{IEEEbiography}{Michael Shell}
%		Biography text here.
%	\end{IEEEbiography}

\begin{IEEEbiographynophoto}{Chun Guo}
	was born in China in 1989. He received
	his BSc from East China Normal University
	in 2011 and his PhD from University of
	Chinese Academy of Sciences in 2017 respectively.
	His research interests include theoretical
	aspects of symmetric cryptography such as
	provable security, generic attacks, and leakage resilience.
\end{IEEEbiographynophoto}

%\newpage


%\vfill


\end{document}